\documentclass[aps,prd,reprint,superscriptaddress,nofootinbib,longbibliography,floatfix]{revtex4-2}

\usepackage{amsmath,amssymb,bm}
\usepackage{graphicx}
\usepackage{booktabs}
\usepackage{siunitx}
\usepackage{xcolor}
\usepackage{hyperref}
\usepackage{microtype}
\usepackage{enumitem}
\usepackage{multirow}
\hypersetup{colorlinks=true,linkcolor=blue,citecolor=blue,urlcolor=blue}
\newcommand{\revtext}[1]{#1}

\newcommand{\revtextC}[1]{#1}
\newcommand{\revtextD}[1]{#1}

\newcommand{\GeV}{\mathrm{GeV}}
\newcommand{\TeV}{\mathrm{TeV}}
\newcommand{\keV}{\mathrm{keV}}
\newcommand{\kms}{\mathrm{km\,s^{-1}}}
\newcommand{\cmsq}{\mathrm{cm}^{2}}
\newcommand{\cmcubeds}{\mathrm{cm}^{3}\,\mathrm{s}^{-1}}

\begin{document}

\title{Solar Capture Tests of Inelastic Dark Matter after the LZ High-Recoil Event}

\author{Mattia Di Mauro}
\email{dimauro.mattia@gmail.com}
\affiliation{Istituto Nazionale di Fisica Nucleare, Sezione di Torino, Via P. Giuria 1, 10125 Torino, Italy}

\author{Halim Shaikh}
\email{halim.shaikh@tum.de}
\affiliation{Technical University of Munich, TUM School of Natural Sciences, Department of Physics, James-Franck-Str. 1, 85748 Garching, Germany}

\date{September 8, 2026}

\begin{abstract}
The LUX-ZEPLIN (LZ) Collaboration has reported a $248\,\keV$ nuclear-recoil candidate for which endothermic dark matter (DM) gives some of the largest local significances. We study Solar-capture constraints on three interpretations: a thermal Higgsino, a thermal pseudo-Dirac fermion with off-diagonal vector interactions, and neutron-philic endothermic spin-dependent scattering through $\mathcal{O}_4$. The canonical full-density thermal Higgsino is excluded: its LZ-preferred splitting near $0.38\,\mathrm{MeV}$ lies well below the splitting required to suppress Solar capture sufficiently to satisfy the IceCube upper limits on DM annihilation in the Sun, $\delta\gtrsim0.51$--$0.56\,\mathrm{MeV}$. For the pseudo-Dirac benchmark parameters that fit the LZ event, we find $C_\odot=1.53\times10^{20}\,\mathrm{s}^{-1}$, while two-state kinetics limits the fixed-orbit annihilation rate to $\Gamma_A\lesssim4.4\times10^{14}\,\mathrm{s}^{-1}$, about $1.7\times10^5$ below the IceCube upper limit for the $b\bar b$ channel. A semi-analytic treatment indicates that re-excitation cycles further cool the captured population, although a full phase-space calculation is required for its final distribution. For neutron-philic $\mathcal{O}_4$ scattering at $m_\chi=1\,\TeV$ and $\delta=300\,\keV$, finite-temperature capture gives $C_\odot=3.98\times10^{17}\,\mathrm{s}^{-1}$, with nuclear-structure uncertainties giving an envelope $(1.77$--$9.72)\times10^{17}\,\mathrm{s}^{-1}$. Under the equilibrium assumption, the upper edge remains about a factor $154$ below the IceCube upper limit for the $b\bar b$ channel. We use $b\bar b$ only as a soft-hadronic proxy because an exact constraint requires the model-specific annihilation spectrum. Solar capture therefore excludes the thermal-Higgsino interpretation but not endothermic explanations generically.
\end{abstract}

\maketitle

\section{Introduction}
\label{sec:intro}

A broad set of astrophysical and cosmological observations demonstrates that most of the matter in the Universe is non-luminous and non-baryonic dark matter (DM) \cite{BertoneHooperSilk2005,BertoneHooper2018,CirelliStrumiaZupan2024,Planck2020}.  Measurements of galaxy rotation curves, gravitational lensing, galaxy clusters, the cosmic microwave background, and the growth of large-scale structure consistently require a matter component that interacts gravitationally but is not described by the known baryonic sector.  Cosmological data imply that DM accounts for approximately $85\%$ of the total matter density, while its microscopic nature remains unknown \cite{Planck2020,CirelliStrumiaZupan2024}.  No particle in the Standard Model (SM) has the required combination of abundance, stability, electrical neutrality, and sufficiently cold cosmological evolution, so particle DM is one of the clearest motivations for physics beyond the SM.  Comprehensive discussions of the particle-DM paradigm, its observational evidence, and the present experimental program can be found in Refs.~\cite{Jungman1996,Feng2010,Arcadi2018,Arcadi2025,CirelliStrumiaZupan2024}.

Weakly interacting massive particles (WIMPs) remain a particularly well-motivated class of candidates \cite{Jungman1996,Feng2010,Arcadi2018,Arcadi2025}.  In the standard thermal freeze-out picture, an annihilation cross section of approximately weak strength naturally produces a relic density close to the observed value, the well-known WIMP-miracle argument.  The same interactions can be tested through direct detection (DD), indirect detection (ID), and collider searches.  Present DD limits exclude large portions of the parameter space of simple WIMP benchmarks; this pressure, and the importance of special regions or interaction structures that suppress the laboratory rate, is especially transparent in recent studies of Higgs-portal, resonant, leptophilic, and broader WIMP constructions \cite{DiMauroArina2023,DiMauroXie2025,KoechlerDiMauro2025,ShaikhDiMauro2026,KongDiMauro2025}.  An important way of preserving thermal freeze-out while suppressing DD is secluded DM: the relic abundance is set mainly by annihilation into lighter dark-sector mediators, while the portal connecting those mediators to the SM can be much smaller than the coupling controlling freeze-out \cite{PospelovRitzVoloshin2008,DiMauroWang2025,DiMauroSecluded2025}.  This separation between the relic-density interaction and the laboratory scattering interaction provides a general route around stringent DD bounds.  Endothermic inelastic scattering, which is the focus of the present work, realizes a different but related strategy by suppressing the kinematically accessible low-energy recoil spectrum rather than simply making the SM portal very small \cite{TuckerSmith2001,TuckerSmith2005}.

The recent extended nuclear-recoil analysis of the LUX-ZEPLIN (LZ) experiment probes precisely such nonstandard signatures.  LZ extends the recoil-energy window to approximately $270\,\keV$ and reports one event with reconstructed nuclear-recoil energy
\begin{equation}
 E_R=248\pm23\,(\mathrm{stat})\pm23\,(\mathrm{sys})\,\keV
\end{equation}
in a $2.84$ tonne-year exposure \cite{LZ2026}.  The largest local significance among the interactions tested by LZ is $3.4\sigma$, reduced to a global significance of $2.6\sigma$ after the look-elsewhere effect.  The event is therefore not evidence for a DM discovery, but its unusually large recoil energy is theoretically informative because it lies far above the region that dominates ordinary elastic spin-independent (SI) WIMP searches.

An important feature of the LZ analysis is that it does not test only the standard elastic SI interaction.  The Collaboration considers elastic and inelastic nonrelativistic operators, including the SI-like $\mathcal{O}_1$ and spin-dependent (SD) $\mathcal{O}_4$ responses, together with a broad set of covariant interactions.  Several of the hypotheses with the highest local significance are endothermic: the inelastic $\mathcal{O}_1^v$ and $\mathcal{O}_4^{s,v}$ cases reach approximately $3.4\sigma$, while the Higgsino-like inelastic $\mathcal{O}_1^s$ interaction reaches approximately $3.3\sigma$ in the high-splitting region \cite{LZ2026,DiMauro2026}.  This is physically natural.  We denote by $\chi_1$ the lighter, stable neutral state that constitutes the Galactic DM population today, and by $\chi_2$ the slightly heavier neutral state with mass $m_{\chi_2}=m_{\chi_1}+\delta$.  In the Higgsino and pseudo-Dirac realizations considered below these are the two nearly degenerate Majorana mass eigenstates.  Endothermic scattering is the excitation process $\chi_1 A\to\chi_2 A$, for which $\delta>0$ must be supplied by the incident kinetic energy.  For a recoil energy $E_R$, the minimum speed of an incident Galactic-halo DM particle $\chi_1$, measured in the detector frame relative to the target nucleus, is
\begin{equation}
 v_{\min}(E_R)=\frac{1}{\sqrt{2m_AE_R}}
 \left(\frac{m_AE_R}{\mu_A}+\delta\right).
 \label{eq:vmin}
\end{equation}
Here $E_R$ is the nuclear recoil energy, $m_A$ is the mass of target nucleus $A$, $m_\chi\equiv m_{\chi_1}$ is the mass of the incoming ground-state DM particle, and $\mu_A=m_\chi m_A/(m_\chi+m_A)$ is the DM--nucleus reduced mass.  The positive quantity $\delta=m_{\chi_2}-m_{\chi_1}$ is the endothermic mass splitting.  Equation~\eqref{eq:vmin} has a minimum at finite recoil energy.  A splitting of a few hundred keV can therefore eliminate the conventional low-energy spectrum and concentrate the observable rate close to the upper LZ recoil window \cite{TuckerSmith2001,TuckerSmith2005,Bramante2016}.

Several recent studies have also emphasized the importance of Solar capture and neutrino-telescope constraints for particle-physics interpretations of the LZ high-energy recoil, finding that their impact can depend strongly on the underlying interaction, electroweak representation, annihilation channel, and halo assumptions \cite{PospelovRamani2026,WuZhangZhu2026,SmirnovGriffithBeacom2026,ElahiSchwaller2026,LeeYoun2026,Langhoff2026,FanHeWangZhao2026}.

The SD response offers an additional possibility that is especially relevant for the Solar test.  The standard nonrelativistic effective theory identifies $\mathcal{O}_4=\mathbf S_\chi\cdot\mathbf S_N$ as the familiar spin--spin interaction, and its nuclear responses have been studied extensively for direct detection and Solar capture \cite{Fitzpatrick2013,Anand2014,Menendez2012,Klos2013,LiangWu2014,CatenaSchwabe2015}.  The naturally abundant spinful xenon isotopes $^{129}$Xe and $^{131}$Xe carry predominantly neutron spin, which is why xenon experiments are especially sensitive to neutron-coupled SD scattering \cite{GarnyIbarraPatoVogl2013}; Solar SD-neutron capture, by contrast, can proceed only on trace odd-neutron isotopes.  A recent dedicated elastic-scattering calculation identified $^{3}$He, $^{13}$C, $^{17}$O, $^{21}$Ne, $^{25}$Mg, and $^{29}$Si as the dominant Solar targets for neutron-only SD scattering \cite{NguyenBlancoLinden2026}.  This target mismatch motivates our third benchmark: it is deliberately chosen to test whether a large neutron-spin response in xenon can coexist with much weaker Solar capture.  Endothermic kinematics strengthens this separation, but it also changes the isotope ranking.  Near $300\,\keV$ the six elastic-era targets lie at or below their stationary-target capture ceilings, while thermally moving nuclei and rarer, heavier odd-neutron isotopes can still capture DM.  We therefore treat the familiar $^{29}$Si endpoint as a diagnostic rather than an exact Solar cutoff and include the heavier open isotopes explicitly.

A number of papers appeared rapidly after the LZ result.  Endothermic phenomenological interpretations were studied in Ref.~\cite{SuYangYang2026}; the possibility of a TeV-scale Higgsino was emphasized independently in Refs.~\cite{FanReece2026,FreeseTheodosopoulos2026,WuZhangZhu2026}, and related ultraviolet realizations were discussed in Ref.~\cite{DuWang2026}.  Alternative interpretations include fermionic DM absorption and inelastic dark-photon scenarios \cite{LouLu2026,Yamashita2026}.  Of particular relevance for the present work is our analysis in Ref.~\cite{DiMauro2026}.  There we did not restrict the discussion to a spectral fit.  We investigated whether the interaction required by the LZ event could be embedded in particle models that also reproduce the observed thermal relic abundance and satisfy complementary ID constraints.  We studied two explicit cases: a nearly pure thermal Higgsino and a minimal pseudo-Dirac fermion with an off-diagonal vector interaction.  In the latter model the same heavy-mediator coefficient controls the endothermic DD process and the early-Universe coannihilation $\chi_1\chi_2\to q\bar q$, leading at $m_\chi\simeq1\,\TeV$ to the thermal target $\sigma_N\simeq6.5\times10^{-43}\,\cmsq$ and an LZ-preferred splitting near $297\,\keV$.  The Higgsino is more predictive: the relic density fixes $m_\chi$ close to $1.1\,\TeV$, while electroweak interactions fix the inelastic $Z$-exchange rate.

Pospelov and Ramani subsequently identified an important constraint that was not included in the first version of Ref.~\cite{DiMauro2026}: gravitational acceleration inside the Sun allows endothermic DM to scatter at velocities much larger than those available in terrestrial detectors \cite{PospelovRamani2026}.  A splitting that strongly suppresses xenon scattering can therefore still allow efficient capture on heavy solar nuclei.  If the captured DM becomes sufficiently concentrated in the Solar interior, annihilation produces high-energy neutrinos that are constrained by IceCube.  Using the ten-year IceCube solar search \cite{IceCubeSolar2025}, Ref.~\cite{PospelovRamani2026} derived a representative limit
\begin{equation}
 \Gamma_A(W^+W^-)<\Gamma_{WW}^{\rm lim}
 =1.5\times10^{19}\,\mathrm{s}^{-1},
 \label{eq:icecubeproxy}
\end{equation}
near the thermal Higgsino mass.  Here $\Gamma_A(W^+W^-)$ is the Solar DM pair-annihilation rate interpreted with the IceCube $W^+W^-$ template, and $\Gamma_{WW}^{\rm lim}$ denotes the corresponding upper limit.  The numerical value quoted above is the $90\%$ confidence-level limit inferred in Ref.~\cite{PospelovRamani2026}.  They obtained a tree-level stalled-orbit endpoint near $506\,\keV$ and a stronger approximately $566\,\keV$ endpoint when loop-induced elastic interactions thermalize the captured population.  Both values are well above the splitting required by the LZ Higgsino interpretation, which is about $370-380\,\keV$.

More recently, Ref.~\cite{SmirnovGriffithBeacom2026} showed that Higgs-coupled minimal DM can reproduce the LZ high-energy recoil while simultaneously satisfying the thermal-relic condition, and argued that Solar-capture constraints do not exclude all electroweak representations: the heavier multiplets can remain compatible with IceCube within the Standard Halo Model, whereas the lighter representations require a sufficiently fast component of the local DM velocity distribution.

The purpose of this paper is to make this Solar constraint explicit and to determine how strongly it depends on the particle realization.  The three benchmarks are chosen to isolate three different physical links in the chain ``LZ scattering $\to$ Solar capture $\to$ post-capture evolution $\to$ annihilation.''  First, the thermal Higgsino is the most predictive case: electroweak quantum numbers fix the dominant inelastic $Z$ interaction and standard thermal freeze-out fixes the mass near $1.1\,\TeV$ \cite{NagataShirai2015,KrallReece2018,Graham2025}.  We independently evaluate its capture, orbital evolution, thermalization, non-equilibrium annihilation, and neutrino signal, paying special attention to the loop-induced elastic SI and SD interactions and to the physical $W^+W^-+ZZ$ annihilation mixture.  Second, we repeat the calculation for the pseudo-Dirac vector model of Ref.~\cite{DiMauro2026}, a standard realization of inelastic DM in which a Dirac fermion is split into two nearby Majorana states \cite{TuckerSmith2001,TuckerSmith2005,DeSimone2010,DallaValleGarcia2025,Foguel2025}.  Here the same off-diagonal structure that produces the LZ recoil also makes the leading annihilation a $\chi_1\chi_2$ process, so the excited-state abundance inside the Sun becomes dynamical.  Third, we study neutron-philic endothermic $\mathcal{O}_4$ scattering, motivated directly by the high LZ significance of this operator and by the known complementarity between xenon neutron-spin responses and Solar SD capture \cite{Fitzpatrick2013,Anand2014,LiangWu2014,CatenaSchwabe2015,BlennowClementzHerreroGarcia2016,NguyenBlancoLinden2026}.  This third benchmark tests a qualitatively different possibility: the Solar constraint can already be strongly weakened at the capture stage because the abundant or standard odd-neutron targets become threshold suppressed and the remaining capture is carried by thermal tails and rare heavy isotopes.

The paper is organized as follows.  Section~\ref{sec:models} summarizes the three models and the interactions relevant to Solar capture.  Section~\ref{sec:solar} develops the capture formalism, specifies the Solar and halo inputs, and discusses both equilibrium and general non-equilibrium evolution.  Sections~\ref{sec:higgsino_results}, \ref{sec:pseudo_results}, and \ref{sec:nsid_results} present the Higgsino, pseudo-Dirac, and neutron-philic SD results, respectively.  Section~\ref{sec:discussion} compares their distinct Solar behaviors and discusses systematic uncertainties, and Section~\ref{sec:conclusion} gives our conclusions.  Appendices~\ref{app:higgsino}, \ref{app:pseudo}, and \ref{app:nsid} contain extended model descriptions and representative ultraviolet completions; the remaining appendices give the numerical capture, orbit, two-state, and neutrino details.

\section{DM models}
\label{sec:models}

The Higgsino and pseudo-Dirac vector models were developed in detail in Ref.~\cite{DiMauro2026}; the underlying Higgsino, inelastic-DM, and pseudo-Dirac constructions have a much broader literature \cite{TuckerSmith2001,TuckerSmith2005,NagataShirai2015,KrallReece2018,DeSimone2010,DallaValleGarcia2025,Foguel2025}.  We add a third benchmark, neutron-philic endothermic SD scattering, because $\mathcal{O}_4$ is one of the highest-significance LZ hypotheses and because the operator provides a clean test of the xenon--Sun target mismatch familiar from SD scattering \cite{Fitzpatrick2013,Anand2014,GarnyIbarraPatoVogl2013,LiangWu2014,CatenaSchwabe2015,BlennowClementzHerreroGarcia2016,NguyenBlancoLinden2026}.  All three models contain two nearby neutral states because the observed endothermic recoil requires the halo state $\chi_1$ to scatter into a heavier state $\chi_2$.  In the Higgsino and minimal pseudo-Dirac cases the pair arises naturally when an approximate Dirac fermion is split into two Majorana eigenstates by small symmetry-breaking masses; in the neutron-philic model the same two-state structure is retained while the transition current is axial.  We use $\chi_1$ for the lighter stable state and $\chi_2$ for the heavier state throughout.  Here we summarize the ingredients that control Solar capture and annihilation; detailed particle content, thermal histories, nuclear matching, and ultraviolet issues are discussed in Appendices~\ref{app:higgsino}, \ref{app:pseudo}, and \ref{app:nsid}.

\subsection{Thermal Higgsino}
\label{subsec:higgsino_model}

In the heavy-gaugino limit of the minimal supersymmetric Standard Model (MSSM), the two neutral Higgsino Weyl fields form an approximate Dirac fermion.  Mixing with heavy bino and wino states generates two nearby neutral Majorana eigenstates.  To keep the notation identical to the other models, we denote the lighter and heavier neutral Higgsino-like states by $\chi_1$ and $\chi_2$, respectively, rather than switching between generic DM and neutralino labels.  Their splitting is
\begin{equation}
 \delta\equiv m_{\chi_2}-m_{\chi_1}
 \simeq m_Z^2\left(\frac{s_W^2}{M_1}+\frac{c_W^2}{M_2}\right),
 \label{eq:higgsino_split}
\end{equation}
where $m_Z$ is the $Z$-boson mass, $M_1$ and $M_2$ are the bino and wino soft masses, and $s_W\equiv\sin\theta_W$ and $c_W\equiv\cos\theta_W$ are the sine and cosine of the weak mixing angle $\theta_W$.  The parameter $\mu$ is the supersymmetric Higgsino mass parameter and $\tan\beta$ is the ratio of the two MSSM Higgs vacuum expectation values.  Equation~\eqref{eq:higgsino_split} is valid up to finite-$\mu$, $\tan\beta$, phase, and threshold corrections \cite{NagataShirai2015,KrallReece2018,Graham2025,FanReece2026}.  The neutral-current interaction is dominantly off diagonal,
\begin{equation}
 \mathcal{L}_{Z\chi\chi}
 \supset \frac{g}{2c_W}Z_{\mu}\bar\chi_2\gamma^{\mu}\chi_1+\mathrm{h.c.},
 \label{eq:higgsino_z}
\end{equation}
where $g$ is the $SU(2)_L$ gauge coupling, $Z_\mu$ is the neutral weak gauge field, and h.c. denotes the Hermitian conjugate.  Thus tree-level DD and solar capture proceed through $\chi_1 A\to\chi_2 A$.  The coherent weak charge is
\begin{equation}
 Q_W=N-\left(1-4s_W^2\right)Z,
 \label{eq:weakcharge}
\end{equation}
where $N$ and $Z$ denote, respectively, the neutron and proton numbers of nucleus $A$; in this equation $Z$ is the nuclear charge, not the $Z$ gauge boson.  In the vector-current convention used consistently here,
\begin{equation}
 \sigma_A^{Z}=\frac{G_F^2\mu_A^2}{2\pi}Q_W^2,
 \qquad
 \sigma_n^{\widetilde H}=\frac{G_F^2\mu_n^2}{2\pi}
 \simeq7.43\times10^{-39}\,\cmsq.
 \label{eq:higgsino_sigman}
\end{equation}
Here $G_F$ is the Fermi constant, $\sigma_A^{Z}$ is the zero-momentum Higgsino--nucleus cross section generated by $Z$ exchange, $\mu_A$ is the DM--nucleus reduced mass defined above, $m_n$ is the neutron mass, $\mu_n=m_\chi m_n/(m_\chi+m_n)$ is the DM--neutron reduced mass, and $\sigma_n^{\widetilde H}$ is the corresponding zero-momentum neutron-normalized cross section.  Appendix~\ref{app:higgsino} discusses the complete Higgsino sector and the interactions relevant for Solar capture.

The thermal relic abundance selects a nearly pure Higgsino mass close to $1.1\,\TeV$.  We use $m_\chi=1.08\,\TeV$ for comparison with Ref.~\cite{PospelovRamani2026} and the reference annihilation rate
\begin{equation}
 \langle\sigma v\rangle_{\rm ann}=1.3\times10^{-26}\,\cmcubeds
 \label{eq:higgsino_sigmav}
\end{equation}
from detailed electroweak calculations \cite{Beneke2015}.  Here $\langle\sigma v\rangle_{\rm ann}$ denotes the velocity-averaged pair-annihilation cross section for the present-day nonrelativistic Higgsino system.  At present-day nonrelativistic velocities the dominant continuum final states are $W^+W^-$ and $ZZ$.  Representative calculations near the thermal Higgsino mass give $\langle\sigma v\rangle_{WW}\simeq8\times10^{-27}\,\cmcubeds$ and $\langle\sigma v\rangle_{ZZ}\simeq5\times10^{-27}\,\cmcubeds$ \cite{Rodd2024}.  Normalizing these two representative continuum rates gives the reference branching rations into $W^+W^-$ and $ZZ$ final states of
\begin{equation}
 B_{WW}^{\rm ref}\simeq0.62,\qquad B_{ZZ}^{\rm ref}\simeq0.38.
 \label{eq:higgsinoBR}
\end{equation}
Here $B_{WW}^{\rm ref}$ and $B_{ZZ}^{\rm ref}$ are not exact model-independent Higgsino branching ratios; they are convenient weights for the two dominant continuum channels at the thermal benchmark.  Sommerfeld enhancement, the charged--neutral and neutral--neutral splittings, and higher-order electroweak effects change the individual rates at the order-ten-percent level, while smaller line and endpoint components are not included in Equation~\eqref{eq:higgsinoBR} \cite{Beneke2015,Rodd2024}.  Both channels nevertheless produce hard neutrino spectra.  The consequences for using the IceCube $W^+W^-$ template are discussed in Section~\ref{subsec:neutrinos} and Section~\ref{subsec:higgsino_br}.

At loop level, the ground-state Higgsino also has elastic SI and SD interactions even when the tree-level diagonal neutral current is absent.  Guided by electroweak one-loop calculations of Higgsino-like neutralino scattering \cite{Hisano2005,Hisano2011,ChenHill2020,Bisal2024}, we adopt the central reference values
\begin{equation}
 \sigma_{\rm el,loop}^{\rm SI,N}=4\times10^{-50}\,\cmsq,
 \qquad
 \sigma_{\rm el,loop}^{\rm SD,p}=5\times10^{-47}\,\cmsq,
 \label{eq:looprefs}
\end{equation}
where $\sigma_{\rm el,loop}^{\rm SI,N}$ is the loop-induced elastic SI cross section per nucleon and $\sigma_{\rm el,loop}^{\rm SD,p}$ is the loop-induced elastic SD cross section on a proton.  We keep the explicit subscript ``loop'' throughout to distinguish these elastic cooling inputs from the tree-level inelastic nucleon and nuclear cross sections, such as $\sigma_n^{\widetilde H}$ and $\sigma_A^{Z}$, that control capture and the LZ recoil signal.  These numbers are benchmark values rather than precision predictions: the SI amplitude is particularly sensitive to cancellations among scalar and spin-two contributions, while the SD cross section receives sizeable radiative corrections.  For direct comparison with Ref.~\cite{PospelovRamani2026}, we also mark $\sigma_{\rm el,loop}^{\rm SI,N}=3.16\times10^{-49}\,\cmsq$.  This number is not an experimental DD limit.  It is the upper edge adopted by Pospelov and Ramani for the cancellation-sensitive heavy-Higgsino SI prediction, based on the heavy-WIMP calculation of Ref.~\cite{ChenHill2020}; their quoted central value is $4\times10^{-50}\,\cmsq$ and the lower side can approach a cancellation zero.  We therefore use $3.16\times10^{-49}\,\cmsq$ only as a reference upper theory value, while scanning a wider range to display the cooling dependence.  These elastic cross sections are irrelevant for producing the LZ high-recoil event but can dominate the secular cooling of a captured Solar population.  Their theoretical origin and uncertainty are analyzed in Appendix~\ref{app:higgsino}.

\subsection{Minimal pseudo-Dirac vector fermion}
\label{subsec:pseudo_model}

The second model is the standard pseudo-Dirac realization of inelastic fermion DM: an approximate Dirac fermion is split into two nearby Majorana eigenstates by small Majorana masses \cite{TuckerSmith2001,TuckerSmith2005,DeSimone2010,DallaValleGarcia2025,Foguel2025}.  This construction is attractive for the LZ event because the vector current of the parent Dirac fermion becomes dominantly off diagonal in the Majorana basis, naturally producing endothermic $\chi_1 A\to\chi_2 A$ scattering while suppressing a diagonal vector interaction.  We start from a Dirac fermion $\Psi$ with small Majorana masses,
\begin{equation}
 \mathcal{L}_{\rm mass}=-m_D\xi\eta-\frac{1}{2}m_L\xi\xi-\frac{1}{2}m_R\eta\eta+\mathrm{h.c.},
 \label{eq:pdm_mass}
\end{equation}
where $\xi$ and $\eta$ are the two left-handed Weyl fields that form the Dirac fermion, $m_D$ is the Dirac mass, and $m_L$ and $m_R$ are the small Majorana masses that break the approximate Dirac-number symmetry.  We assume $|m_L|,|m_R|\ll m_D$.  The two Majorana eigenstates have $\delta\simeq m_L+m_R$.  A vector mediator couples to the Dirac current and to quarks,
\begin{equation}
 \mathcal{L}\supset g_\chi V_\mu\bar\Psi\gamma^\mu\Psi
 +g_qV_\mu\sum_q\bar q\gamma^\mu q.
 \label{eq:pdm_vector}
\end{equation}
Here $V_\mu$ is a new vector mediator, $g_\chi$ is its coupling to DM, $g_q$ is the universal vector coupling to quarks, and the sum runs over kinematically relevant quark flavors.  After diagonalization the vector current is off diagonal, while diagonal Majorana vector currents vanish.  For a heavy mediator,
\begin{equation}
 \mathcal{L}_{\rm eff}=i\frac{g_\chi g_q}{m_V^2}
 \left(\bar\chi_2\gamma_\mu\chi_1\right)\sum_q\bar q\gamma^\mu q,
 \label{eq:pdm_contact}
\end{equation}
where $m_V$ is the vector-mediator mass and the expression assumes momentum transfers well below $m_V$.  \revtextC{We take $m_V>m_\chi$ throughout, as the contact-operator relic calculation below presupposes; this also closes $\chi_1\chi_1\to VV$ through $\chi_2$ exchange, which would otherwise provide a diagonal annihilation channel free of the two-state bottleneck of Section~\ref{sec:pseudo_results}.}  The isoscalar per-nucleon cross section is
\begin{equation}
 \sigma_N=\frac{\mu_N^2}{\pi}\left(\frac{3g_\chi g_q}{m_V^2}\right)^2.
 \label{eq:pdm_sigma}
\end{equation}
where $\mu_N=m_\chi m_N/(m_\chi+m_N)$ is the DM--nucleon reduced mass and $m_N$ denotes the nucleon mass.  The same coefficient controls the early-Universe coannihilation $\chi_1\chi_2\to q\bar q$.  In the heavy-mediator, six-flavor limit,
\begin{equation}
 \langle\sigma v\rangle_{\rm eff}\simeq\frac{m_\chi^2}{\mu_N^2}\sigma_N c,
 \label{eq:direct_relic}
\end{equation}
where $c$ is the speed of light and $\langle\sigma v\rangle_{\rm eff}$ is the effective thermally averaged coannihilation cross section entering freeze-out.  Equation~\eqref{eq:direct_relic} by itself fixes the scattering normalization required for thermal freeze-out at a chosen $m_\chi$; it does not determine the inelastic splitting.  In Ref.~\cite{DiMauro2026} we combined this thermal target with the LZ high-recoil rate: for $m_\chi=1\,\TeV$, the relic-density condition selects $\sigma_N\simeq6.5\times10^{-43}\,\cmsq$, and the LZ recoil calculation then selects $\delta\simeq297\,\keV$.  This combination defines the thermal LZ benchmark
\begin{equation}
\begin{aligned}
 m_\chi&=1\,\TeV,\qquad \delta=297\,\keV,\\
 \sigma_N&=6.5\times10^{-43}\,\cmsq.
\end{aligned}
 \label{eq:pdm_benchmark}
\end{equation}
For universal quark couplings the six open flavors have almost equal branching fractions, approximately $1/6$ each; the exact mass-corrected values used in the code are given in Appendix~\ref{app:pseudo}.  The crucial solar feature is that the leading annihilation process needs one $\chi_1$ and one $\chi_2$.  Capture therefore cannot be converted into a neutrino flux without following the two internal populations.

\subsection{Neutron-philic endothermic spin-dependent fermion}
\label{subsec:nsid_model}

Our third benchmark again contains a lighter Majorana state $\chi_1$ and an excited Majorana state $\chi_2$ separated by $\delta=m_{\chi_2}-m_{\chi_1}>0$, but the transition interaction is axial rather than vector.  The model combines three well-studied ingredients: two-state endothermic DM \cite{TuckerSmith2001,TuckerSmith2005}, the nonrelativistic spin--spin operator $\mathcal O_4$ and its nuclear responses \cite{Fitzpatrick2013,Anand2014,Menendez2012,Klos2013}, and neutron-dominated SD scattering, whose complementarity between xenon and Solar capture has been studied in Refs.~\cite{LiangWu2014,CatenaSchwabe2015,NguyenBlancoLinden2026}.  We use the two-state structure because a positive splitting is precisely what moves the observable xenon spectrum toward the high-recoil LZ event.  A convenient simplified interaction is
\begin{equation}
 \mathcal{L}_{Z'}\supset g_\chi Z'_\mu\left(\bar\chi_2\gamma^\mu\gamma^5\chi_1+\mathrm{h.c.}\right)
 +Z'_\mu\sum_q g_q^A\bar q\gamma^\mu\gamma^5q,
 \label{eq:nsid_lagrangian}
\end{equation}
where $Z'_\mu$ is a massive spin-one mediator, $g_\chi$ is the off-diagonal axial coupling in the DM sector, and $g_q^A$ is the axial coupling to quark flavor $q$.  For momentum transfer $q\ll m_{Z'}$, where $m_{Z'}$ is the mediator mass, integrating out $Z'$ gives the four-fermion operator
\begin{equation}
 \mathcal{L}_{\rm eff}^{\rm SD}=\sum_q G_q
 \left(\bar\chi_2\gamma^\mu\gamma^5\chi_1\right)
 \left(\bar q\gamma_\mu\gamma^5q\right),
 \qquad G_q\equiv\frac{g_\chi g_q^A}{m_{Z'}^2}.
 \label{eq:nsid_contact}
\end{equation}
The nucleon axial coefficient is $a_N=\sum_q g_q^A\Delta q^{(N)}$, where $\Delta q^{(N)}$ is the contribution of quark $q$ to the spin of nucleon $N=p,n$.  We choose the neutron-philic limit $|a_p|\ll|a_n|$.  The leading nonrelativistic transition operator is then
\begin{equation}
 \mathcal{O}_4=\mathbf{S}_\chi\cdot\mathbf{S}_N,
 \label{eq:nsid_o4}
\end{equation}
where $\mathbf{S}_\chi$ and $\mathbf{S}_N$ are the DM and nucleon spin operators.  This is the standard SD operator tested by LZ \cite{LZ2026,Fitzpatrick2013,Anand2014}.

We define the neutron-normalized zero-momentum cross section by
\begin{equation}
 \sigma_n^{\rm SD}=\frac{3\mu_n^2}{\pi}
 \left(\frac{g_\chi a_n}{m_{Z'}^2}\right)^2,
 \label{eq:nsid_sigman}
\end{equation}
with $\mu_n$ the DM--neutron reduced mass.  In the one-body neutron-only limit, a nucleus of spin $J_A$ and neutron spin expectation value $\langle S_n\rangle_A$ has
\begin{equation}
 \sigma_A^{\rm SD}(0)=\sigma_n^{\rm SD}
 \frac{\mu_A^2}{\mu_n^2}\frac{4}{3}\frac{J_A+1}{J_A}
 \langle S_n\rangle_A^2.
 \label{eq:nsid_sigmaA}
\end{equation}
Finite momentum transfer is controlled by the SD nuclear structure functions.  For xenon we use the $^{129}$Xe and $^{131}$Xe spin information from Refs.~\cite{Menendez2012,Klos2013}; because the event lies at unusually large momentum transfer, our recoil-level normalization is quoted as a transparent public estimate rather than as a replacement for the official LZ profile likelihood.

At $m_\chi=1\,\TeV$ and $\delta=300\,\keV$, LZ Supplemental Table S8 gives local significances $3.4\sigma$ for $\mathcal{O}_4^s$ and $3.3\sigma$ for $\mathcal{O}_4^v$ \cite{LZ2026}.  To attach a transparent cross-section scale to this detector-level result, our public recoil calculation asks which neutron-normalized $\mathcal O_4$ cross section gives approximately one accepted event in the $2.84$ tonne-year exposure.  We fold the natural $^{129}$Xe and $^{131}$Xe abundances, the finite-momentum spin response used below, and the public LZ nuclear-recoil efficiency over the extended recoil window.  This procedure gives the representative normalization
\begin{align}
 m_\chi&=1\,\TeV,\qquad \delta=300\,\keV,\\
 \sigma_n^{\rm SD}&\simeq3.5\times10^{-39}\,\cmsq.
 \label{eq:nsid_benchmark}
\end{align}
with $v_{\min}(248\,\keV)\simeq710\,\kms$ for $^{129}$Xe.  The exact cross-section normalization is more sensitive than the significance to the high-$q$ xenon spin response, so Equation~\eqref{eq:nsid_benchmark} should be read as the central normalization of our exposed nuclear-response prescription.  The Solar conclusion below is not based on an exact kinematic closure.  Instead, the same LZ normalization is propagated through a finite-temperature Solar calculation, and the dominant uncertainty is the poorly known high-$q$ spin response of the rare heavy odd-neutron isotopes.

We do not attempt to specify the cosmological history responsible for the DM relic abundance in this scenario.  In particular, the abundance need not be controlled by the small visible-sector couplings entering the interaction in Equation~\eqref{eq:nsid_contact}.  It can instead be set entirely within the hidden sector, for example through annihilations into additional dark-sector states or through other secluded interactions that are essentially independent of the coupling responsible for scattering on nuclei \cite{PospelovRitzVoloshin2008,EscuderoWitteHooper2017,DiMauroWang2025}.  The subsequent evolution and possible decay of these hidden-sector states can be arranged consistently without affecting the direct-detection phenomenology considered here.  Since these additional ingredients are not required for either the interpretation of the LZ event or the calculation of Solar capture, we leave the relic-density mechanism unspecified and do not introduce a numerical cosmological benchmark.  A more detailed discussion of possible ultraviolet realizations and of the associated dark-sector structure is given in Appendix~\ref{app:nsid}.  The Solar conclusion derived below depends only on the capture rate and on the general bound $\Gamma_A\leq C_\odot/2$, and is therefore independent of the mechanism responsible for setting the relic abundance.

\section{Solar capture, thermalization, and neutrino signal}
\label{sec:solar}

The solar-neutrino constraint contains three physically distinct steps: gravitational capture, post-capture evolution of the orbital distribution, and annihilation into final states that yield high-energy neutrinos.  For an ordinary elastic WIMP these steps are often compressed into the equilibrium relation $\Gamma_A=C_\odot/2$.  For endothermic and multi-state DM that shortcut can fail.  We therefore keep the three stages separate.  The capture formalism follows the standard treatment of WIMP capture in the Sun, generalized to inelastic scattering as in Refs.~\cite{NussinovWangYavin2009,MenonMorrisPierceWeiner2010,BlennowClementzHerreroGarcia2016,Blennow2018,Catena2018,PospelovRamani2026}; the nuclear responses used for the SD cases are described explicitly below.

\subsection{Solar structure, elemental composition, and halo boundary conditions}
\label{subsec:solarinputs}

The capture probability depends on the number density $n_A(r)$ of each Solar nucleus and on the gravitational potential through $v_{\rm esc}(r)$.  We use the public BS05(AGS,OP) Standard Solar Model (SSM), the same SSM family adopted by Ref.~\cite{Bahcall2005,PospelovRamani2026}.  The Solar model used in our calculation specifies the radial coordinate $x=r/R_\odot$, enclosed mass $M(r)/M_\odot$, temperature $T(r)$, density $\rho(r)$, and the H, $^4$He, $^{12}$C, $^{14}$N, and $^{16}$O mass fractions.  The tabulated Solar-model quantities are evaluated between radial grid points using shape-preserving cubic interpolation, with the density interpolated logarithmically, rather than replacing the Solar structure by an analytic profile.  The central values are
\begin{equation}
 T_c=1.548\times10^7\,\mathrm{K},\qquad
 \rho_c=150.5\,\mathrm{g\,cm^{-3}},
\end{equation}
corresponding to $T_c\simeq1.33\,\keV$.  Here $T_c\equiv T(0)$ and $\rho_c\equiv\rho(0)$ are the central Solar temperature and mass density.

The escape speed is derived from the same enclosed-mass profile,
\begin{equation}
 v_{\rm esc}^2(r)=2G\left[\frac{M_\odot}{R_\odot}
 +\int_r^{R_\odot}dr'\,\frac{M(r')}{r'^2}\right],
 \label{eq:vesc_ssm}
\end{equation}
rather than fitted independently.  Here $G$ is Newton's constant, $M_\odot$ and $R_\odot$ are the Solar mass and radius, and $M(r)$ is the mass enclosed inside radius $r$.  The resulting values are $v_{\rm esc}(0)\simeq1385\,\kms$ and $v_{\rm esc}(R_\odot)=617.7\,\kms$.  At $r=0.1R_\odot$ and $0.2R_\odot$ we obtain approximately $1322$ and $1195\,\kms$, respectively.  These numbers illustrate why endothermic capture probes splittings that are inaccessible to terrestrial experiments.

For every species,
\begin{equation}
 n_A(r)=\frac{\rho(r)X_A(r)}{A m_u}.
 \label{eq:solar_numberdensity}
\end{equation}
where $X_A(r)$ is the mass fraction of species $A$, $A$ is its mass number, and $m_u$ is the atomic mass unit; thus $n_A(r)$ is the local number density of that nuclear species.  The radial H, He, C, N, and O abundances are taken directly from BS05(AGS,OP).  We additionally follow Ne, Mg, Si, S, Ca, Ti, Cr, Fe, Ni, Zn, Ge, Pb, and U.  \revtextD{All of these species, including Ca, Ti and Cr, enter the coherent capture sums; at the pseudo-Dirac benchmark Ca alone contributes rather more than S.}  These species are not individually tabulated in the public BS05 file used here, so we assign AGS-like normalizations and, following the common-heavy-element-profile approximation used in semi-analytic Solar-capture studies, let them track the mild radial diffusion trend of oxygen.  This is more realistic than the constant-core abundances used in our exploratory calculation while keeping the treatment transparent.  The central reference values are summarized in Table~\ref{tab:solarcomp}.  Zn, Ge, Pb, and U have tiny abundances but are retained because very heavy nuclei can remain kinematically active near the extreme endothermic endpoint; their contribution is shown explicitly rather than hidden in the total.

\begin{table*}[t]
\caption{Central mass fractions used in the BS05(AGS,OP)-based capture calculation.  H, He, C, N, and O are read directly from the Solar table.  The remaining heavy-element values are the adopted surface normalizations multiplied by the common oxygen radial-profile factor at the Solar center.}
\label{tab:solarcomp}
\begin{ruledtabular}
\begin{tabular}{lccccccc}
Species & H & He & C & N & O & Ne & Mg\\
$X_A(0)$ & $0.36462$ & $0.62029$ & $7.79\times10^{-6}$ & $3.73\times10^{-3}$ & $6.06\times10^{-3}$ & $1.38\times10^{-3}$ & $7.76\times10^{-4}$\\
\hline
Species & Si & S & Ca & Ti & Cr & Fe & Ni\\
$X_A(0)$ & $7.36\times10^{-4}$ & $3.43\times10^{-4}$ & $7.14\times10^{-5}$ & $3.48\times10^{-6}$ & $1.85\times10^{-5}$ & $1.44\times10^{-3}$ & $7.82\times10^{-5}$\\
\hline
Species & Zn & Ge & Pb & U & & & \\
$X_A(0)$ & $1.92\times10^{-6}$ & $2.65\times10^{-7}$ & $9.46\times10^{-9}$ & $5.57\times10^{-11}$ & & & \\
\end{tabular}
\end{ruledtabular}
\end{table*}

For the Galactic DM distribution we use $\rho_\chi=0.3\,\GeV\,\mathrm{cm}^{-3}$, $v_0=220\,\kms$, Solar speed $v_\odot=232\,\kms$, and Galactic escape speed $v_{\rm esc}^{\rm Gal}=544\,\kms$, matching Ref.~\cite{PospelovRamani2026}.  Defining
\begin{equation}
 u_{\pm}=v_{\rm esc}^{\rm Gal}\pm v_\odot,
 \qquad
 E_{\pm}(u)=\exp\left[-\frac{(u\pm v_\odot)^2}{v_0^2}\right],
\end{equation}
the Solar-frame speed distribution is
\begin{equation}
\begin{aligned}
 f_\odot(u)&=\frac{u}{\sqrt{\pi}v_0v_\odot N_{\rm esc}}\\
 &\quad\times
 \begin{cases}
 E_-(u)-E_+(u), & u<u_-,\\
 E_-(u)-e^{-z^2}, & u_-\leq u<u_+,\\
 0, & u\geq u_+.
 \end{cases}
\end{aligned}
 \label{eq:fsun}
\end{equation}
where
\begin{equation}
 N_{\rm esc}=\operatorname{erf}(z)-\frac{2z}{\sqrt{\pi}}e^{-z^2},
 \qquad z=\frac{v_{\rm esc}^{\rm Gal}}{v_0}.
\end{equation}
The variable $u$ is the DM speed far from the Sun, $f_\odot(u)$ is the normalized Solar-frame speed probability density, $v_0$ is the halo velocity-dispersion parameter, $v_\odot$ is the Solar speed through the halo, $v_{\rm esc}^{\rm Gal}$ is the Galactic escape speed, and $N_{\rm esc}$ normalizes the truncated distribution.  The truncation is important for consistency, although gravitational acceleration makes the inner-Solar capture problem much less sensitive to the terrestrial high-velocity tail than the LZ event itself.

For the neutron-philic SD benchmark the relevant Solar composition is qualitatively different from the coherent Higgsino and pseudo-Dirac cases.  Ref.~\cite{NguyenBlancoLinden2026} identifies $^{3}$He, $^{13}$C, $^{17}$O, $^{21}$Ne, $^{25}$Mg, and $^{29}$Si as the dominant targets for \emph{elastic} SD-neutron capture and omits $^{57}$Fe because its small magnetic moment, uncertain neutron spin, and low isotopic abundance make it negligible in that problem.  Near a $300\,\keV$ endothermic threshold the hierarchy changes: the six elastic-era targets are strongly phase-space suppressed, while heavier trace odd-neutron isotopes possess substantially larger stationary-target ceilings.  We therefore retain $^{21}$Ne, $^{25}$Mg, and $^{29}$Si together with $^{33}$S, $^{43}$Ca, $^{47,49}$Ti, $^{53}$Cr, $^{57}$Fe, $^{61}$Ni, $^{67}$Zn, and $^{73}$Ge in the finite-temperature benchmark calculation.  Spin inputs are taken from Ref.~\cite{NguyenBlancoLinden2026} for the first group and from the zero-momentum spin compilation of Ref.~\cite{Bednyakov2005} or odd-group-model estimates for the rarer heavy nuclei.  The retained set is not intended as an exhaustive isotope catalog; it contains the abundance-relevant additions through $^{73}$Ge identified by the endothermic threshold hierarchy in our Solar composition model.  Still heavier odd-neutron species have rapidly falling elemental abundances and poorly controlled spin responses, and we treat their omission as part of the residual composition/nuclear-response uncertainty discussed below.  Because the endothermic benchmark probes substantial momentum transfer, we also use the finite-momentum review of Ref.~\cite{BednyakovSimkovic2006} to motivate treating the heavy-isotope momentum dependence as a leading uncertainty rather than as a precisely known response.  The resulting response uncertainty is propagated explicitly in Section~\ref{sec:nsid_results}.

\subsection{Inelastic gravitational capture}
\label{subsec:capture}

Consider a DM particle with asymptotic speed $u$ relative to the Sun.  Conservation of energy in the Solar gravitational potential gives the local speed
\begin{equation}
 w^2(r,u)=u^2+v_{\rm esc}^2(r).
 \label{eq:wsolar}
\end{equation}
where $w(r,u)$ is the DM speed at Solar radius $r$ after gravitational acceleration from the asymptotic speed $u$\footnote{Taking the gravitational potential to vanish at infinity, $\Phi(\infty)=0$, energy conservation gives $\frac{1}{2}m_\chi u^2=\frac{1}{2}m_\chi w^2+m_\chi\Phi(r)$.  Since the local escape velocity is defined by $v_{\rm esc}^2(r)=-2\Phi(r)$, one obtains $w^2=u^2+v_{\rm esc}^2$.  Thus an incoming halo particle with $u>0$ has positive total energy and is initially unbound, even though gravity accelerates it to $w>v_{\rm esc}$ inside the Sun.  Capture requires scattering to reduce its total energy below zero.}.  For endothermic scattering $\chi_1 A\to\chi_2 A$, part of the relative kinetic energy is converted into the internal mass splitting.  The reaction is possible only when
\begin{equation}
 \frac{1}{2}\mu_Aw^2>\delta.
 \label{eq:threshold}
\end{equation}
This simple condition explains the exceptional reach of the Sun.  Near the Solar center $v_{\rm esc}\sim1380\,\kms$, so the kinetic energy available to a TeV particle scattering on Fe or Ni is much larger than in a terrestrial detector.  Heavy nuclei are doubly important at large $\delta$: their larger reduced mass raises the maximum splitting accessible at fixed collision speed (equivalently, it lowers the minimum relative speed required for a fixed splitting), while their coherent charge can enhance SI matrix elements.

The final relative speed after the inelastic transition is
\begin{equation}
 w'=\sqrt{w^2-\frac{2\delta}{\mu_A}},
\end{equation}
where $w'$ is the relative speed after paying the excitation energy $\delta$, i.e.~the DM speed after the collision $\chi_1 A\to\chi_2 A$.  Two-body kinematics gives the allowed recoil interval
\begin{equation}
 E_R^{\pm}=\frac{\mu_A^2}{2m_A}\left(w\pm w'\right)^2.
 \label{eq:recoilrange}
\end{equation}
where $E_R^-$ and $E_R^+$ are the minimum and maximum recoil energies allowed by two-body inelastic kinematics.  
A scattering event does not necessarily imply Solar capture. The outgoing particle is gravitationally bound only if its total energy satisfies
\begin{equation}
E_{\rm out}=
\frac{1}{2}m_\chi w'^2
+
m_\chi\Phi(r)
<0,
\end{equation}
or equivalently $w'<v_{\rm esc}(r)$.  Otherwise, the scattered particle remains unbound and eventually escapes the Sun.

Since the internal excitation itself removes energy from translational motion, the binding condition is
\begin{equation}
 E_R+\delta>\frac{1}{2}m_\chi u^2,
 \label{eq:capture_condition}
\end{equation}
so the physical lower recoil boundary is
\begin{equation}
 E_R^{\rm cap}=\max\left[E_R^-,\frac{1}{2}m_\chi u^2-\delta\right].
 \label{eq:recoil_min_capture}
\end{equation}
where $E_R^{\rm cap}$ is the smallest recoil energy that both satisfies the inelastic kinematics and leaves the outgoing DM state gravitationally bound.  When $E_R^{\rm cap}>E_R^+$ the scattering may be kinematically allowed but cannot leave the outgoing state gravitationally bound.

The capture rate for element $A$ is evaluated in the standard Gould form \cite{Gould1987a,Gould1987b,Jungman1996},
\begin{equation}
 \frac{dC_A}{dV}=\frac{\rho_\chi}{m_\chi}\int_0^\infty du\,
 \frac{f_\odot(u)}{u}\,w\,\Omega_A^-(w),
 \label{eq:gould}
\end{equation}
where $dC_A/dV$ is the capture rate per unit Solar volume contributed by nuclear species $A$, $\rho_\chi/m_\chi$ is the local halo number density, and $\Omega_A^-(w)$ is the scattering rate per incident DM particle into gravitationally bound final states.  Explicitly,
\begin{equation}
 \Omega_A^-(w)=n_A(r)w\int_{E_R^{\rm cap}}^{E_R^+}dE_R\,
 \frac{d\sigma_A}{dE_R}.
 \label{eq:omega}
\end{equation}
with $d\sigma_A/dE_R$ the differential DM--nucleus cross section.  The total Solar capture rate is
\begin{equation}
 C_\odot=4\pi\sum_A\int_0^{R_\odot}dr\,r^2\frac{dC_A}{dV}.
 \label{eq:ctotal}
\end{equation}
For the Higgsino we use
\begin{equation}
 \frac{d\sigma_A^{\widetilde H}}{dE_R}=\frac{G_F^2m_A}{4\pi w^2}Q_W^2F_A^2(q),
 \label{eq:higgsino_diff}
\end{equation}
where $q=\sqrt{2m_AE_R}$ is the momentum transfer and $F_A(q)$ is the Helm nuclear form factor.  For the pseudo-Dirac isoscalar vector interaction
\begin{equation}
 \frac{d\sigma_A^{\rm PD}}{dE_R}=\frac{m_A\sigma_A}{2\mu_A^2w^2}F_A^2(q),\qquad
 \sigma_A=\sigma_N\frac{\mu_A^2}{\mu_N^2}A^2.
 \label{eq:pdm_diff}
\end{equation}
Here $\sigma_A$ is the zero-momentum coherent cross section on nucleus $A$, $\sigma_N$ is the isoscalar per-nucleon cross section, and $\mu_N$ is the DM--nucleon reduced mass.  We use a Helm form factor in both coherent cases.  For the neutron-philic SD model we instead write
\begin{equation}
 \frac{d\sigma_A^{\rm SD}}{dE_R}=\frac{m_A\sigma_A^{\rm SD}(0)}{2\mu_A^2w^2}
 F_{\rm SD,A}^2(q),
 \label{eq:nsid_diff}
\end{equation}
where $F_{\rm SD,A}^2(q)$ is the normalized SD structure factor, $F_{\rm SD,A}^2(0)=1$, and $\sigma_A^{\rm SD}(0)$ is defined in Equation~\eqref{eq:nsid_sigmaA}.  The crucial point for Solar capture is independent of the detailed form factor.  For a Solar nucleus treated as stationary, a collision contributes only when both the endothermic threshold in Equation~\eqref{eq:threshold} and the binding condition in Equation~\eqref{eq:capture_condition} can be satisfied.  This stationary-target form is adequate away from a sharp endpoint, but it is not an exact statement for a thermal Solar plasma.  For the neutron-philic benchmark, which lies within a few keV of the stationary $^{29}$Si endpoint, we instead average over the nuclear Maxwell--Boltzmann velocity distribution and impose binding on the outgoing DM speed in the Solar frame, as described in Equation~\eqref{eq:nsid_thermalcapture}.  Appendix~\ref{app:capture} gives numerical convergence tests and discusses the residual SSM and nuclear-response approximations.

\subsection{From capture to annihilation: equilibrium and the general non-equilibrium case}
\label{subsec:equilibrium}

For a one-species population, the most general one-zone evolution equation relevant here is
\begin{equation}
 \frac{dN}{dt}=C(t)-C_E(t)N-C_A(t)N^2.
 \label{eq:generalN}
\end{equation}
Here $N(t)$ is the total number of captured DM particles, $C(t)$ is the capture rate, $C_E(t)$ is the one-particle evaporation coefficient, and $C_A(t)$ is the two-particle annihilation coefficient.  Thus the three terms describe capture, evaporation, and annihilation.  At TeV masses evaporation is negligible, $C_E\simeq0$, but Equation~\eqref{eq:generalN} is useful because it makes clear that equilibrium is an assumption rather than a consequence of capture.  For constant $C$, $C_E$, and $C_A$, with $N(0)=0$, the exact one-zone solution is
\begin{equation}
 N(t)=\frac{2C\tanh(\lambda t/2)}{\lambda+C_E\tanh(\lambda t/2)},
 \qquad
 \lambda=\sqrt{C_E^2+4CC_A}.
 \label{eq:general_constant_solution}
\end{equation}
The quantity $\lambda$ is the inverse relaxation timescale defined in Equation~\eqref{eq:general_constant_solution}.  The late-time population is $N_\infty=2C/(\lambda+C_E)$ and the instantaneous annihilation rate is always
\begin{equation}
 \Gamma_A(t)=\frac{1}{2}C_A N^2(t).
 \label{eq:gamma_general}
\end{equation}
A nonzero evaporation rate therefore lowers both the equilibrium population and the asymptotic annihilation signal.  For the TeV masses considered here $C_E$ is completely negligible.  Equation~\eqref{eq:general_constant_solution} then reduces to
\begin{equation}
 N(t)=\sqrt{\frac{C}{C_A}}\tanh\left(\frac{t}{\tau_{\rm eq}}\right),\qquad
 \tau_{\rm eq}=\frac{1}{\sqrt{CC_A}},
\end{equation}
and
\begin{equation}
 \Gamma_A(t)=\frac{C}{2}\tanh^2\left(\frac{t}{\tau_{\rm eq}}\right).
 \label{eq:gammaeq}
\end{equation}
Here $\tau_{\rm eq}$ is the capture--annihilation equilibration time and $t_\odot=4.57\,\mathrm{Gyr}$ is the Solar age.  The commonly used replacement $\Gamma_A=C/2$ is therefore a limiting result, valid only when $t_\odot/\tau_{\rm eq}\gg1$.  A large capture rate by itself does not guarantee a large present-day neutrino signal.

Computing the initial Solar capture rate is not sufficient to determine the annihilation signal. After the first capturing collision, a DM particle can remain on an extended orbit that crosses only part of the Sun. Subsequent scatterings can progressively reduce its orbital energy and drive the captured population toward more compact orbits, thereby increasing its spatial density and the probability of annihilation. It is therefore necessary to follow, at least approximately, the post-capture orbital evolution when translating the capture rate into an annihilation rate and a neutrino signal. This is particularly important for endothermic DM, for which additional scatterings can become kinematically suppressed as the particles lose energy.

The annihilation coefficient depends on the spatial distribution,
\begin{equation}
 C_A(t)=\langle\sigma v\rangle\frac{\int d^3r\,n_\chi^2(r,t)}{\left[\int d^3r\,n_\chi(r,t)\right]^2}.
 \label{eq:CAgeneral}
\end{equation}
where $n_\chi(r,t)$ is the captured-DM number-density profile and the denominator equals $N^2(t)$.  For the transparent one-zone approximation used in our plots,
\begin{equation}
 C_A\simeq\frac{\langle\sigma v\rangle}{V_{\rm eff}},\qquad
 V_{\rm eff}=\frac{4\pi}{3}R_\chi^3.
 \label{eq:caeff}
\end{equation}
where $V_{\rm eff}$ is the effective annihilation volume and $R_\chi$ is the one-zone effective radius used to characterize the spatial extent of the captured population.  When capture or orbital cooling is time dependent, the general implementation is to integrate the population and orbital dynamics simultaneously,
\begin{align}
 \frac{dN}{dt}&=C(t)-C_E(t)N-C_A[n_\chi(t)]N^2,\\
 \frac{d\boldsymbol{\theta}_{\rm orb}}{dt}&=\boldsymbol{F}_{\rm scat}
 \left[\boldsymbol{\theta}_{\rm orb},t\right],
 \label{eq:numerical_noneq}
\end{align}
where $\boldsymbol{\theta}_{\rm orb}$ denotes the chosen description of the orbital distribution and $\boldsymbol{F}_{\rm scat}$ is the orbit-averaged scattering functional that gives the collision-driven evolution of those orbital variables.  In our one-radius treatment $\boldsymbol{\theta}_{\rm orb}=R_\chi$, while a full calculation would evolve the distribution in orbital energy and angular momentum.  At each step $C_A$ is evaluated from Equation~\eqref{eq:CAgeneral} and the neutrino source is obtained from Equation~\eqref{eq:gamma_general}; no equilibrium substitution is required.  This procedure is essential whenever the cooling time is comparable with the Solar age, $C(t)$ varies, or more than one internal state participates in annihilation.  Our Higgsino plots use fixed-radius solutions to expose this dependence transparently, whereas the pseudo-Dirac analysis below solves the coupled two-state population equations explicitly.

\begin{figure}[t]
\centering
\includegraphics[width=\columnwidth]{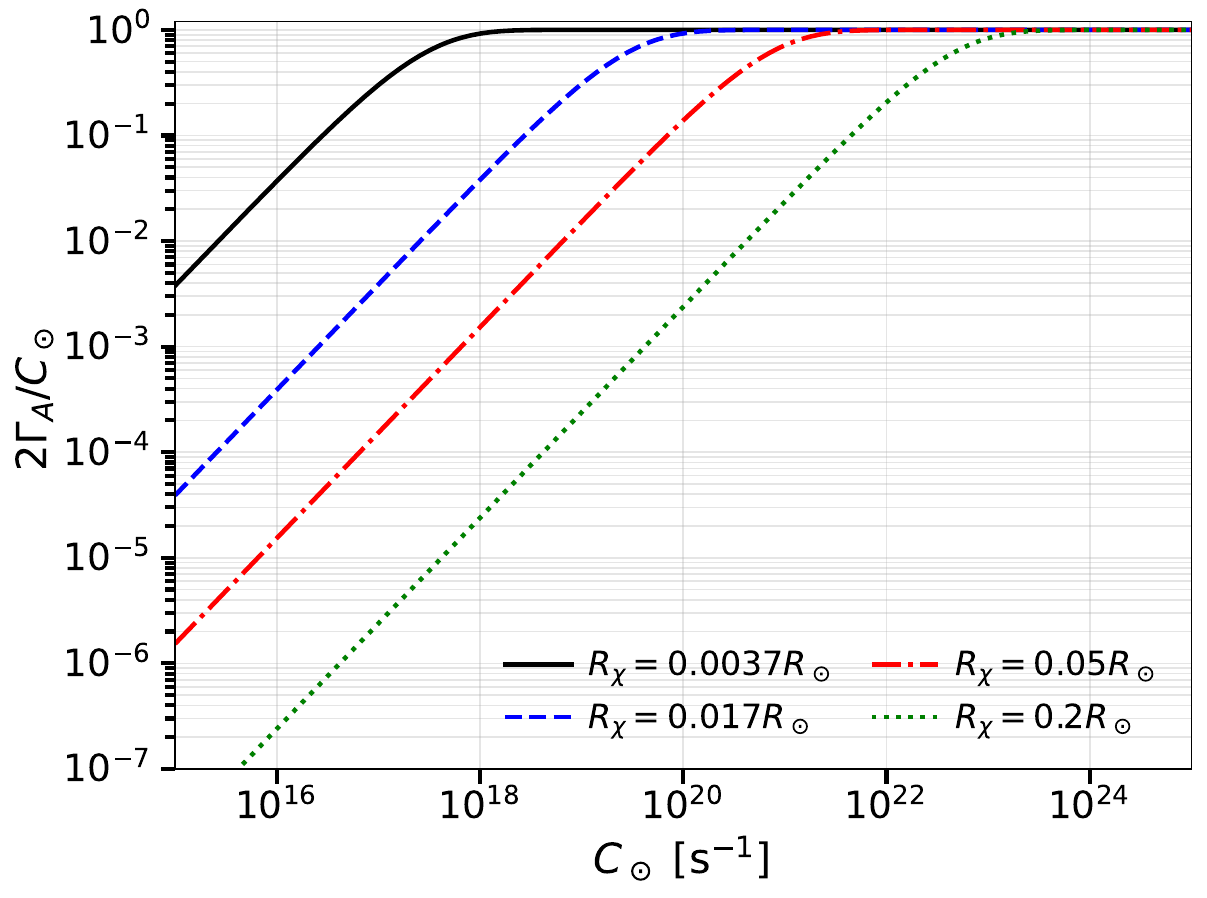}
\caption{Present-day suppression factor $2\Gamma_A/C_\odot$ as a function of the Solar capture rate $C_\odot$.  The black solid, blue dashed, red dash-dotted, and green dotted curves correspond to the labeled one-zone effective radii $R_\chi$.}
\label{fig:noneq}
\end{figure}

Figure~\ref{fig:noneq} illustrates the physical point.  It shows $2\Gamma_A/C_\odot$ for several assumed radii $R_{\chi}$. More spatial compact thermalized populations reach unity at much smaller capture rates, whereas an extended population can remain far from equilibrium even when capture itself is large.  This figure is particularly useful for understanding why the Higgsino and pseudo-Dirac cases cannot be judged from $C_\odot$ alone.

\subsection{Post-capture inelastic cooling and elastic thermalization}
\label{subsec:thermalization}

Once DM is gravitationally captured by the Sun, its subsequent evolution is governed by the competition between inelastic transitions, which progressively reduce the orbital energy while they remain kinematically accessible, and elastic scattering, which eventually drives the surviving population toward thermalization with the Solar medium. This post-capture evolution is important because the annihilation rate depends not only on the number of captured DM particles, but also on their spatial distribution inside the Sun.

The first successful scattering of an incoming Galactic-halo DM particle with a Solar nucleus can remove enough kinetic energy to convert its initially unbound trajectory into a gravitationally bound orbit around the Sun. Capture, however, does not in general imply immediate thermalization. After the first scattering, the DM particle can remain on a highly eccentric and spatially extended orbit, with an outer turning point that is a sizable fraction of $R_\odot$. During each subsequent passage through the dense Solar interior, the particle encounters Solar nuclei again and therefore has additional opportunities to scatter and lose orbital energy.

For endothermic DM, this early post-capture evolution can be dominated by further inelastic up-scattering. In each allowed transition, part of the DM kinetic energy is transferred to the recoiling nucleus, while an additional amount $\delta$ is required to excite the DM particle to the heavier state. Repeated inelastic scatterings therefore progressively reduce the orbital energy and contract the orbit. As this contraction proceeds, however, the maximum velocity reached by the DM particle during a Solar crossing also decreases. The inelastic process then becomes kinematically inaccessible once the available center-of-mass kinetic energy is no longer sufficient to overcome the mass splitting, as determined by Equation~\eqref{eq:threshold}. Because the corresponding threshold depends on the target-nucleus mass, different Solar elements cease to support up-scattering at different stages of the orbital evolution: lighter nuclei typically become inaccessible first, while heavier nuclei can continue to induce inelastic transitions down to more compact orbits. Consequently, inelastic cooling does not by itself necessarily drive the captured population to thermal equilibrium, but can instead terminate at a finite characteristic orbital size \cite{Blennow2018,PospelovRamani2026}.

\begin{figure*}
\centering
\includegraphics[width=1.59\columnwidth]{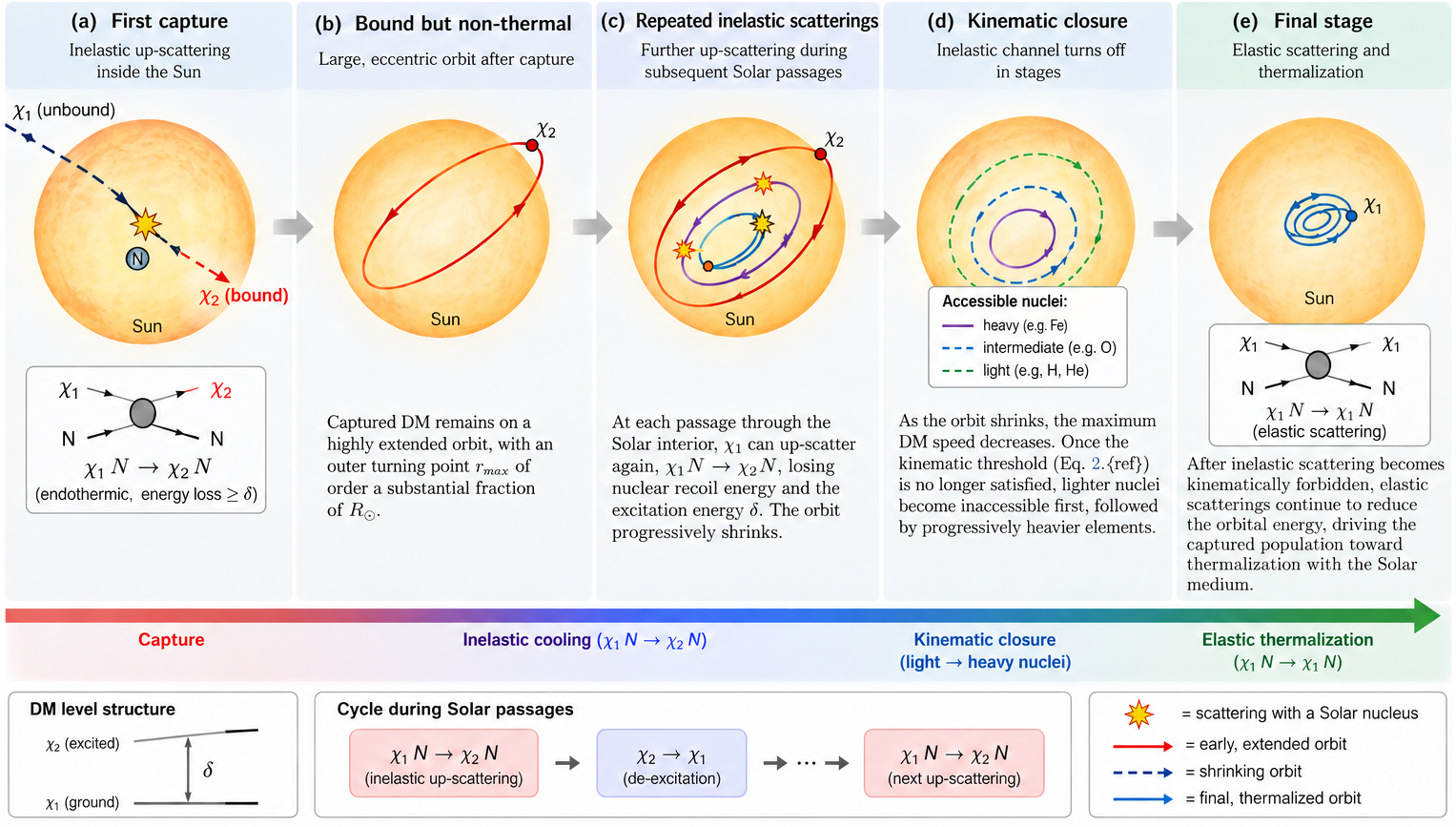}
\caption{Schematic illustration of the post-capture evolution of endothermic dark matter in the Sun. Panel (a) shows the initial capture of an incoming Galactic-halo particle $\chi_1$ through the endothermic transition $\chi_1 N \rightarrow \chi_2 N$. Panel (b) illustrates the resulting gravitationally bound but non-thermal orbit. Panel (c) shows the subsequent orbital contraction produced by repeated inelastic up-scatterings $\chi_1 N \rightarrow \chi_2 N$, followed by de-excitation $\chi_2 \rightarrow \chi_1$. In panel (d), the shrinking orbit progressively reduces the maximum DM velocity reached inside the Sun, causing the inelastic channel to become kinematically inaccessible first for lighter and subsequently for heavier nuclei. Panel (e) illustrates the final stage, in which elastic scattering $\chi_1 N \rightarrow \chi_1 N$ can continue to reduce the orbital energy and drive the captured population toward thermal equilibrium with the Solar medium.}
\label{fig:scheme}
\end{figure*}

This process is shown in Figure~\ref{fig:scheme}.

The origin of the stalling condition can be understood directly from the orbital energy of a captured DM particle. Consider, for simplicity, a nearly radial bound orbit characterized by its apocenter $r_{\rm max}$, namely the largest radius reached by the particle along the orbit. At the apocenter the velocity vanishes in the radial-orbit approximation, and the total specific orbital energy is therefore
\begin{equation}
\varepsilon_{\rm orb} = \Phi(r_{\rm max}),
\end{equation}
where $\Phi(r)$ is the Solar gravitational potential. At a generic Solar radius $r$, conservation of orbital energy gives
\begin{equation}
\frac{1}{2}v^2(r;r_{\rm max}) + \Phi(r)
=
\Phi(r_{\rm max}),
\end{equation}
so that
\begin{equation}
v(r;r_{\rm max})
=
\sqrt{
v_{\rm esc}^2(r)
-
v_{\rm esc}^2(r_{\rm max})
},
\label{eq:radialv}
\end{equation}
where we have used $v_{\rm esc}^2(r) = -2\Phi(r)$ and $v(r;r_{\rm max})$ denotes the instantaneous DM speed at Solar radius $r$ for a nearly radial orbit with apocenter $r_{\rm max}$.

The largest speed along the orbit is reached near the Solar center. Setting $r=0$ in Equation~\eqref{eq:radialv} gives
\begin{equation}
v^2(0;r_{\rm max})
=
v_{\rm esc}^2(0)
-
v_{\rm esc}^2(r_{\rm max}).
\label{eq:center_orbit_speed}
\end{equation}
As subsequent scatterings remove orbital energy, the apocenter moves inward and $r_{\rm max}$ decreases. The difference in gravitational potential between the apocenter and the Solar center therefore becomes smaller, and consequently the maximum speed $v(0;r_{\rm max})$ reached during each Solar passage also decreases. Eventually this speed becomes too small to satisfy the kinematic threshold for further endothermic scattering.
For a Solar nucleus $A$ taken to be at rest, further endothermic scattering becomes kinematically impossible anywhere along the orbit once
\begin{equation}
\frac{1}{2}\mu_A v^2(0;r_{\rm max}) < \delta.
\label{eq:stallcriterion}
\end{equation}
The corresponding stationary-target kinematic closure radius is determined implicitly by
\begin{equation}
v_{\rm esc}^2(0)
-
v_{\rm esc}^2(r_A^{\rm kin})
=
\frac{2\delta}{\mu_A}.
\label{eq:stallradius}
\end{equation}
Here $r_A^{\rm kin}$ is the formal radial-orbit apocenter at which further up-scattering on species $A$ first becomes kinematically forbidden if the nucleus is taken to be at rest. This equation is only a threshold diagnostic; it does not by itself define the final DM distribution. A trace very-heavy element may remain kinematically accessible at a smaller radius but still be dynamically irrelevant if its abundance or form factor makes its scattering optical depth negligible, while thermally moving nuclei can reopen an endothermic transition close to threshold. The physical evolution is therefore controlled by the abundance-weighted collision rates and by the orbital phase-space distribution. We retain the full element decomposition in the capture calculation and treat the additional pseudo-Dirac re-excitation issue explicitly in Section~\ref{subsec:pseudo_reexcitation}.

The orbital geometry matters because a captured particle spends most of its time near the apocenter but acquires its largest speed and encounters the largest Solar density near the center. The nearly radial approximation introduced above provides a simple semi-analytic description of this evolution. A realistic captured population has a distribution in orbital energy and angular momentum; we discuss the resulting uncertainty in Section~\ref{sec:discussion}.

For the Higgsino there is a second cooling stage. Once tree-level inelastic scattering is kinematically blocked, loop-induced elastic SI and SD processes remain available because they do not need to overcome the mass splitting. Although the elastic cross sections are extraordinarily small, a captured particle crosses the Sun an enormous number of times over $4.57$ Gyr. The relevant quantity is therefore not whether one elastic collision is probable on a single orbit, but the integrated secular energy loss over the Solar age.

For an elastic process we write schematically
\begin{equation}
\frac{d\sigma_{A,\rm el}^{\rm loop}}{dq^2}
=
\frac{
A^2 \sigma_{\rm el,loop}^{\rm SI,N} F_A^2(q)
+
\sigma_{A,\rm el,loop}^{\rm SD}
}{
4\mu_n^2 v^2
}.
\label{eq:elastic_loop_diff}
\end{equation}
where $q$ is the momentum transfer, $F_A(q)$ is the SI nuclear form factor, $\sigma_{\rm el,loop}^{\rm SI,N}$ is the loop-induced nucleon-normalized SI input, and $\sigma_{A,\rm el,loop}^{\rm SD}$ denotes the loop-induced effective SD nuclear response for species $A$ obtained from the proton-level input $\sigma_{\rm el,loop}^{\rm SD,p}$. The explicit ``loop'' label prevents either quantity from being confused with the tree-level inelastic nuclear cross sections used in the capture calculation. Equation~\eqref{eq:elastic_loop_diff} is schematic and is used only to display the ingredients of the orbit-averaged slowing calculation; the numerical implementation treats the SI and SD channels separately.
The energy-weighted slowing cross section is
\begin{equation}
\sigma_{A,\rm slow}^{\rm loop}(v)
=
\int_0^{4\mu_A^2 v^2} dq^2\,
\frac{E_R}{E_\chi}
\frac{d\sigma_{A,\rm el}^{\rm loop}}{dq^2},
\label{eq:slowcross}
\end{equation}
with
\begin{equation}
E_R = \frac{q^2}{2m_A},
\qquad
E_\chi = \frac{m_\chi v^2}{2}.
\end{equation}
Here $v \equiv v(r;r_{\rm max})$ is the instantaneous speed of the bound DM particle along the orbit defined in Equation~\eqref{eq:radialv}, $E_\chi = m_\chi v^2/2$ is its instantaneous kinetic energy, and $\sigma_{A,\rm slow}^{\rm loop}$ weights each collision by the fractional kinetic-energy loss $E_R/E_\chi$. The orbit-averaged specific-energy loss is then integrated to evolve $r_{\rm max}(t)$; Appendix~\ref{app:orbit} gives the explicit equations. This treatment makes the loop uncertainty transparent: changing $\sigma_{\rm el,loop}^{\rm SI,N}$ or $\sigma_{\rm el,loop}^{\rm SD,p}$ changes the cooling time and hence $R_\chi$, which in turn changes $C_A$ through Equation~\eqref{eq:caeff}. It is this chain, rather than capture itself, that produces the spread between the stalled-orbit and fully thermalized IceCube endpoints.

Thermalization is stronger than merely becoming gravitationally bound or suffering several collisions. It means that repeated scattering erases the memory of the original capture orbit and drives the spatial distribution toward an approximately isothermal form,
\begin{equation}
n_\chi(r)
\propto
\exp\left[
-\frac{
m_\chi \left(\Phi(r)-\Phi(0)\right)
}{
T_\chi
}
\right],
\label{eq:isothermal_distribution}
\end{equation}
where $\Phi(r)$ is the Solar gravitational potential and $T_\chi$ is the effective temperature of the thermalized DM population. For efficient thermal contact, $T_\chi$ is of order the central Solar temperature. Expanding the potential near the center for an approximately constant core density gives a Gaussian distribution with characteristic scale
\begin{equation}
R_{\rm therm}
\simeq
\sqrt{
\frac{3T_c}{2\pi G\rho_c m_\chi}
}
\simeq
0.0037R_\odot
\sqrt{
\frac{1\,\TeV}{m_\chi}
}.
\label{eq:rtherm}
\end{equation}
Here $R_{\rm therm}$ denotes the characteristic thermal radius of the captured DM distribution in the constant-density core approximation. Ref.~\cite{PospelovRamani2026} finds that a Higgsino population localized within approximately $0.017R_\odot$ already reaches capture--annihilation equilibrium over the Solar age. Our orbit calculation reproduces this hierarchy and is used below to scan the loop-induced elastic rates continuously rather than selecting only two limiting cases.

\subsection{Neutrino flux, branching fractions, and IceCube mapping}
\label{subsec:neutrinos}

For annihilation channel $f$, the neutrino flux at Earth is
\begin{equation}
 \frac{d\Phi_\nu}{dE_\nu}=\frac{\Gamma_A}{4\pi D_\odot^2}
 \sum_f B_f\frac{dN_{\nu,f}}{dE_\nu}.
 \label{eq:nuflux}
\end{equation}
where $D_\odot$ is the Sun--Earth distance, $B_f$ is the annihilation branching fraction into final state $f$, and $dN_{\nu,f}/dE_\nu$ is the propagated neutrino yield per annihilation as a function of neutrino energy $E_\nu$.  The yield includes primary decays, electroweak radiation, charged- and neutral-current interactions in the Sun, tau regeneration, flavor oscillations, and propagation to Earth \cite{Blennow2008,Charon2020,Bauer2021}.  IceCube constrains the convolution of this flux with its energy- and direction-dependent response, not $\Gamma_A$ in isolation.

For direct comparison with Ref.~\cite{PospelovRamani2026} we use Equation~\eqref{eq:icecubeproxy} for the Higgsino.  The approximation is good in this case because both $W^+W^-$ and $ZZ$ generate hard electroweak showers.  We nevertheless quantify the branching uncertainty conservatively.  If one discards all neutrinos from the $ZZ$ branch and counts only the $B_{WW}^{\rm ref}\simeq0.62$ component, the corresponding upper limit on the \emph{total} annihilation rate weakens to
\begin{equation}
 \Gamma_A<\Gamma_{A,\rm cons}^{\rm lim}
 =\frac{\Gamma_{WW}^{\rm lim}}{B_{WW}^{\rm ref}}
 \simeq2.44\times10^{19}\,\mathrm{s}^{-1}.
 \label{eq:conservativeBR}
\end{equation}
This construction is intentionally more conservative than a physical $WW+ZZ$ spectral combination because it assigns zero IceCube sensitivity to $ZZ$.

The annihilation-channel dependence is especially important for the other two benchmarks.  At $m_\chi\simeq1\,\TeV$, the $b\bar b$ limit is weaker than the $W^+W^-$ limit because the neutrino spectrum is softer.  A direct reading of the $b\bar b$ and $W^+W^-$ curves in both the spin-dependent and spin-independent panels of Figs.~5 and 7 of Ref.~\cite{IceCubeSolar2025} gives the representative ratio $\Gamma_{b\bar b}^{\rm lim}/\Gamma_{WW}^{\rm lim}\simeq5$.  Using the $W^+W^-$ normalization in Equation~\eqref{eq:icecubeproxy}, we therefore adopt
\begin{equation}
 \Gamma_{b\bar b}^{\rm lim}\simeq7.5\times10^{19}\,\mathrm{s}^{-1}
 \label{eq:icecubebbproxy}
\end{equation}
as the IceCube upper limit for the $b\bar b$ channel at $m_\chi\simeq1\,\TeV$ used in our nominal non-Higgsino comparisons.  This conversion follows the channel dependence of the published IceCube curves; it is not a detector-level recast of either model.

The physical annihilation spectra of the two benchmarks must be kept distinct from this experimental template.  The pseudo-Dirac vector model annihilates through an approximately democratic $u,d,s,c,b,t$ mixture, whereas the LZ scattering operator of the neutron-philic benchmark does not by itself determine its annihilation final state.  Neither model is therefore literally a pure $b\bar b$ source.  In the dense Solar core, the ordinary light-hadron component of $u,d,s$ jets is strongly degraded before decay, while heavy flavors yield harder neutrinos.  At TeV DM masses electroweak radiation also generates a hard neutrino component even for light-fermion primaries \cite{RottSiegalGaskinsBeacom2013,BernalMartinAlboPalomaresRuiz2013,IbarraTotzauerWild2014,Bauer2021}.  The ten-year IceCube analysis includes these electroweak corrections and shows that they are important for the high-mass hadronic limits \cite{IceCubeSolar2025}.  We therefore use the IceCube $b\bar b$ upper limit as a nominal soft-hadronic proxy and the published $W^+W^-$ upper limit as a hard comparison.  A channel-exact constraint requires the branching-fraction-weighted annihilation spectrum to be propagated through the Sun and convolved with the IceCube detector response.
The spectral uncertainty is discussed in Appendix~\ref{app:neutrino}.

\section{Results I: thermal Higgsino}
\label{sec:higgsino_results}

\subsection{Solar capture and the fully equilibrated reference limit}
\label{subsec:higgsino_capture}

We first isolate the part of the calculation that depends only on capture.  We take a thermal Higgsino with $m_\chi=1.08\,\TeV$, the local halo and Solar inputs of Section~\ref{sec:solar}, and the tree-level inelastic weak interaction in Equation~\eqref{eq:higgsino_sigman}.  A halo particle arrives at infinity with speed $u$ in the Solar rest frame and accelerates to the local speed $w(r,u)=\sqrt{u^2+v_{\rm esc}^2(r)}$ at Solar radius $r$.  Capture means that an inelastic collision $\chi_1 A\to\chi_2 A$ removes enough kinetic energy that the outgoing particle is gravitationally bound.  The rate $C_\odot$ therefore measures the injection of Higgsinos into bound Solar orbits; by itself it does \emph{not} imply an annihilation rate or capture--annihilation equilibrium.

\begin{figure}[t]
\centering
\includegraphics[width=\columnwidth]{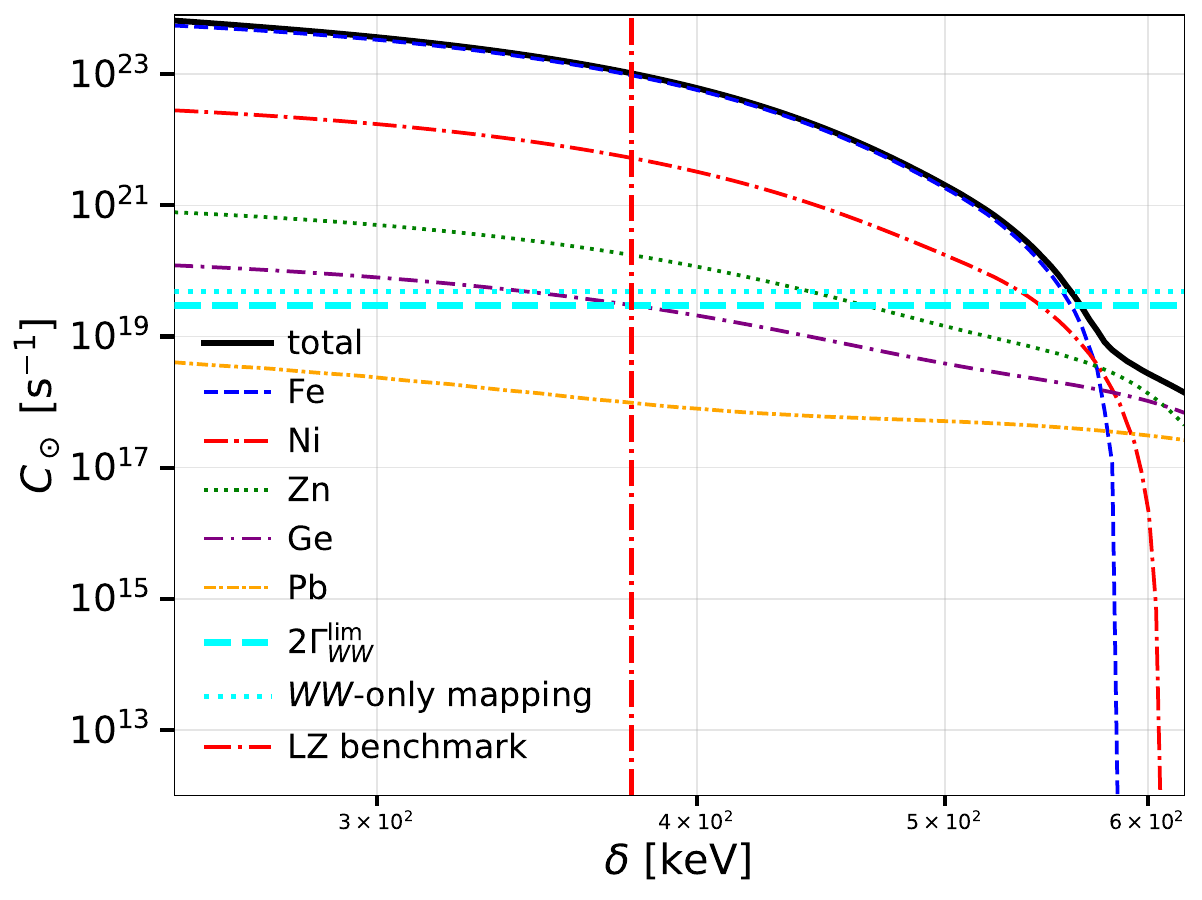}
\caption{Solar capture rate for a $1.08\,\TeV$ Higgsino as a function of $\delta$.  The black solid curve is the total rate and the colored curves are the elemental contributions indicated in the legend.  The cyan dashed and dotted horizontal lines mark $2\Gamma_{WW}^{\rm lim}$ and the conservative $WW$-only mapping, while the red dash-dotted vertical line marks the LZ benchmark.}
\label{fig:hcap}
\end{figure}

Figure~\ref{fig:hcap} shows the resulting capture rate as a function of the neutral-state splitting $\delta$.  The total rate falls steeply as the inelastic threshold approaches the maximum kinetic energy available in the Solar potential.  At the Higgsino LZ benchmark, $\delta\simeq377\,\keV$, we obtain
\begin{equation}
 C_\odot(377\,\keV)\simeq\revtextD{1.03}\times10^{23}\,\mathrm{s}^{-1}.
 \label{eq:c377}
\end{equation}
\revtextD{At this splitting Fe supplies approximately $93.4\%$ of the capture rate and Ni approximately $5.1\%$, followed by Cr at $0.95\%$, Ca at $0.19\%$, Zn at $0.17\%$ and Ti at $0.12\%$.}  Around $\delta=506\,\keV$, Fe and Ni provide about $89.7\%$ and $9.2\%$, respectively\revtextD{, and Ca, Ti and Cr have already closed}.  Very close to the ultimate endpoint, for example at $566\,\keV$, the contribution of still heavier nuclei becomes more important: Fe contributes about $49\%$, Ni $27\%$, Zn $16\%$, and Ge $6.6\%$.  This progression has a simple kinematic origin: increasing $\delta$ successively removes lighter nuclei because their DM--nucleus reduced masses are too small to supply the excitation energy.

The capture-only curve shown in Figure \ref{fig:hcap} provides a useful limiting case.  If annihilation is fast enough that the number of bound particles has reached steady state, evaporation is negligible, and every relevant annihilation final state is constrained with the adopted IceCube template, Equation~\eqref{eq:generalN} reduces to $\Gamma_A=C_\odot/2$.  The largest splitting for which this equilibrium signal reaches the IceCube bounds for the $WW$ channel is then
\begin{equation}
 \delta_{\rm lim}^{\rm therm}=564.8\,\keV.
 \label{eq:hthermendpoint}
\end{equation}
We call this the fully thermalized reference endpoint because a compact thermal distribution maximizes the annihilation coefficient $C_A$ for a fixed captured population.  It is not assumed a priori in the remainder of the analysis.  The central question of Sections~\ref{subsec:higgsino_loopres} and \ref{subsec:higgsino_lzcomparison} is whether the captured Higgsino population actually becomes compact enough, within the Solar age, for the equilibrium approximation to be justified.

\subsection{Post-capture stalling, elastic cooling, and non-equilibrium}
\label{subsec:higgsino_loopres}

As explained in Sec.~\ref{subsec:thermalization}, the first successful endothermic collision captures an incoming halo particle, but capture is only the beginning of the Solar evolution.  The process $\chi_1A\to\chi_2A$ converts an unbound halo trajectory into a gravitationally bound, typically eccentric orbit whose apocenter can still be a sizeable fraction of $R_\odot$.  The subsequent dynamics has a simple physical sequence.  First, the captured excited state can undergo the exothermic transition $\chi_2A\to\chi_1A$.  Once the particle is again in the ground state, a later collision can re-excite it through $\chi_1A\to\chi_2A$ if the relative kinetic energy is large enough to pay the splitting $\delta$.  Repeated state-changing collisions transfer recoil energy to Solar nuclei and progressively reduce the orbital energy, so the orbit contracts.  Over a complete up--down cycle the excitation energy paid in the endothermic step is returned in the exothermic step; the secular cooling is therefore controlled by the net energy transferred to the nuclei rather than by removing $\delta$ permanently from the DM system.

The inelastic stage cannot continue indefinitely.  As the orbit contracts, the apocenter $r_{\rm max}$ decreases and the maximum speed reached during the central Solar passage also decreases.  For the radial-orbit diagnostic this maximum speed is $v_c(r_{\rm max})$ from Equation~\eqref{eq:center_orbit_speed}.  Endothermic up-scattering on a stationary nucleus $A$ is possible only while the available center-of-mass kinetic energy satisfies $\mu_Av_c^2(r_{\rm max})/2\geq\delta$.  The equality defines the kinematic closure radius $r_A^{\rm kin}$ in Equation~\eqref{eq:stallradius}.  Its meaning is concrete: if one follows a captured particle while collisions shrink its orbit, then at $r_{\rm max}=r_A^{\rm kin}$ even the fastest part of that orbit, near the Solar center, has exactly the energy needed for $\chi_1A\to\chi_2A$.  For $r_{\rm max}<r_A^{\rm kin}$, an up-scatter on a stationary target $A$ is impossible anywhere along that orbit.  This is the sense in which the collision at the equality is the last kinematically possible up-scattering on species $A$.

There is no single universal closure radius because each nucleus has a different reduced mass $\mu_A$.  Lighter targets become inaccessible first, while heavier nuclei can remain kinematically open on more compact orbits.  Kinematic openness, however, is not the same as dynamical importance: the actual cooling rate also depends on the Solar abundance of the isotope, its nuclear form factor, the collision optical depth accumulated along the orbit, and the thermal velocity of the target nuclei.  Thermal motion smooths the stationary-target threshold and can reopen collisions close to nominal closure.  We therefore use $r_A^{\rm kin}$ only as a marker of the available phase space, not as the final radius of the captured population.

For the deliberately conservative no-elastic Higgsino construction of Ref.~\cite{PospelovRamani2026}, uranium is used as the last extremely heavy stationary target that can support tree-level inelastic cooling.  With our BS05 potential, $r_U^{\rm kin}=0.200R_\odot$ at $\delta=500\,\keV$, $0.201R_\odot$ at $506\,\keV$, and $0.203R_\odot$ at $513\,\keV$.  We denote this scale by $r_{\rm stall,U}\equiv r_U^{\rm kin}$.  Uranium contributes negligibly to the capture rate; here it is used only to define an intentionally extended post-capture configuration when no elastic cooling is allowed.  The numerical no-elastic curve retains the full $\delta$ dependence $r_U^{\rm kin}(\delta)$.  Because thermally moving nuclei can sustain some scattering below the stationary-target closure scale, this prescription is conservative for the neutrino signal: it keeps the Higgsino population more extended and suppresses annihilation.

The evolution changes qualitatively if loop-induced elastic SI or SD scattering is present.  Elastic collisions do not need to excite $\chi_1$ into $\chi_2$ and therefore remain possible after the tree-level inelastic channel has become threshold suppressed.  A single elastic collision is extremely rare, but a bound Higgsino crosses the Sun an enormous number of times over $t_\odot=4.57\,\mathrm{Gyr}$; cross sections around $10^{-50}$--$10^{-47}\,\cmsq$ can therefore produce a sizeable cumulative energy loss.  We evolve this second cooling stage with the orbit-averaged slowing equation of Section~\ref{subsec:thermalization} and Appendix~\ref{app:orbit}.  The velocity $v$ entering that kernel is the instantaneous speed of the already-bound particle, $v(r;r_{\rm max})$ in Equation~\eqref{eq:radialv}, rather than the asymptotic halo speed $u$ or the local pre-capture speed $w$.

The physical competition can now be stated directly.  Tree-level inelastic collisions first cool the newly captured Higgsino but eventually switch off as the orbit contracts.  Loop-induced elastic collisions can continue shrinking the orbit beyond that point.  Annihilation acts simultaneously, but its efficiency is controlled by how compact the captured population has become: in the one-zone description $C_A\propto R_\chi^{-3}$.  Thus the same capture rate can correspond either to a weak neutrino signal, if the population remains on extended stalled orbits, or to capture--annihilation equilibrium, if elastic cooling compresses it into the Solar core.  

\begin{figure}[t]
\centering
\includegraphics[width=\columnwidth]{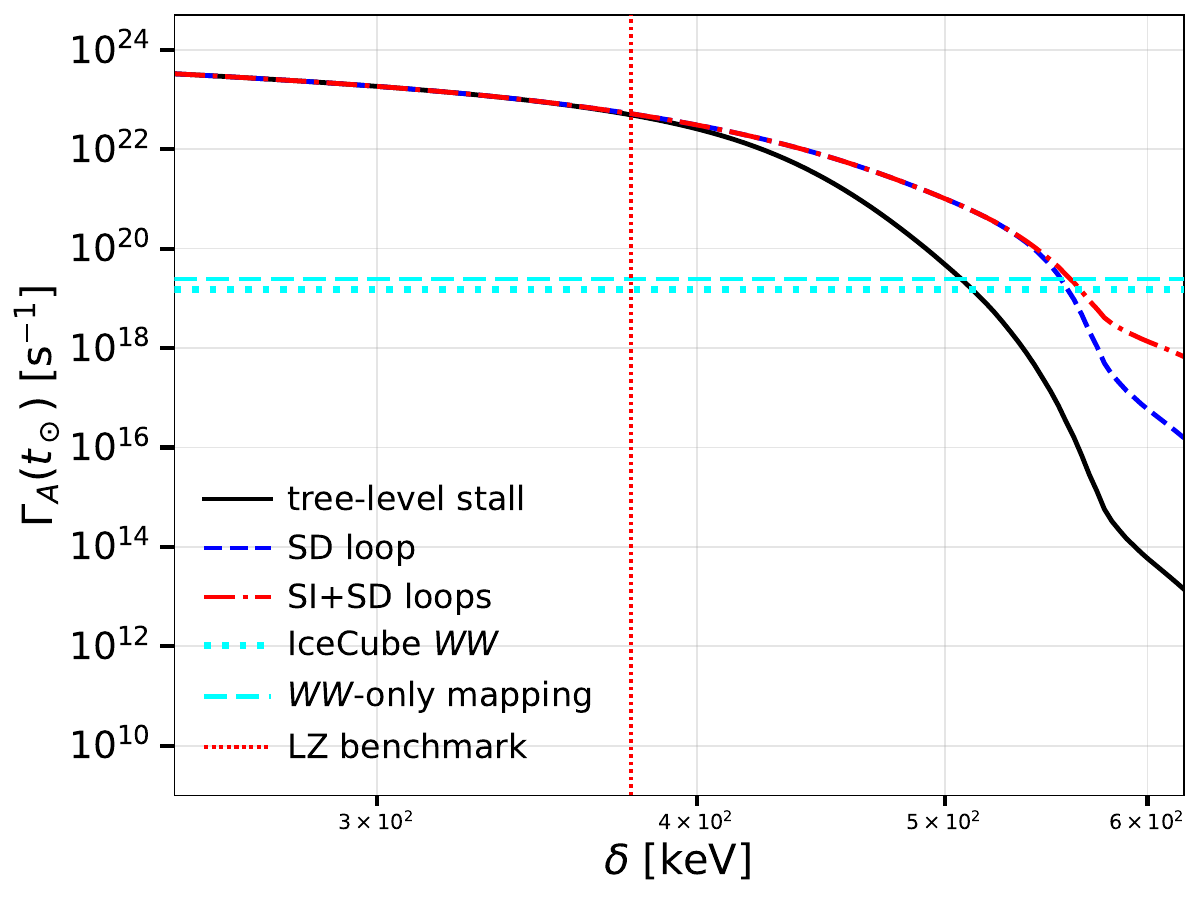}
\caption{Present Solar Higgsino annihilation rate as a function of $\delta$.  The black solid, blue dashed, and red dash-dotted curves correspond to tree-level inelastic stalling, SD-loop cooling, and SI+SD-loop cooling, respectively.  The cyan horizontal lines show the two IceCube $WW$ mappings, and the red vertical line marks the LZ benchmark.}
\label{fig:hann}
\end{figure}

Figure~\ref{fig:hann} isolates this competition by using the same capture calculation for all curves and varying only the post-capture cooling assumption.
The figure compares three deliberately distinct assumptions.  The black curve switches off the loop-induced elastic processes and follows the $\delta$-dependent tree-level inelastic stall described above.  This is an extreme lower-cooling benchmark, not the physical expectation for a pure Higgsino.  The blue curve keeps the central SD loop interaction but sets the SI loop interaction to zero.  The red curve uses the central SI and SD benchmark cross sections in Equation~\eqref{eq:looprefs}.  For the last two cases we evolve the orbit from the representative endpoint stall scale toward the thermal radius.  The resulting endpoints are
\begin{align}
 \delta_{\rm lim}^{\rm no\,elastic}&\simeq512.7\,\keV,\\
 \delta_{\rm lim}^{\rm SD}&\simeq558.6\,\keV,\\
 \delta_{\rm lim}^{\rm SI+SD}&\simeq564.8\,\keV.
 \label{eq:higgsino_endpoints}
\end{align}
For the SD-only benchmark the effective radius after one Solar age is $R_\chi\simeq0.0228R_\odot$, whereas the central SI+SD case reaches the thermal radius $R_{\rm therm}\simeq0.0037R_\odot$.

The distinction between equilibrium and non-equilibrium is important here.  At any fixed effective radius the exact one-species solution is Equation~\eqref{eq:gamma_general}, not $C_\odot/2$.  The equilibration time is $\tau=(C_\odot C_A)^{-1/2}$, so enlarging the spatial distribution reduces $C_A$ and lengthens $\tau$.  The black curve therefore gives a smaller annihilation rate even though its capture rate is identical to the other curves.  Conversely, once elastic cooling makes $R_\chi$ sufficiently small, $t_\odot/\tau\gg1$ and the result becomes insensitive to the exact radius: the annihilation rate saturates at $C_\odot/2$.  This is why the SD-only and fully thermalized splitting endpoints differ by only a few keV despite their different characteristic radii.

\subsection{Loop uncertainty and comparison with the LZ Higgsino region}
\label{subsec:higgsino_lzcomparison}

\begin{figure}[t]
\centering
\includegraphics[width=\columnwidth]{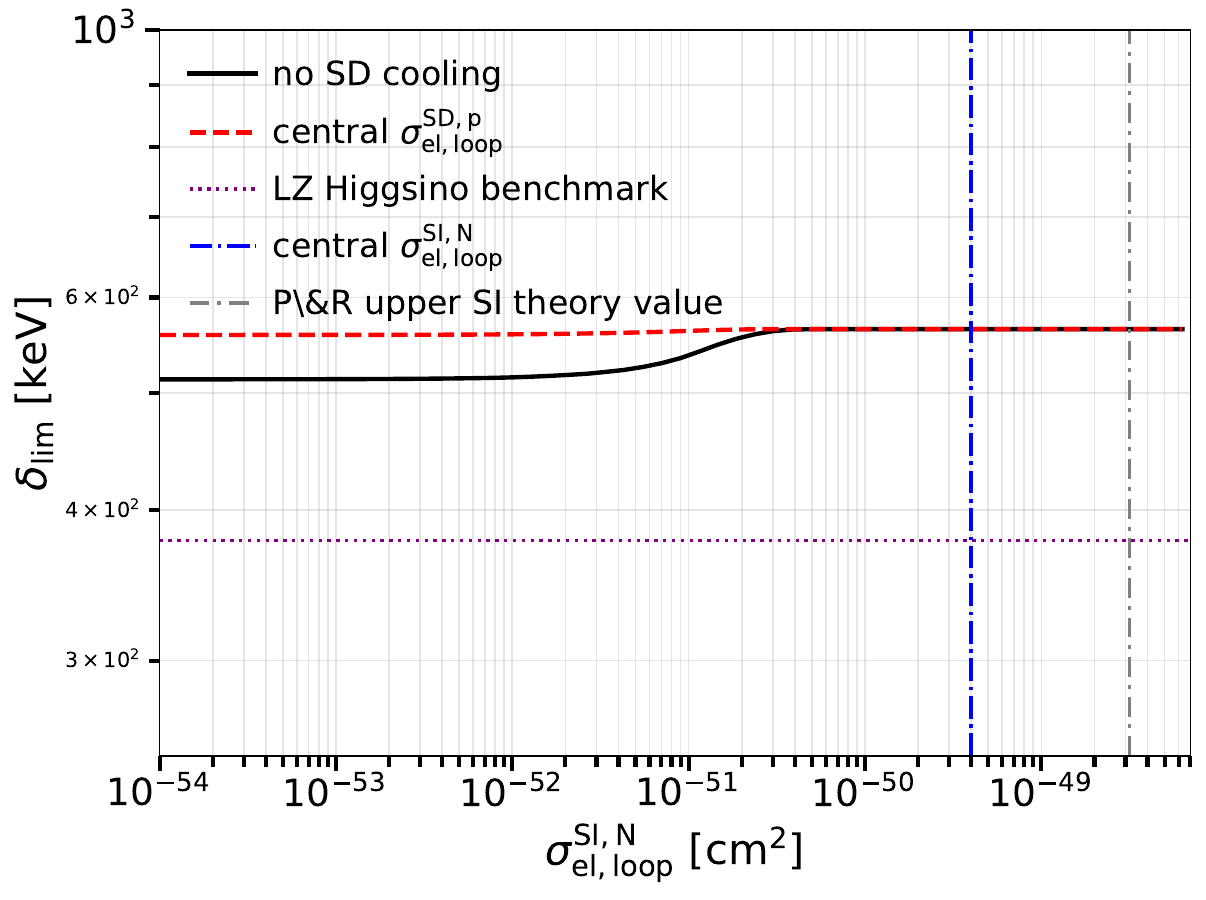}
\caption{Higgsino exclusion endpoint $\delta_{\rm lim}$ as a function of the loop-induced SI nucleon cross section.  The black solid curve has no SD cooling and the red dashed curve uses the central SD-loop value.  The purple horizontal line marks the LZ benchmark; the blue and gray vertical lines mark the central and upper reference SI values.}
\label{fig:hloop}
\end{figure}

The loop-induced elastic cross sections are the principal particle-physics uncertainty in the post-capture evolution.  Figure~\ref{fig:hloop} has therefore been simplified to show only the two cases needed to expose this uncertainty: no SD cooling and the central SD benchmark.  For each case we vary the SI cross section continuously.  The vertical markers indicate the central SI reference value and the upper SI theory reference adopted by Pospelov and Ramani; the horizontal marker gives the splitting near which the Higgsino reproduces the LZ event.  The fully thermalized and purely inelastic limits are discussed in the text rather than added as further lines.

For the central SD benchmark, we determine the SI cross section required to cool the captured Higgsino population to a given orbital radius within the age of the Sun. Specifically, fixing $\sigma_{\rm el,loop}^{\rm SD,p}=5\times10^{-47}\,\cmsq$, we solve for the value of $\sigma_{\rm el,loop}^{\rm SI,N}$ such that the orbit-averaged cooling time from the representative initial apocenter $r_{\rm max}=0.20R_\odot$ to a chosen final radius equals the Solar age, $t_\odot=4.57\,{\rm Gyr}$,
\begin{equation}
t_{\rm cool}\left(0.20R_\odot\rightarrow r_f;\sigma_{\rm el,loop}^{\rm SI,N},\sigma_{\rm el,loop}^{\rm SD,p}\right)=t_\odot.
\end{equation}
Taking $r_f=0.017R_\odot$, namely the characteristic radius below which capture--annihilation equilibrium is reached over the Solar age according to Ref.~\cite{PospelovRamani2026}, gives
\begin{equation}
\sigma_{\rm el,loop}^{\rm SI,N}\simeq6.0\times10^{-52}\,\cmsq.
\label{eq:si017}
\end{equation}
Requiring instead the orbit to contract all the way to the thermal radius, $r_f=R_{\rm therm}\simeq0.0037R_\odot$, gives
\begin{equation}
\sigma_{\rm el,loop}^{\rm SI,N}\simeq7.6\times10^{-51}\,\cmsq.
\label{eq:sitherm}
\end{equation}
These values should therefore be interpreted as the minimum SI loop cross sections, for the adopted central SD contribution, required in our orbit treatment to reach the corresponding radii within the Solar lifetime. The benchmark value $4\times10^{-50}\,\cmsq$ lies comfortably above both thresholds and therefore leads to efficient thermalization.
Reducing the SI cross section by a factor of ten still gives $R_\chi\simeq0.0066R_\odot$ and essentially the thermal endpoint; a factor-of-one-hundred reduction gives $R_\chi\simeq0.0189R_\odot$ and $\delta_{\rm lim}\simeq560.8\,\keV$.  If the SI amplitude approaches its cancellation zero, the central SD interaction alone still yields $\delta_{\rm lim}\simeq558.6\,\keV$.  Only a simultaneous suppression of both elastic channels approaches the tree-level stalled-orbit result near $512.7\,\keV$.

The shape of Figure~\ref{fig:hloop} also explains why the loop calculation matters without making the final conclusion fragile.  The SI amplitude of a nearly pure Higgsino contains cancellations among scalar and spin-two electroweak contributions, so its precise value can vary strongly with matching assumptions and higher-order effects \cite{Hisano2005,Hisano2011,ChenHill2020}.  The SD interaction has a different perturbative structure and also receives sizeable radiative corrections \cite{Hisano2011,Bisal2024}.  Our scan therefore treats Equation~\eqref{eq:looprefs} as a benchmark rather than an exact prediction.  Nevertheless, the entire interpolation from a compact thermal population to the deliberately extreme no-elastic case remains well above the terrestrial splitting required by LZ.

\begin{figure}[t]
\centering
\includegraphics[width=\columnwidth]{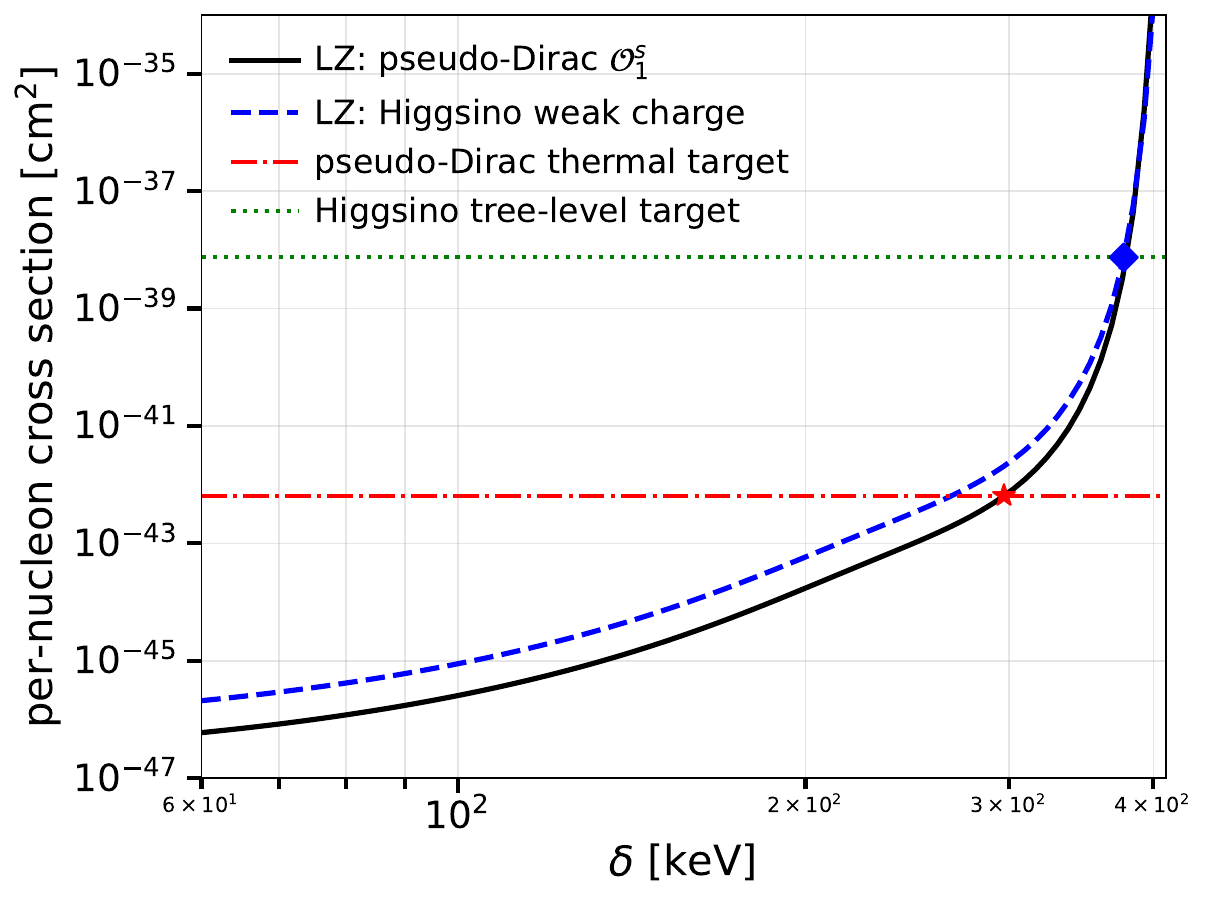}
\caption{Per-nucleon cross section required for approximately one high-recoil LZ event as a function of $\delta$.  The black solid curve is the pseudo-Dirac target and the blue dashed curve is the Higgsino target.  The red dash-dotted and green dotted horizontal lines are the corresponding thermal-model cross sections.}
\label{fig:lztargets}
\end{figure}

For completeness, our public-information recoil calculation with the Higgsino weak-charge normalization in Equation~\eqref{eq:higgsino_sigman} reproduces the LZ event near
\begin{equation}
 \delta_{\rm LZ}\simeq377.1\,\keV,
 \label{eq:lz377}
\end{equation}
with a deliberately broad halo/efficiency stress range of approximately $337$--$427\,\keV$.  Figure~\ref{fig:lztargets} shows how this Higgsino point and the pseudo-Dirac thermal benchmark arise from the same public recoil calculation.  The important comparison is therefore between a terrestrial Higgsino region below roughly $0.4\,\mathrm{MeV}$ and a Solar-neutrino endpoint above $0.5\,\mathrm{MeV}$ even in the least efficient post-capture-cooling limit.

Figure~\ref{fig:lztargets} makes the terrestrial normalization of the two coherent models explicit.  For each splitting, the black and blue curves give the nucleon cross section required to keep the high-recoil LZ event rate near one.  Intersecting those curves with the horizontal thermal-model cross sections selects the representative pseudo-Dirac and Higgsino splittings used in the Solar analysis.  The steep rise of both LZ target curves at large $\delta$ is the terrestrial counterpart of the same endothermic threshold that controls Solar capture, but the Sun accesses much larger collision speeds through gravitational acceleration.

\subsection{The $WW$ and $ZZ$ final states and the use of the IceCube $WW$ bound}
\label{subsec:higgsino_br}

A pure-Higgsino neutral pair annihilates to both $W^+W^-$ and $ZZ$.  Near the thermal mass, representative present-day continuum rates are approximately
\begin{align}
 \langle\sigma v\rangle_{WW}&\simeq8\times10^{-27}\,\cmcubeds,\\
 \langle\sigma v\rangle_{ZZ}&\simeq5\times10^{-27}\,\cmcubeds.
 \label{eq:wwzzrates}
\end{align}
corresponding to the reference fractions in Equation~\eqref{eq:higgsinoBR} \cite{Rodd2024}.  The full electroweak problem contains nearly degenerate neutral and charged states and Sommerfeld enhancement, so the exact channel fractions depend weakly on the splittings and on the electroweak treatment \cite{Beneke2015,Rodd2024}.  This uncertainty does not make either channel soft: both inject TeV-scale electroweak bosons and produce hard neutrino spectra after propagation in the Sun.

Using the IceCube upper limit for the $W^+W^-$ channel as a proxy is therefore a reasonable first estimate, but it should not be interpreted as an exact mixed-channel likelihood.  To quantify the maximum plausible weakening without a detector simulation, we take the deliberately pessimistic limit of assigning zero signal to the entire reference $ZZ$ component.  Using $B_{WW}^{\rm ref}\simeq0.62$ in Equation~\eqref{eq:conservativeBR} weakens the limit on the total annihilation rate by approximately $1/B_{WW}^{\rm ref}\simeq1.6$.  Repeating the capture crossing gives
\begin{equation}
 \delta_{\rm lim}^{\rm cons}\simeq560.0\,\keV
\end{equation}
for a thermalized population and approximately $507\,\keV$ in the no-elastic case.  Even this unphysical worst-case treatment remains far above the LZ region.  A full $WW+ZZ$ IceCube likelihood can only interpolate between this conservative construction and the hard-channel proxy, so the conclusion is robust.

\subsection{Assessment of the Pospelov--Ramani constraint}

Our independent calculation therefore supports the main conclusion of Ref.~\cite{PospelovRamani2026}.  The fully thermalized endpoint agrees at the approximately $1$--$2\,\keV$ level despite different numerical implementations.  The precise SD-only thermalization efficiency is more model dependent because it is sensitive to orbital averaging and to the loop calculation, but the extreme tree-level inelastic-stall limit still gives $\delta_{\rm lim}\simeq512.7\,\keV$.  Neither the loop uncertainty nor the $WW/ZZ$ branching uncertainty can lower the bound to the $\delta\simeq377\,\keV$ region in which the thermal Higgsino reproduces the LZ event in our public-information recoil calculation.  The thermal Higgsino interpretation of the event is therefore excluded under the standard assumptions of a full-density thermal relic and the adopted local halo density.

\section{Results II: pseudo-Dirac vector dark matter}
\label{sec:pseudo_results}

\subsection{Solar capture at the LZ thermal benchmark}

\begin{figure}[t]
\centering
\includegraphics[width=\columnwidth]{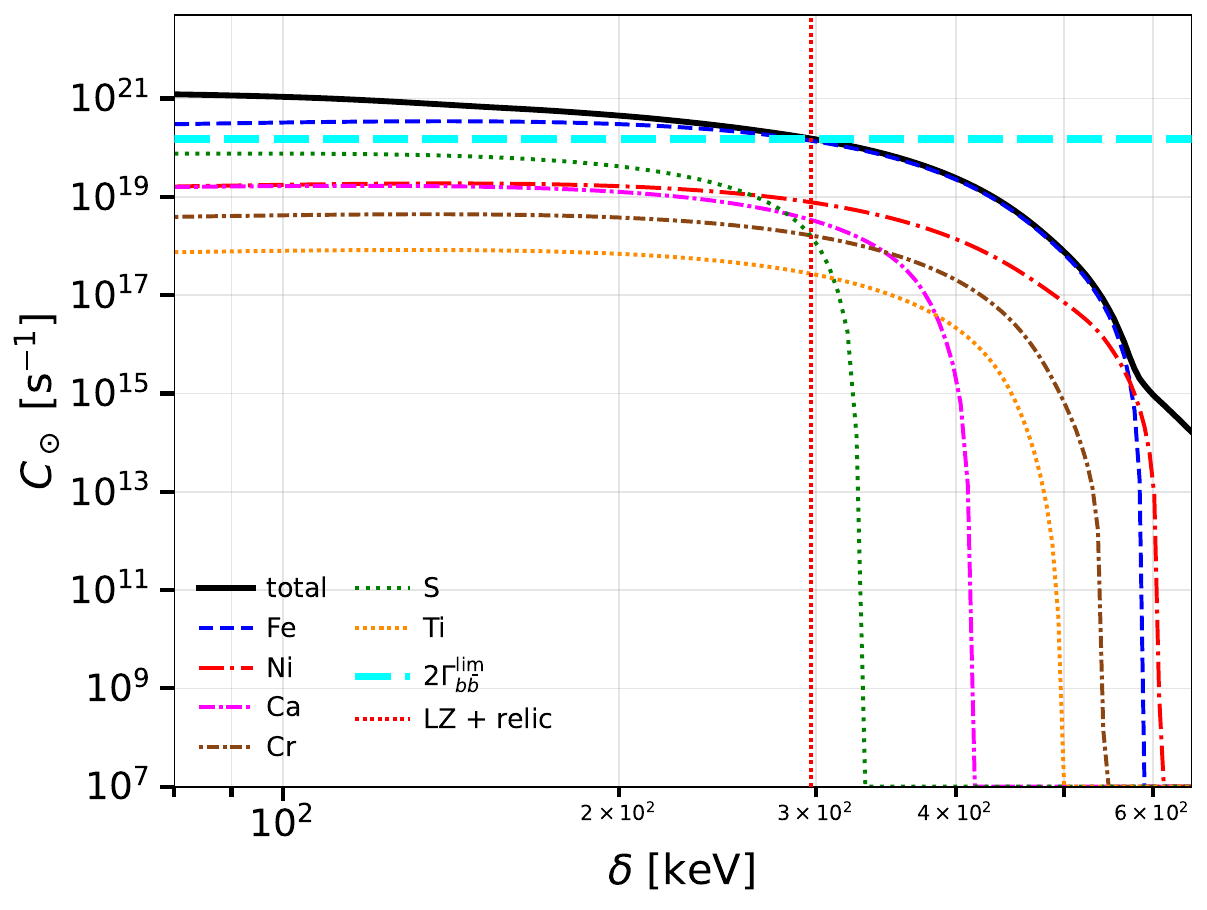}
\caption{Solar capture rate $C_\odot$ for the pseudo-Dirac vector model as a function of the mass splitting $\delta$.  The black solid curve is the total capture rate, while the colored curves show the separate Fe, Ni, Ca, Cr, S, and Ti contributions.  The cyan dashed horizontal line marks the equilibrium capture scale $C_\odot=2\Gamma_{b\bar b}^{\rm lim}$ associated with the IceCube upper limit for the $b\bar b$ channel, and the red dotted vertical line marks the LZ-plus-relic benchmark at $\delta=297\,\keV$.}
\label{fig:pdcap}
\end{figure}

The pseudo-Dirac vector interaction is much weaker at the nucleon level than the Higgsino $Z$ exchange, but Solar capture remains sizeable.  The numerical comparison is explicit from Equations~\eqref{eq:higgsino_sigman} and \eqref{eq:pdm_benchmark}: $\sigma_n^{\widetilde H}=7.43\times10^{-39}\,\cmsq$ for the Higgsino, whereas $\sigma_N=6.5\times10^{-43}\,\cmsq$ for the pseudo-Dirac benchmark, a ratio of $1.14\times10^4$.  \revtextD{The corresponding Solar capture rates at their LZ splittings are nevertheless $1.03\times10^{23}\,\mathrm{s}^{-1}$ and $1.53\times10^{20}\,\mathrm{s}^{-1}$, respectively, only a factor of about $674$ apart.}  The reason is that capture is not controlled by the nucleon cross section alone: the pseudo-Dirac benchmark has the smaller splitting, $297\,\keV$ rather than $377\,\keV$, and its coherent isoscalar interaction still accesses the heavy Fe/Ni targets that dominate high-$\delta$ Solar capture.  Figure~\ref{fig:pdcap} shows this elemental decomposition.  At the thermal LZ benchmark of Equation~\eqref{eq:pdm_benchmark}, the truncated-halo calculation gives
\begin{equation}
 C_\odot(297\,\keV)=\revtextD{1.53\times10^{20}}\,\mathrm{s}^{-1}.
 \label{eq:pdcap297}
\end{equation}
\revtextD{The elemental decomposition is highly concentrated: Fe contributes approximately $90.3\%$ and Ni $5.1\%$, followed by Ca at $2.2\%$, Cr at $1.1\%$, S at $1.0\%$ and Ti at $0.18\%$, with lighter species negligible at this splitting.  Ca, Ti and Cr lie further above their thresholds than S at this splitting, which is why Ca outweighs it here even though S is the more abundant element.}  Thus the benchmark is not capture-starved.  Indeed, if one incorrectly applies the one-species equilibrium relation, $C_\odot/2$ is already of the same order as the IceCube hard-channel limit.

Figure~\ref{fig:pdcap} should be read first as a capture calculation and only then as an input to the annihilation problem.  The elemental curves show that Fe and Ni dominate at the LZ-plus-relic point, while the smaller Ca, Cr, S, and Ti terms close at lower splittings as their endothermic thresholds are reached.  The cyan line is not a prediction for the pseudo-Dirac neutrino signal.  It is the capture rate $C_\odot=2\Gamma_{b\bar b}^{\rm lim}$ that would reproduce the IceCube $b\bar b$ upper limit only if the captured population behaved as a single annihilating species in full equilibrium.  The total capture rate at $\delta=297\,\keV$ lies essentially at this diagnostic scale, so capture suppression alone is not what makes the model viable.  The crucial next question is whether both internal states coexist in sufficient numbers to support the coannihilation process $\chi_1\chi_2\to q\bar q$.  This is why the state-transition dynamics in the following subsections must be treated explicitly.

\subsection{Kinematic closure radii are not a final Solar distribution}
\label{subsec:pseudo_orbits}

The first capture event is the endothermic transition $\chi_1 A\to\chi_2 A$.  Because the vector interaction is off diagonal, a later collision can convert the excited state back through the exothermic process $\chi_2 A\to\chi_1 A$, and a subsequent collision can excite it again whenever the relative kinetic energy is sufficient.  It is therefore important to distinguish a \emph{kinematic closure radius} from a physical final distribution.  We use the radial-orbit construction of Section~\ref{subsec:thermalization} only as a transparent diagnostic.  For stationary nuclei, Equation~\eqref{eq:stallradius} defines the formal apocenter $r_A^{\rm kin}$ below which endothermic scattering on a given species $A$ is forbidden for a radial orbit.  It does not state that the DM population stops there.

\begin{figure}[t]
\centering
\includegraphics[width=\columnwidth]{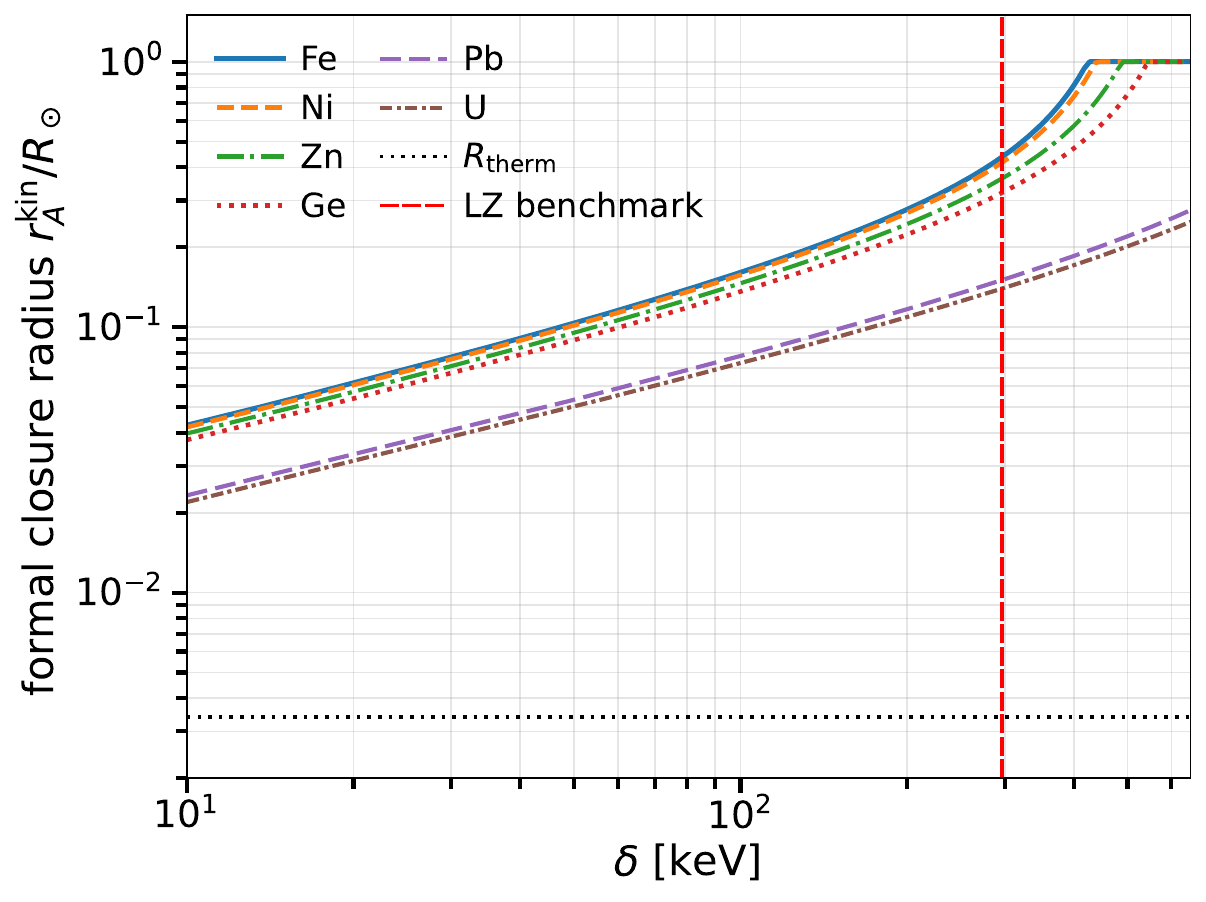}
\caption{Stationary-target kinematic closure radius $r_A^{\rm kin}/R_\odot$ for pseudo-Dirac up-scattering as a function of $\delta$.  The colored curves correspond to Fe, Ni, Zn, Ge, Pb, and U and are obtained by saturating the endothermic threshold for a radial orbit.  The black dotted horizontal line shows the reference thermal radius $r_{\rm therm}$, while the red dashed vertical line marks the $\delta=297\,\keV$ LZ-plus-relic benchmark.}
\label{fig:pdstall}
\end{figure}

Using the stationary-target approximation and the nearly radial-orbit kinematics introduced above, we determine the formal kinematic closure radius for each nuclear species from Equation~\eqref{eq:stallradius}, evaluated with the BS05 Solar potential. At the LZ benchmark $\delta=297\,\keV$, we obtain
\begin{align}
r_{\rm Fe}^{\rm kin}&=0.438R_\odot, & r_{\rm Ni}^{\rm kin}&=0.418R_\odot, &
r_{\rm Zn}^{\rm kin}&=0.364R_\odot,\\
r_{\rm Ge}^{\rm kin}&=0.321R_\odot, & r_{\rm Pb}^{\rm kin}&=0.151R_\odot, &
r_{\rm U}^{\rm kin}&=0.140R_\odot.
\label{eq:pdkinradii}
\end{align}
These radii should be interpreted only as stationary-target kinematic thresholds: they identify the apocenter below which the maximum DM speed along the assumed radial orbit is no longer sufficient to induce endothermic up-scattering on a given species. The previously tempting scale near $0.418R_\odot$ is therefore specifically the Ni stationary-target closure radius, not a demonstrated ``stopping radius.'' Less abundant heavier nuclei remain kinematically available deeper in the Sun. Their dynamical importance depends on abundance, form-factor suppression, the time spent in the relevant part of the orbit, and the velocity distribution of the Solar nuclei. Figure~\ref{fig:pdstall} makes this distinction explicit by plotting the individual closure radii instead of combining them into one effective stall radius.

This point is also where the pseudo-Dirac problem differs from the conservative Higgsino construction of Ref.~\cite{PospelovRamani2026}.  Pospelov and Ramani use uranium as the last possible stationary-nucleus inelastic target to obtain a conservative Higgsino tree-level compactification scale.  That construction is useful for the Higgsino endpoint, but it is not a calculation of the two-state pseudo-Dirac phase-space distribution.  A full treatment should evolve orbital energy and angular momentum and include thermally moving nuclei, as done in Monte Carlo studies of inelastic DM in the Sun \cite{Blennow2018,BlennowErratum2019}.  Those studies emphasize that inelastic populations need not become stationary or isothermal over the Solar age.

At the vertical LZ benchmark in Figure~\ref{fig:pdstall}, the Fe and Ni curves intersect at approximately $0.44R_\odot$ and $0.42R_\odot$, whereas the much rarer Pb and U targets remain kinematically open down to approximately $0.15R_\odot$ and $0.14R_\odot$.  The black horizontal thermal-radius line lies far below all of these stationary-target scales.  The figure therefore makes two points visually clear.  First, there is no unique ``closure radius'' shared by all nuclei.  Second, none of the colored curves gives the final spatial distribution: they indicate only where a particular stationary target ceases to allow endothermic up-scattering on a radial orbit.  Actual orbital evolution depends on the transition rates, thermal nuclear motion, and repeated up- and down-scattering, which are addressed next.

\subsection{Thermal nuclear motion and re-excitation}
\label{subsec:pseudo_reexcitation}

The internal-state evolution cannot be inferred from the stationary-target threshold alone.  A Solar nucleus has a thermal velocity distribution, so the relative collision speed is $v_{\rm rel}=|\mathbf v_\chi-\mathbf v_A|$.  Following the standard treatment of inelastic Solar scattering \cite{Blennow2018}, the local rate contains the thermal average
\begin{equation}
 dR_A=n_A(r)f_A(\mathbf v_A;r)\,\sigma_{ij}(v_{\rm rel})v_{\rm rel}\,d^3v_A,
 \label{eq:pdm_local_thermal_rate}
\end{equation}
where $f_A$ is the Maxwell--Boltzmann distribution at the local Solar temperature and $i\to j$ denotes either $1\to2$ or $2\to1$.  We denote by $\Gamma_{ij}(r_{\rm max})$ the orbit-averaged transition rate per captured DM particle from internal state $i$ to state $j$ on a trial radial orbit of apocenter $r_{\rm max}$.  We therefore evaluate
\begin{align}
 \Gamma_{ij}(r_{\rm max})={}&
 \frac{4}{T(r_{\rm max})}\int_0^{r_{\rm max}}\frac{dr}{v_\chi(r;r_{\rm max})}\nonumber\\
 &\times\sum_A n_A(r)\left\langle\sigma_{ij}v_{\rm rel}\right\rangle_{T(r)}.
 \label{eq:pdm_orbit_transition}
\end{align}
The factor of four averages over a complete radial period.  The purpose of Equation~\eqref{eq:pdm_orbit_transition} is deliberately narrower than that of a phase-space simulation.  After capture, a particle produced as $\chi_2$ can down-scatter exothermically, $\chi_2\to\chi_1$, and thereby enter the ground state.  The question is whether a later collision can reverse this transition and repopulate $\chi_2$.  For a particle kept on a specified orbit, the dimensionless quantity $\Gamma_{12}t_\odot$ is the expected number of thermally assisted re-excitations over the Solar age: $\Gamma_{12}t_\odot\ll1$ would justify neglecting re-excitation, while $\Gamma_{12}t_\odot\gg1$ means that repeated $\chi_1\to\chi_2$ transitions are possible and must be included.  The second rate, $\Gamma_{21}$, controls how quickly an excited particle is returned to $\chi_1$.  Therefore re-excitation may be dynamically important even when the instantaneous excited-state fraction remains small if $\Gamma_{21}\gg\Gamma_{12}$.

At $r_{\rm max}=r_{\rm Ni}^{\rm kin}=0.418R_\odot$ and $\delta=297\,\keV$ we obtain
\begin{align}
 \Gamma_{12}&=1.10\times10^{-10}\,\mathrm{s}^{-1},\\
 \Gamma_{21}&=1.92\times10^{-7}\,\mathrm{s}^{-1},
 \label{eq:pdm_transition_rates_ni}
\end{align}
so $\Gamma_{12}t_\odot\simeq1.6\times10^7\gg1$.  A ground-state particle placed on such an extended orbit therefore has many opportunities to be promoted back to $\chi_2$ during the Solar age, and the approximation $\Gamma_{12}=0$ is not self-consistent near the Fe/Ni kinematic scales.  This does \emph{not} mean that a large excited population accumulates.  At the same orbit, $\Gamma_{12}/\Gamma_{21}\simeq5.7\times10^{-4}$: once a particle is re-excited, exothermic down-scattering returns it to $\chi_1$ about $1.8\times10^3$ times faster than up-scattering repopulates $\chi_2$.  Re-excitation is therefore frequent over long times but the excited state is short lived.  Treating the nuclei as stationary gives instead $\Gamma_{12}\simeq4.6\times10^{-13}\,\mathrm{s}^{-1}$ at the same apocenter.  Solar nuclear thermal motion therefore enhances the threshold-sensitive up-scattering rate by more than two orders of magnitude; Fe gives the largest thermally assisted contribution, followed by Ni.

\begin{figure}[t]
\centering
\includegraphics[width=\columnwidth]{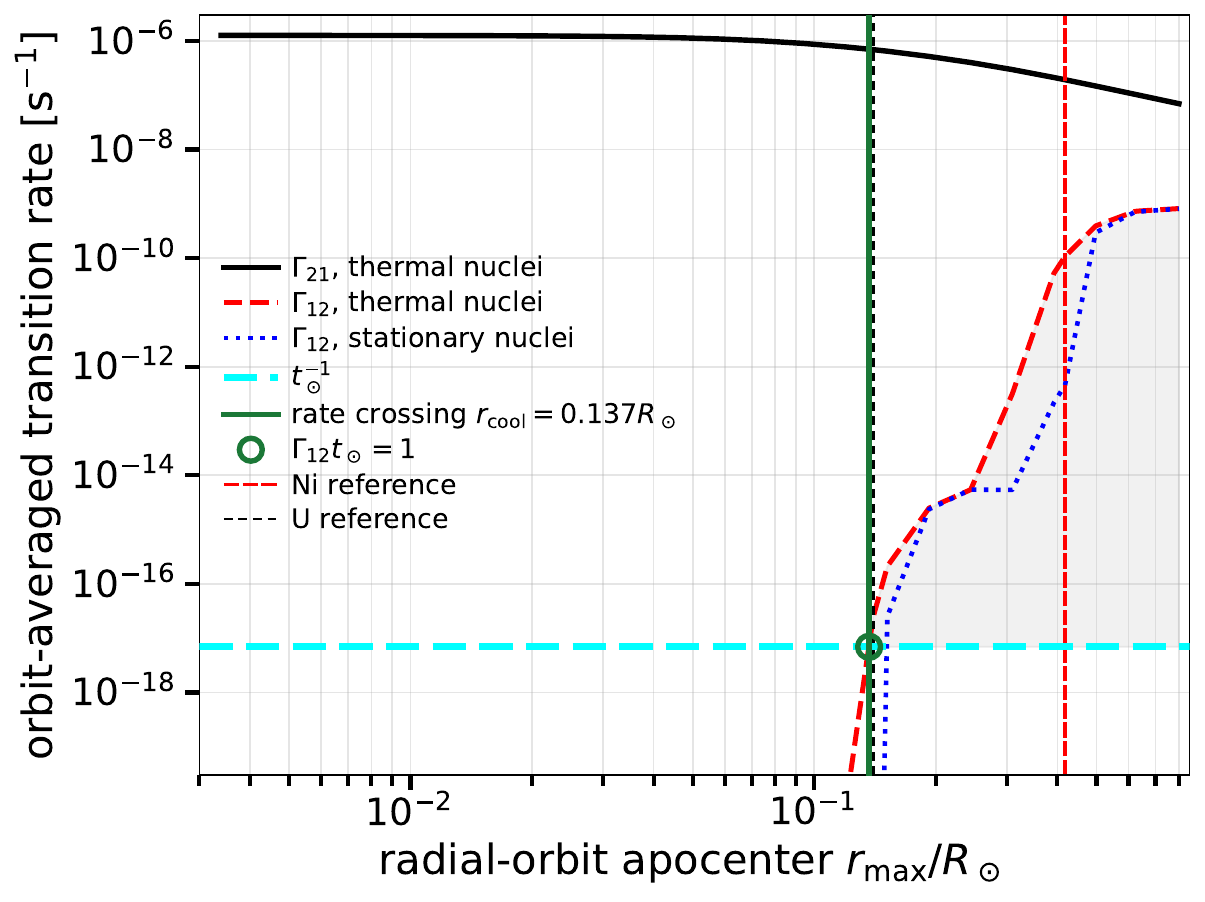}
\caption{Orbit-averaged pseudo-Dirac state-transition rates at $\delta=297\,\keV$ as a function of the radial-orbit apocenter $r_{\max}/R_\odot$.  The black solid curve is the thermally averaged de-excitation rate $\Gamma_{21}$, the red dashed curve is the thermally averaged re-excitation rate $\Gamma_{12}$, and the blue dotted curve is the stationary-target estimate of $\Gamma_{12}$.  The cyan horizontal line is $t_\odot^{-1}$; the green marker and vertical line denote the thermal $\Gamma_{12}=t_\odot^{-1}$ crossing, and the blue and red vertical lines show the U and Ni stationary-target closure radii.}
\label{fig:pddown}
\end{figure}

As the orbit becomes more compact, $\Gamma_{12}$ falls extremely rapidly.  In the same fixed-orbit diagnostic it drops below one thermally assisted up-scatter per Solar age at $r_{\rm max}\simeq0.137R_\odot$.  Its numerical proximity to the stationary-target U closure scale in Equation~\eqref{eq:pdkinradii} is accidental.  Section~\ref{subsec:pseudo_cooling} uses this crossing only as a characteristic scale in a semi-analytic cooling estimate; it is not identified with an exact stopping radius or a reconstructed final distribution.  Figure~\ref{fig:pddown} is intended to answer three distinct questions before any annihilation calculation is attempted.  First, comparing the thermally averaged up-scattering rate $\Gamma_{12}$ with $t_\odot^{-1}$ determines whether a particle on a given trial orbit is likely to be re-excited at all during the Solar age.  Second, comparing the thermal-nucleus and stationary-nucleus $\Gamma_{12}$ curves quantifies how strongly nuclear thermal motion reopens an otherwise threshold-suppressed transition.  Third, comparing $\Gamma_{21}$ with $\Gamma_{12}$ determines whether re-excited particles accumulate or are quickly returned to the ground state.
For the LZ benchmark the conclusion is therefore clear. Thermally assisted re-excitation cannot be neglected while the captured DM population remains on sufficiently extended orbits, because the larger orbital velocities and the thermal motion of Solar nuclei can reopen the endothermic transition $\chi_1 A\rightarrow\chi_2 A$ even close to the stationary-target closure threshold. However, once an excited state is produced, the reverse process $\chi_2 A\rightarrow\chi_1 A$ proceeds much more efficiently and rapidly removes the population from $\chi_2$. As a result, re-excitation affects the post-capture evolution and must be included in the kinetics, but it does not generate a large steady-state excited-state abundance: the instantaneous fraction of captured particles in $\chi_2$ remains small.
Figure~\ref{fig:pddown} is a kinetics and timescale diagnostic; the separate annihilation consequence is shown in Figure~\ref{fig:pdann}.

The hierarchy visible in Figure~\ref{fig:pddown} is the key kinetic result.  On extended Fe/Ni-scale orbits the red thermal $\Gamma_{12}$ curve lies well above both the stationary-target result and $t_\odot^{-1}$, so Solar nuclear motion makes re-excitation unavoidable over the Solar age.  At the same time the black $\Gamma_{21}$ curve remains orders of magnitude larger than $\Gamma_{12}$, which means that an excited particle is returned to $\chi_1$ much faster than the ground state is repopulated into $\chi_2$.  Moving to smaller apocenters suppresses $\Gamma_{12}$ sharply, and the green crossing near $0.137R_\odot$ marks the trial-orbit scale where the expected number of re-excitations over $t_\odot$ falls to unity.  Thus the figure explains how re-excitation can be dynamically frequent without producing a large instantaneous excited-state abundance.

If the population genuinely reaches local thermal equilibrium in the core, the situation is different.  Detailed balance then gives an equilibrium excited fraction of order
\begin{equation}
 \frac{n_2}{n_1}\sim e^{-\delta/T_c},
\end{equation}
and for $\delta=297\,\keV$ and $T_c\simeq1.3\,\keV$,
\begin{equation}
 \frac{n_2}{n_1}\sim 6.0\times10^{-100}.
 \label{eq:boltzmannratio}
\end{equation}
Thus thermally assisted re-excitation matters for nonthermal eccentric orbits near threshold, whereas a fully thermalized compact population contains essentially no excited state.  These are two different dynamical regimes and should not be conflated.

The role of re-excitation is particularly important for the pseudo-Dirac model because its leading annihilation process is the coannihilation $\chi_1\chi_2\to q\bar q$, so the annihilation rate depends directly on the excited-state population. This differs from the Higgsino case, where annihilation of the ground state $\chi_1\chi_1$ into electroweak final states remains efficient after de-excitation and the principal post-capture uncertainty is instead the spatial compactification produced by loop-induced elastic scattering.

\subsection{Semi-analytic estimate of re-excitation cooling}
\label{subsec:pseudo_cooling}

The previous section showed that thermally assisted re-excitation $\chi_1 A\rightarrow\chi_2 A$ can remain active for captured pseudo-Dirac DM on sufficiently extended orbits, while the reverse de-excitation process $\chi_2 A\rightarrow\chi_1 A$ is typically much faster and keeps the excited-state fraction small. These repeated transitions can affect not only the relative populations of $\chi_1$ and $\chi_2$, but also the orbital energy of the captured DM. It is therefore important to estimate whether successive up-scattering and down-scattering cycles produce a net secular cooling of the orbit and drive the captured population toward smaller radii.
A simple way to understand this effect is first to consider a stationary Solar nucleus. One complete cycle, $\chi_1\rightarrow\chi_2\rightarrow\chi_1$, changes the DM orbital energy by the sum of the kinetic energies transferred to the recoiling nuclei,
\begin{equation}
\Delta E_{\rm cycle}=-\left(E_R^{\uparrow}+E_R^{\downarrow}\right),
\label{eq:pdm_cycle_loss}
\end{equation}
because the mass-splitting energy $\delta$ absorbed in the endothermic up-scattering is returned to kinetic energy in the subsequent exothermic transition. The splitting therefore cancels over a complete cycle, whereas the recoil energy transferred to the Solar nuclei represents a net loss from the DM orbital motion. Repeated cycles consequently tend to reduce the orbital energy and contract the orbit.
For thermally moving nuclei the situation is more complicated because the nuclei themselves carry kinetic energy. The energy exchange is then stochastic, and an individual collision can either cool or heat the DM particle. Equation~\eqref{eq:pdm_cycle_loss} should therefore not be interpreted as an exact event-by-event relation for the finite-temperature scattering kernel. Since the Solar-core temperature is of order keV, while the characteristic orbital and recoil energies relevant here are of order MeV, the stationary-target result nevertheless provides a useful semi-analytic estimate of the direction and magnitude of the mean secular energy drift.

To quantify its possible size, we use the same radial-orbit approximation as in the fixed-orbit scan, take Fe as a representative cycling target, and estimate the recoil loss from central collisions.  Using this representative Fe recoil-loss prescription, the estimated orbital-energy loss per complete $\chi_1\to\chi_2\to\chi_1$ cycle is approximately $1.4\,\mathrm{MeV}$ at $r_{\rm max}=0.62R_\odot$, $1.1\,\mathrm{MeV}$ at the Ni reference radius $r_{\rm max}=0.418R_\odot$, and $0.5\,\mathrm{MeV}$ at $r_{\rm max}=0.14R_\odot$. The transition rate itself includes the thermal motion of the Solar nuclei, whereas these recoil losses are evaluated with the simplified stationary-target, central-collision prescription described above. 
To estimate how these repeated transitions modify the orbit, we translate the energy lost in each $\chi_1\rightarrow\chi_2\rightarrow\chi_1$ cycle into a secular change of the orbital apocenter. In the simplified radial-orbit description, the magnitude of the DM binding energy is\footnote{The binding energy is the energy required to bring the captured DM particle from its bound orbit to infinity with zero velocity. For a nearly radial orbit, the velocity vanishes at the apocenter $r_{\rm max}$, so the orbital energy is $E_{\rm orb}=m_\chi\Phi(r_{\rm max})<0$. Therefore $E_b\equiv-E_{\rm orb}=-m_\chi\Phi(r_{\rm max})=\frac{1}{2}m_\chi v_{\rm esc}^2(r_{\rm max})$, using $v_{\rm esc}^2(r)=-2\Phi(r)$.}
\begin{equation}
E_b(r_{\rm max})=\frac{1}{2}m_\chi v_{\rm esc}^2(r_{\rm max}),
\end{equation}
which increases as the orbit contracts and $r_{\rm max}$ moves inward. We take the thermally averaged up-scattering rate $\Gamma_{12}(r_{\rm max})$ to set the characteristic rate of complete excitation--de-excitation cycles, with the up-scattering step acting as the bottleneck. Since each complete cycle removes on average an orbital energy $\langle E_R^{\uparrow}+E_R^{\downarrow}\rangle$, the corresponding average increase of the binding energy is
\begin{equation}
\frac{dE_b}{dt}=\Gamma_{12}(r_{\rm max})\left\langle E_R^{\uparrow}+E_R^{\downarrow}\right\rangle.
\end{equation}
Using $dE_b/dt=(dE_b/dr_{\rm max})(dr_{\rm max}/dt)$ then gives the effective one-dimensional cooling equation
\begin{align}
\frac{dr_{\rm max}}{dt}&=-\frac{\Gamma_{12}(r_{\rm max})\,\langle E_R^{\uparrow}+E_R^{\downarrow}\rangle}{\left|dE_b/dr_{\rm max}\right|}.
\label{eq:pdm_cooling_track}
\end{align}
The negative sign explicitly shows that energy loss contracts the orbit, $dr_{\rm max}/dt<0$. At large $r_{\rm max}$, thermally assisted up-scattering is relatively efficient and the orbit contracts rapidly. As $r_{\rm max}$ decreases, however, the maximum DM speed reached inside the Sun decreases and the thermal tail of the Solar nuclei becomes increasingly important for overcoming the endothermic threshold. The rate $\Gamma_{12}$ therefore falls rapidly and the orbital evolution slows down.

As the orbit contracts, the thermally averaged up-scattering rate $\Gamma_{12}(r_{\rm max})$ decreases rapidly because increasingly energetic Solar nuclei are required to overcome the endothermic threshold. The inverse rate, $\tau_{12}=1/\Gamma_{12}$, gives the characteristic waiting time for one additional $\chi_1\rightarrow\chi_2$ transition. We define a useful cooling scale by requiring this waiting time to become equal to the Solar age,
\begin{equation}
\Gamma_{12}(r_{\rm cool})t_\odot=1,
\qquad
r_{\rm cool}\simeq0.137R_\odot.
\label{eq:pdm_stall}
\end{equation}
Thus, at $r_{\rm cool}$, a captured particle would require on average one Solar age to undergo one further up-scattering. Below this radius the characteristic scattering time is even longer, so continued inelastic cooling becomes extremely inefficient. The quantity $r_{\rm cool}$ is therefore not a sharp final radius, but a convenient diagnostic of where the secular orbital contraction effectively stalls.
Integrating Equation~\eqref{eq:pdm_cooling_track} makes this slowdown explicit. Within the simplified radial-orbit treatment, particles initially on $0.3$--$0.8R_\odot$ orbits reach approximately $0.30R_\odot$ after $4\times10^{4}\,{\rm yr}$, $0.20R_\odot$ after $1.1\times10^{7}\,{\rm yr}$, $0.15R_\odot$ after $1.0\times10^{8}\,{\rm yr}$, and $0.139R_\odot$ after $3.8\times10^{8}\,{\rm yr}$. The rapid increase of the cooling time toward smaller radii reflects the strong suppression of $\Gamma_{12}$ near the endothermic threshold. In particular, reaching $0.139R_\odot$ does not imply that the particle will continue inward at the same rate: $\Gamma_{12}$ decreases steeply over the last few percent in radius, and at $r_{\rm cool}\simeq0.137R_\odot$ the mean waiting time for the next up-scattering has already grown to the age of the Sun.

For completeness, applying the same continuous-capture one-zone prescription to the resulting radial fractions gives an illustrative present-day annihilation rate of order
\begin{equation}
 \Gamma_A(t_\odot)\sim4\times10^{10}\,\mathrm{s}^{-1},
 \label{eq:pdm_gamma_today}
\end{equation}
well below the fixed-orbit maximum.  We use this number only as evidence that the secular cooling tendency strengthens, rather than weakens, the two-state suppression.  The robust exclusion test does not rely on it: the fixed-orbit calculation already scans the extended orbits on which re-excitation is largest and gives the conservative envelope of Equation~\eqref{eq:pdm_gamma_fo_max}.  A full Monte Carlo in orbital energy, angular momentum, target species, and internal state is required before assigning a precise final radius or present-day $\Gamma_A$.

Two additional checks support use of this estimate as a diagnostic.  In the sampled bound orbits, the maximum kinetic-energy gain from a single exothermic down-scatter is much smaller than the several-MeV binding energy, so state-changing ejection is not expected to dominate.  In addition, the orbital period is hours whereas transition times are much longer, validating the orbit-averaged treatment.  These checks do not remove the need for a phase-space calculation; they only show that the semi-analytic cooling picture is dynamically plausible.

\subsection{General two-state evolution and the fixed-orbit re-excitation calculation}
\label{subsec:twostate}

The minimal pseudo-Dirac model must be described by coupled internal-state populations.  We denote by $N_1(t)$ and $N_2(t)$ the numbers of captured ground- and excited-state particles.  Neglecting evaporation and taking capture to populate predominantly $\chi_2$,
\begin{align}
 \frac{dN_1}{dt}&=\Gamma_{21}N_2-\Gamma_{12}N_1-C_{12}N_1N_2,\label{eq:N1}\\
 \frac{dN_2}{dt}&=C_\odot+\Gamma_{12}N_1-\Gamma_{21}N_2-C_{12}N_1N_2,\label{eq:N2}
\end{align}
with
\begin{equation}
 C_{12}=\frac{\langle\sigma v\rangle_{12}}{V_{12}^{\rm eff}},
 \qquad
 \Gamma_A=C_{12}N_1N_2.
 \label{eq:pdmGamma}
\end{equation}
Here $V_{12}^{\rm eff}$ is the effective overlap volume of the two state distributions.  A full solution requires $\Gamma_{12}$, $\Gamma_{21}$, and $V_{12}^{\rm eff}$ to evolve together with the orbital-energy and angular-momentum distributions.  We do not claim to have performed that Monte Carlo calculation.  We instead perform a \emph{fixed-orbit re-excitation calculation}, whose meaning is the following.  We choose a trial radial orbit with apocenter $r_{\rm max}$ and, only for the purpose of evaluating the state kinetics, keep that orbit fixed.  On it we compute the thermally averaged up- and down-scattering rates $\Gamma_{12}$ and $\Gamma_{21}$, solve for the corresponding two-state populations in the quasi-steady approximation below, and map the same characteristic scale to a one-zone overlap volume.  We repeat the procedure over a broad range of $r_{\rm max}$.  This is not a claim that the particles actually stop at any of the trial radii.  Rather, it is a controlled conditional calculation that asks whether repeated re-excitation can make the $\chi_1\chi_2$ coannihilation signal approach the IceCube limit anywhere over the relevant range of orbital scales.  Section~\ref{subsec:pseudo_cooling} then uses a deliberately simplified energy-loss model to test the direction and possible size of the secular orbital drift.  If the fixed-orbit rate remains small throughout the scan, the viability conclusion does not depend on assigning a precise final orbit. 

For the benchmark, all annihilation rates found in this fixed-orbit calculation are tiny compared with $C_\odot/2$, so the total captured population is well approximated by $N_1+N_2\simeq C_\odot t_\odot$.  Because $\Gamma_{21}$ is fast compared with secular accumulation, the excited state is quasi-steady,
\begin{equation}
 N_2\simeq\frac{C_\odot+\Gamma_{12}N_1}{\Gamma_{21}+C_{12}N_1},
 \label{eq:pdm_N2_reexc}
\end{equation}
with $N_1\simeq C_\odot t_\odot$.  The corresponding fixed-orbit annihilation rate is
\begin{equation}
 \Gamma_A^{\rm FO}\simeq
 C_{12}(C_\odot t_\odot)
 \frac{C_\odot+\Gamma_{12}(C_\odot t_\odot)}
 {\Gamma_{21}+C_{12}(C_\odot t_\odot)}.
 \label{eq:pdm_gamma_reexc}
\end{equation}
Appendix~\ref{app:twostate} gives the relation to the zero-re-excitation analytic limit and discusses the approximations.

The correction is numerically large relative to the old $\Gamma_{12}=0$ limit but still far too small to produce an IceCube constraint.  At the Ni kinematic reference radius, the zero-re-excitation result using the orbit-averaged $\Gamma_{21}$ is $\Gamma_A\simeq7.0\times10^6\,\mathrm{s}^{-1}$, whereas Equation~\eqref{eq:pdm_gamma_reexc} gives
\begin{equation}
 \Gamma_A^{\rm FO}(r_{\rm Ni}^{\rm kin})\simeq1.1\times10^{14}\,\mathrm{s}^{-1}.
 \label{eq:pdm_gamma_ni_reexc}
\end{equation}
At this reference orbit, including re-excitation therefore raises the two-state annihilation rate by a factor of approximately $1.6\times10^7$ relative to the artificial $\Gamma_{12}=0$ limit.
Scanning fixed radial-orbit apocenters from the thermal scale to $0.8R_\odot$, the largest rate occurs near $r_{\rm max}\simeq0.62R_\odot$ and is
\begin{equation}
 \Gamma_{A,\max}^{\rm FO}\simeq\revtextD{4.4}\times10^{14}\,\mathrm{s}^{-1}
 \simeq\revtextD{5.8}\times10^{-6}\Gamma_{b\bar b}^{\rm lim}.
 \label{eq:pdm_gamma_fo_max}
\end{equation}
The role of re-excitation is therefore twofold and should be stated precisely.  It is \emph{dynamically important}: omitting it changes the two-state annihilation rate by many orders of magnitude and is inconsistent for extended Fe/Ni-scale orbits.  It is nevertheless \emph{not large enough to produce an IceCube exclusion}: rapid down-scattering keeps the excited population small, with $N_2/N_1\simeq5.7\times10^{-4}$ at the Ni reference radius and only about $6.9\times10^{-3}$ near the maximum of the fixed-orbit scan.  \revtextC{Consequently even the maximum of the scan remains lower than the IceCube upper limit for the $b\bar b$ channel used in our nominal comparison by a factor of about \revtextD{$1.7\times10^5$}, and the maximum is itself an envelope: it belongs to an extended trial orbit that the semi-analytic cooling diagnostic contracts well within the Solar age (Section~\ref{subsec:pseudo_cooling}), and the present-day rate of Equation~\eqref{eq:pdm_gamma_today} lies four further orders of magnitude below it.}  Since $2\Gamma_{A,\max}^{\rm FO}/C_\odot\simeq\revtextD{5.7}\times10^{-6}$, the approximation $N_1+N_2\simeq C_\odot t_\odot$ is internally consistent over the entire scan.

\begin{figure}[t]
\centering
\includegraphics[width=\columnwidth]{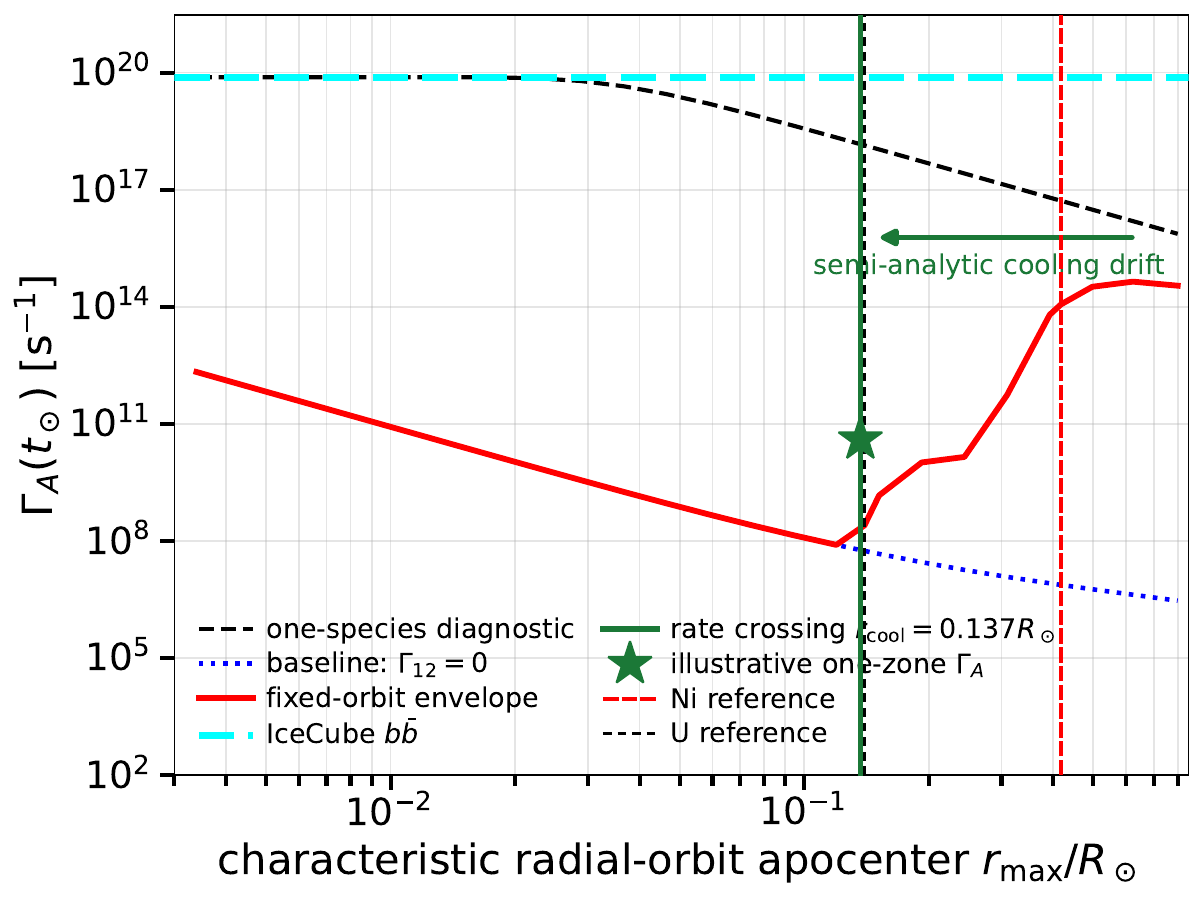}
\caption{Present-day pseudo-Dirac annihilation rate $\Gamma_A(t_\odot)$ at $\delta=297\,\keV$ as a function of the characteristic radial-orbit apocenter.  The black dashed curve is the one-species/equal-population diagnostic, the blue dotted curve is the $\Gamma_{12}=0$ baseline, and the red solid curve is the two-state fixed-orbit result including re-excitation.  The cyan horizontal line is the IceCube $b\bar b$ upper limit; the green vertical line and star show the semi-analytic cooling diagnostics, and the blue and red vertical lines mark the U and Ni closure radii.}
\label{fig:pdann}
\end{figure}

Figure~\ref{fig:pdann} translates the transition-rate hierarchy into the annihilation signal.  At compact trial radii the red two-state curve lies close to the blue $\Gamma_{12}=0$ baseline because endothermic re-excitation is strongly suppressed.  On more extended orbits the rapid increase of $\Gamma_{12}$ raises the red curve by many orders of magnitude, with the fixed-orbit envelope reaching its maximum near $r_{\rm max}\simeq0.62R_\odot$.  Even at that maximum it remains more than five orders of magnitude below the IceCube $b\bar b$ upper limit shown by the cyan line.  The green cooling direction points toward smaller apocenters, while the green star is the illustrative one-zone rate obtained from the semi-analytic cooling estimate in Eq.~\ref{eq:pdm_gamma_today}.  The star lies far below the fixed-orbit maximum, which is why we use the red envelope, rather than the semi-analytic final-radius estimate, as the conservative result for the IceCube comparison.

\subsection{A state-fraction-independent annihilation ceiling and the IceCube margin}

The fixed-orbit calculation discussed above predicts a very small excited-state population, and therefore a strongly suppressed annihilation rate. It is useful, however, to construct a complementary bound that does not rely on the calculated partition between $\chi_1$ and $\chi_2$. This allows us to ask a more conservative question: even if the two internal states were populated in the way that maximizes annihilation, how compact would the captured DM population have to become before reaching the IceCube upper limit?
The total number of DM particles retained in the Sun cannot exceed the number accumulated through capture over the Solar age. Neglecting both ejection and annihilation therefore gives the conservative upper bound
\begin{equation}
N_1+N_2\leq C_\odot t_\odot.
\end{equation}
For a fixed total population, the product $N_1N_2$, which controls $\chi_1\chi_2$ annihilation, is maximized when the two states are equally populated. Hence
\begin{equation}
N_1N_2\leq\frac{(N_1+N_2)^2}{4}\leq\frac{(C_\odot t_\odot)^2}{4}.
\end{equation}
This result removes the uncertainty associated with the internal-state fraction: it gives the largest annihilation rate compatible with a given total captured population, independently of the detailed two-state kinetics.
In a one-zone description with effective overlap volume
\begin{equation}
V_{12}^{\rm eff}=\frac{4\pi}{3}R_{\rm eff}^3,
\end{equation}
the annihilation rate is
\begin{equation}
\Gamma_A=\frac{\langle\sigma v\rangle_{12}}{V_{12}^{\rm eff}}N_1N_2,
\end{equation}
and therefore satisfies the population ceiling
\begin{equation}
\Gamma_A\leq\Gamma_A^{\rm pop}(R_{\rm eff})=\frac{\langle\sigma v\rangle_{12}}{V_{12}^{\rm eff}}\frac{(C_\odot t_\odot)^2}{4}.
\label{eq:pdm_population_ceiling}
\end{equation}
Because this expression assumes both the largest possible retained population and the maximally favorable partition $N_1=N_2$, it deliberately overestimates the annihilation rate. We can therefore use it to determine how small the effective overlap radius would have to become before IceCube could constrain the model even under these extreme assumptions. Setting Equation~\eqref{eq:pdm_population_ceiling} equal to the adopted IceCube $b\bar b$ upper limit gives
\begin{equation}
R_{\rm crit}=0.0370R_\odot.
\label{eq:pdm_Rcrit}
\end{equation}
Within the one-zone mapping, any population with $R_{\rm eff}>R_{\rm crit}$ necessarily remains below the IceCube $b\bar b$ upper limit even if $N_1=N_2$. Both the Ni and U kinematic reference radii are substantially larger than $R_{\rm crit}$. The semi-analytic cooling scale $r_{\rm cool}\simeq0.137R_\odot$ is also larger than this value, indicating that the estimated secular contraction does not drive the benchmark into the region where even the equal-population ceiling would become constrained.
The opposite limit also illustrates why this bound should not be interpreted as a prediction for the final Solar distribution. A fully thermalized population would have a characteristic radius much smaller than $R_{\rm crit}$, but in that regime the excited-state abundance is exponentially suppressed according to Equation~\eqref{eq:boltzmannratio}. The compactness of the spatial distribution and the relative occupancy of the two states therefore affect the annihilation rate in different ways. Equation~\eqref{eq:pdm_population_ceiling} deliberately removes the second effect by maximizing $N_1N_2$, whereas the explicit two-state calculation retains the actual state kinetics.

\begin{figure}[t]
\centering
\includegraphics[width=\columnwidth]{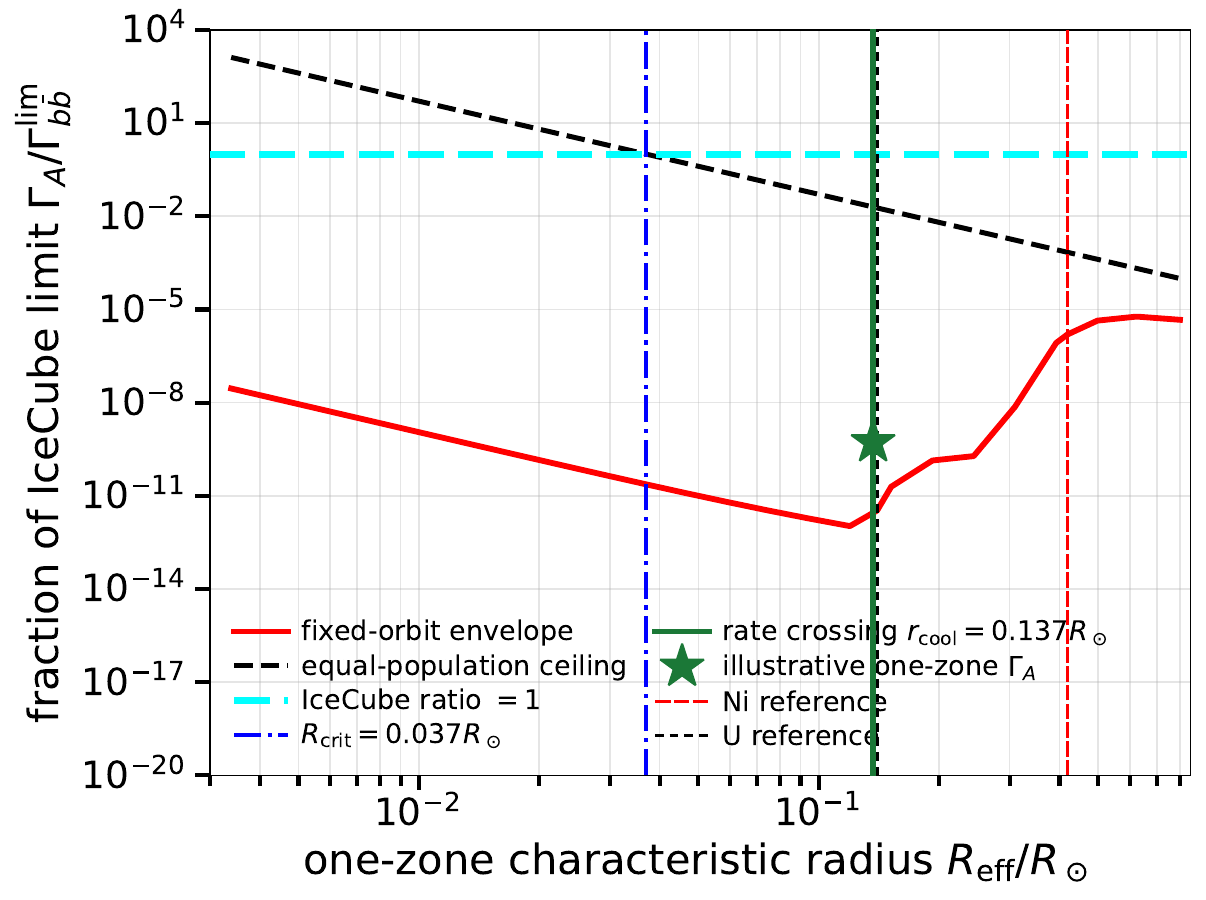}
\caption{Pseudo-Dirac annihilation rate normalized to the IceCube upper limit for the $b\bar b$ channel, $\Gamma_A/\Gamma_{b\bar b}^{\rm lim}$, as a function of the one-zone characteristic radius. The black dashed curve shows the equal-population ceiling obtained by maximizing $N_1N_2$ at fixed total captured population, while the red solid curve shows the fixed-orbit two-state result using the calculated internal-state kinetics. The cyan horizontal line marks the IceCube upper limit, the blue dash-dotted vertical line gives $R_{\rm crit}$, the green line and star indicate the semi-analytic cooling diagnostics, and the remaining vertical reference lines show the U and Ni closure radii.}
\label{fig:pdbound}
\end{figure}

Figure~\ref{fig:pdbound} shows these two complementary estimates through the dimensionless ratio $\Gamma_A/\Gamma_{b\bar b}^{\rm lim}$. The horizontal value one corresponds to the adopted IceCube upper limit. The black dashed curve shows the equal-population ceiling of Equation~\eqref{eq:pdm_population_ceiling}; its crossing at $R_{\rm crit}=0.0370R_\odot$ gives the characteristic one-zone radius below which this deliberately maximal configuration would exceed the limit. The red curve instead uses the calculated up- and down-scattering rates and therefore includes the suppression of the excited-state population found in the fixed-orbit two-state treatment. It remains many orders of magnitude below the IceCube limit throughout the scan.
The figure therefore separates the two main uncertainties in a transparent way. The black curve asks how large the annihilation rate could become if the internal-state populations were arranged to maximize annihilation, independently of the actual two-state kinetics. The red curve instead uses the calculated state populations within the fixed-orbit approximation. The fact that both the semi-analytic cooling scale and the relevant kinematic reference radii lie to the right of $R_{\rm crit}$, while the explicit two-state result remains far below unity, shows that the pseudo-Dirac benchmark retains a large margin with respect to the adopted IceCube constraint.

\subsection{Annihilation branching fractions and the IceCube template}

For universal vector couplings the leading coannihilation is to quarks.  Including the quark-mass phase-space factor, our benchmark gives
\begin{align}
 B_{u\bar u}&=0.16668, & B_{d\bar d}&=0.16668, & B_{s\bar s}&=0.16668,\\
 B_{c\bar c}&=0.16668, & B_{b\bar b}&=0.16668, & B_{t\bar t}&=0.16662.
 \label{eq:pdmBR}
\end{align}
The model therefore does not predict a pure $b\bar b$ neutrino spectrum.  Its contact interaction gives an approximately democratic six-flavor $q\bar q$ mixture once all thresholds are open.  The light-flavor tree-level cascade is strongly softened by Solar interactions, whereas charm, bottom, and especially top provide harder prompt components; electroweak radiation additionally generates a hard tail for TeV-scale light-fermion primaries \cite{IbarraTotzauerWild2014,Bauer2021}.  There is consequently no single ``dominant IceCube channel'' that can be assigned without a branching-fraction-weighted propagation of the complete mixture.  For the present robustness test we use the IceCube upper-limit curve for the $b\bar b$ channel as a soft-hadronic proxy and the $WW$ upper-limit curve as a deliberately hard comparison.  The fixed-orbit maximum remains below these two templates by factors of approximately $1.7\times10^5$ and $3.4\times10^4$, respectively.  The conclusion is therefore insensitive to the channel mapping at the precision relevant here; a full spectrum propagation is needed only for a precision experimental limit.  The dominant remaining theoretical task for the Solar dynamics is a full $E$--$L$ phase-space evolution of the two internal states.

\subsection{Take-home message for the pseudo-Dirac model}

For the pseudo-Dirac interpretation, we have tested explicitly whether the large Solar capture rate can nevertheless produce an observable neutrino signal. The relevant IceCube comparison for our nominal soft-hadronic proxy is $\Gamma_{b\bar b}^{\rm lim}=7.5\times10^{19}\,{\rm s}^{-1}$. A treatment based only on stationary Solar nuclei would suggest that inelastic up-scattering closes near the Ni reference radius, $r_{\rm Ni}^{\rm kin}\simeq0.418R_\odot$. We find, however, that this is not a physical stopping radius: the thermal motion of Solar nuclei efficiently reopens the endothermic transition and repeatedly repopulates $\chi_2$. At the Ni reference orbit, including this thermally assisted re-excitation gives $\Gamma_A\simeq1.2\times10^{14}\,{\rm s}^{-1}$ still about $6\times10^5$ below the adopted IceCube $b\bar b$ upper limit. 
Scanning over the fixed radial orbits gives a maximum annihilation rate $\Gamma_A^{\rm max}\simeq4.4\times10^{14}\,{\rm s}^{-1}$,
which remains approximately $1.7\times10^5$ below the same IceCube limit. Thus, although thermal re-excitation increases the annihilation rate substantially compared with a calculation in which re-excitation is neglected, the explicit two-state kinetics keeps the signal far below the present neutrino constraint because the excited state is rapidly depleted by the much faster exothermic transition $\chi_2\rightarrow\chi_1$.
We have also checked whether further orbital contraction could invalidate this conclusion. Repeated excitation--de-excitation cycles can cool the captured population below the stationary-target Ni closure scale, but the up-scattering rate decreases rapidly as the orbit contracts. Our semi-analytic treatment gives a characteristic cooling scale $r_{\rm cool}\simeq0.137R_\odot$, where the mean waiting time for an additional up-scattering becomes comparable to the Solar age. Since this calculation does not evolve the full distribution in orbital energy and angular momentum, we do not interpret $r_{\rm cool}$ as a precise final radius.
Finally, we constructed a deliberately conservative bound that does not use the calculated $\chi_1$--$\chi_2$ population ratio. Maximizing the annihilation rate by imposing $N_1=N_2$, the IceCube $b\bar b$ limit is reached only if the effective overlap radius becomes smaller than $R_{\rm crit}\simeq0.037R_\odot$.
This scale is substantially smaller than the semi-analytic cooling scale. The fixed-orbit calculation and the state-fraction-independent ceiling therefore lead to the same qualitative conclusion: thermal re-excitation and orbital cooling must be included in the Solar evolution, but neither effect brings the pseudo-Dirac benchmark close to the current IceCube constraint. A full phase-space calculation would refine the final Solar distribution, but is not required to establish the large margin found in the explicit calculations presented here.

\section{Results III: neutron-philic endothermic spin-dependent dark matter}
\label{sec:nsid_results}

The third benchmark provides a useful counterexample to the two previous Solar histories. In the Higgsino case, DM is captured efficiently and the captured population can also annihilate efficiently. In the minimal pseudo-Dirac vector model, capture is still efficient, but the annihilation rate is strongly suppressed by the two-state population dynamics. In the neutron-philic SD model, by contrast, the suppression already occurs at the level of Solar capture itself: only a small subset of Solar nuclei can satisfy both the endothermic kinematics and the required neutron-spin structure.
The reduction of the capture rate has two distinct origins. First, for the chosen mass splitting, the endothermic threshold kinematically excludes all Solar targets lighter than approximately $A\simeq30$, independently of the coupling strength. Second, the neutron-philic isospin structure suppresses the contribution of heavy odd-proton isotopes that would otherwise remain kinematically accessible. The surviving capture rate is therefore controlled by a small set of relatively rare heavy odd-neutron isotopes.

\subsection{LZ normalization and the inelastic $\mathcal{O}_4$ region}

LZ directly tests inelastic $\mathcal{O}_4$ at $m_\chi=1\,\TeV$.  Supplemental Table S8 gives a local preference of $3.4\sigma$ for $\mathcal{O}_4^s$ and $3.3\sigma$ for $\mathcal{O}_4^v$ at $\delta=300\,\keV$, rising to $3.4\sigma$ for both at $350\,\keV$ \cite{LZ2026}.  The preference is therefore an experimental property of the full LZ detector-level likelihood and does not depend on our simplified rate normalization.

\begin{figure}[t]
\centering
\includegraphics[width=\columnwidth]{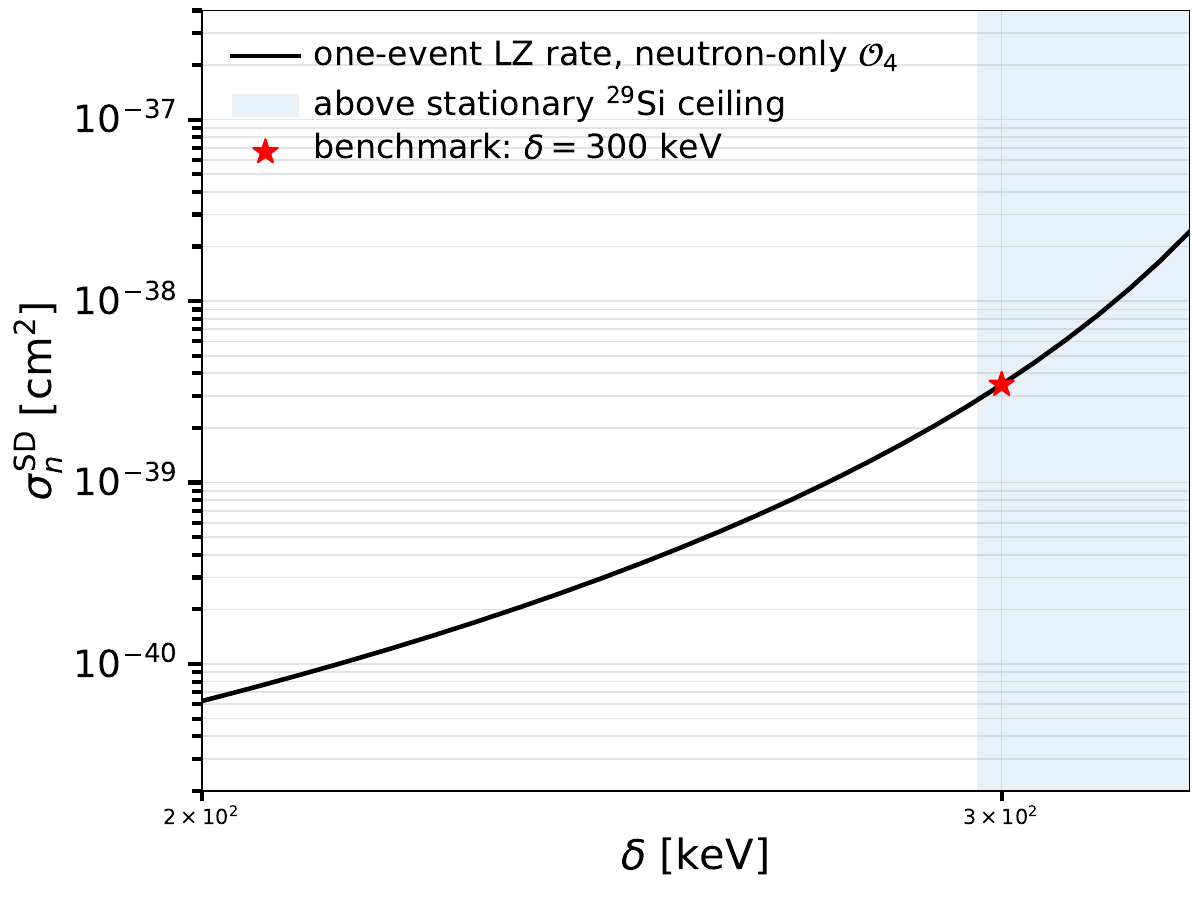}
\caption{Neutron-normalized spin-dependent cross section $\sigma_n^{\rm SD}$ required to reproduce approximately one high-recoil LZ event for the neutron-philic $\mathcal O_4$ interaction, shown as a function of $\delta$ at $m_\chi=1\,\TeV$.  The black curve is the recoil-level LZ normalization, the red star marks the benchmark at $\delta=300\,\keV$, and the blue shaded region begins above the stationary-target $^{29}$Si capture ceiling.}
\label{fig:nsidlz}
\end{figure}

Figure~\ref{fig:nsidlz} shows the neutron-only cross section required to obtain approximately one accepted event in our public recoil-level calculation.  The estimate uses the $2.84$ tonne-year LZ exposure, the public nuclear-recoil efficiency across the extended energy window, natural $^{129}$Xe and $^{131}$Xe abundances, the neutron spin expectation values $\langle S_n\rangle_{129}=0.329$ and $\langle S_n\rangle_{131}=-0.272$, and a finite-$q$ SD response calibrated to the standard xenon oscillator scale \cite{Menendez2012,Klos2013}.  At the benchmark,
\begin{equation}
 \sigma_n^{\rm SD}\simeq3.5\times10^{-39}\,\cmsq,
 \qquad v_{\min}(248\,\keV)\simeq710\,\kms.
 \label{eq:nsid_lzresult}
\end{equation}
The velocity follows directly from Equation~\eqref{eq:vmin}; the cross section is the one-event normalization of our exposed recoil prescription and is consequently more dependent on the finite-$q$ xenon response and detector efficiency.  We use the official LZ significance to establish the quality of the $\mathcal{O}_4$ hypothesis and Equation~\eqref{eq:nsid_lzresult} only as the central normalization for the Solar calculation.

The black curve in Figure~\ref{fig:nsidlz} rises with $\delta$ because a larger splitting pushes the recoil signal further into the high-velocity tail, so a larger neutron-normalized cross section is required to keep the accepted LZ event rate fixed.  The red star identifies the benchmark used in the Solar calculation.  The blue shading begins at the $^{29}$Si stationary-target ceiling, only a few keV below the star.  The shading is not an excluded region; it marks where a calculation restricted to stationary $^{29}$Si would incorrectly suggest that this target has closed.  The proximity of the benchmark to that boundary motivates the finite-temperature and heavier-isotope analysis of the next two subsections.

\subsection{Stationary-target ceilings and the endothermic isotope hierarchy}
\label{subsec:nsidclosure}

A useful first diagnostic is the largest mass splitting that can be captured through a single collision with a Solar nucleus taken to be at rest. This quantity is not a capture rate and should not be interpreted as a physical cutoff in the real, finite-temperature Sun. It answers only the following kinematic question: for a target nucleus $A$, does there exist a Solar radius, an incoming halo speed, and a scattering configuration for which the endothermic transition is allowed and the outgoing DM particle remains gravitationally bound?

The endothermic transition is kinematically accessible only if
\begin{equation}
v_{\rm rel}^2>\frac{2\delta}{\mu_A},
\label{eq:nsid_relative_threshold}
\end{equation}
where $v_{\rm rel}$ is the DM--nucleus relative speed and $\mu_A$ is the corresponding reduced mass. For a stationary Solar nucleus, $v_{\rm rel}=w$, where $w$ is the local DM speed in the Solar frame defined in Equation~\eqref{eq:wsolar}. Satisfying Equation~\eqref{eq:nsid_relative_threshold} is necessary but not sufficient for Solar capture. After the collision, the outgoing state $\chi_2$ must also remain gravitationally bound. Denoting its post-scattering speed in the Solar frame by $v_{\chi_2}^{\rm out}$, the additional condition is
\begin{equation}
v_{\chi_2}^{\rm out}<v_{\rm esc}(r).
\end{equation}

For each nuclear species $A$, we define $\delta_{\rm cap}^{\rm stat}(A)$ as the largest splitting for which these two conditions can be satisfied simultaneously in the stationary-target approximation. Numerically, we determine this quantity by scanning the physical halo-speed interval, the Solar radius, and the complete kinematically allowed recoil range. Since the largest local DM speeds are reached in the deepest part of the Solar potential, the maximal splitting is obtained close to the Solar center.

For the six odd-neutron nuclei emphasized in the elastic SD-neutron analysis of Ref.~\cite{NguyenBlancoLinden2026}, we obtain
\begin{align}
\delta_{\rm cap}^{\rm stat}(^{3}{\rm He})&=29.9\,\keV, &
\delta_{\rm cap}^{\rm stat}(^{13}{\rm C})&=130.8\,\keV,\\
\delta_{\rm cap}^{\rm stat}(^{17}{\rm O})&=171.7\,\keV, &
\delta_{\rm cap}^{\rm stat}(^{21}{\rm Ne})&=212.9\,\keV,\\
\delta_{\rm cap}^{\rm stat}(^{25}{\rm Mg})&=254.4\,\keV, &
\delta_{\rm cap}^{\rm stat}(^{29}{\rm Si})&=296.3\,\keV.
\label{eq:nsidceilings}
\end{align}

These values have a direct interpretation. For example, in the stationary-target approximation, $^{29}{\rm Si}$ can capture the DM particle only for $\delta\lesssim296.3\,\keV$. For a larger splitting, there is no allowed combination of incoming halo speed and scattering kinematics for which the transition occurs and the outgoing $\chi_2$ is left on a bound Solar orbit. The increase of $\delta_{\rm cap}^{\rm stat}$ with nuclear mass is expected because heavier targets have a larger reduced mass and therefore provide more center-of-mass kinetic energy to overcome the endothermic mass splitting while still permitting sufficient energy transfer for capture.

This observation is particularly important for the LZ benchmark, $\delta=300\,\keV$. Among the six nuclei listed above, the largest ceiling is that of $^{29}{\rm Si}$, which lies only $3.7\,\keV$ below the benchmark. Therefore, if one keeps only this elastic-era isotope set and assumes stationary Solar targets, none of these nuclei can capture the $\delta=300\,\keV$ benchmark. The calculated capture rate would consequently appear to develop an abrupt cutoff close to the LZ-preferred splitting. This apparent threshold is not a physical property of the Sun: it is produced by the simultaneous use of stationary targets and an isotope set selected for elastic SD scattering.

It is useful to distinguish this purely kinematic suppression from the effect of the neutron-philic coupling structure. At $\delta=300\,\keV$, all nuclei lighter than approximately $A\simeq30$ are already closed by the endothermic kinematics, independently of their spin structure or of the values of the proton and neutron couplings. For example, the stationary-target ceilings are $9.9\,\keV$ for $^{1}{\rm H}$, $29.9\,\keV$ for $^{3}{\rm He}$, $130.8\,\keV$ for $^{13}{\rm C}$, $171.7\,\keV$ for $^{17}{\rm O}$, and $296.3\,\keV$ for $^{29}{\rm Si}$. Thus hydrogen and the lighter odd-neutron targets are removed by kinematics alone.

The condition $a_p\simeq0$ acts differently. It suppresses the response of heavy odd-proton isotopes that remain kinematically accessible at $\delta=300\,\keV$. For example, $^{55}{\rm Mn}$ and $^{59}{\rm Co}$ have stationary-target ceilings of approximately $576\,\keV$ and $621\,\keV$, respectively, and are therefore well above the LZ benchmark. In a generic SD interaction these nuclei could contribute efficiently to Solar capture. In the neutron-philic scenario their contribution is instead strongly suppressed, leaving the capture rate primarily sensitive to heavy odd-neutron isotopes.

This shows why the isotope set relevant for elastic SD-neutron scattering is not sufficient for the endothermic problem. At the LZ benchmark, $\delta=300\,\keV$, the lighter odd-neutron nuclei are close to or already above their stationary-target kinematic ceilings; for example, $^{29}{\rm Si}$ has $\delta_{\rm cap}^{\rm stat}=296.3\,\keV$ and is therefore closed in this approximation. Heavier odd-neutron isotopes instead have larger reduced masses and correspondingly larger capture ceilings, with $\delta_{\rm cap}^{\rm stat}>300\,\keV$. They therefore remain kinematically open at the LZ benchmark. As a result, these heavier isotopes can contribute significantly to Solar capture even when their abundances are smaller than those of the lighter targets.

Kinematic accessibility alone, however, does not determine the capture rate. The contribution of each isotope is additionally weighted by its Solar abundance and radial distribution, isotopic fraction, nuclear spin matrix element, finite-momentum-transfer response, and the available scattering phase space. Moreover, Solar nuclei are not stationary. Their thermal velocities modify $v_{\rm rel}$ and can reopen endothermic scattering even when the corresponding stationary-target configuration is nominally closed. The purpose of the ceiling calculation is therefore only to identify which nuclear species can potentially contribute and should be retained before the full dynamical weights are evaluated.

The example of $^{57}{\rm Fe}$ illustrates why a heavier isotope can become relevant despite its small isotopic fraction. In our central Solar composition,
\begin{equation}
X_{\rm Fe}(0)f_{57}\simeq3.05\times10^{-5},
\qquad
X_{\rm Si}(0)f_{29}\simeq3.46\times10^{-5}.
\label{eq:nsid_fe_si_abundance}
\end{equation}
The effective abundance of $^{57}{\rm Fe}$ is therefore comparable to that of $^{29}{\rm Si}$, while its larger nuclear mass leaves substantially more endothermic phase space available at $\delta=300\,\keV$. The spin inputs are taken from the available SD nuclear-structure calculations where possible and from odd-group-model estimates otherwise. For the latter we adopt a broad response range because both the zero-momentum spin contents and the finite-$q$ responses of these rare isotopes are less well controlled than for xenon \cite{Bednyakov2005,BednyakovSimkovic2006}.

\begin{figure}[t]
\centering
\includegraphics[width=\columnwidth]{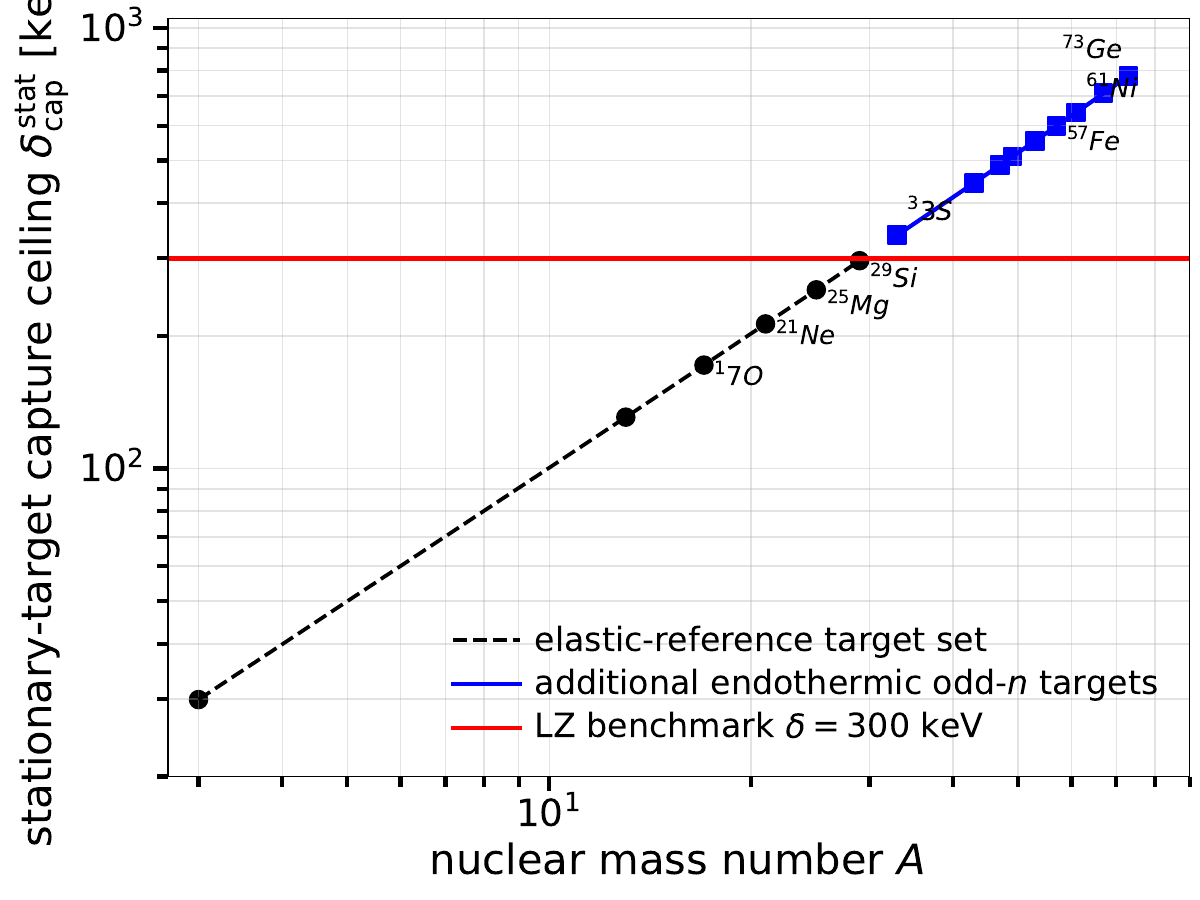}
\caption{Stationary-target endothermic capture ceiling $\delta_{\rm cap}^{\rm stat}$ as a function of nuclear mass number $A$ for $m_\chi=1\,\TeV$. Black circles connected by the dashed line denote the odd-neutron isotopes in the elastic-reference target set, while blue squares denote the additional heavier odd-neutron isotopes retained for the endothermic calculation. Selected isotope labels are shown next to the points. The red horizontal line marks the benchmark splitting $\delta=300\,\keV$.}
\label{fig:nsidceil}
\end{figure}

Figure~\ref{fig:nsidceil} should therefore be interpreted as a map of which targets remain kinematically accessible, rather than as a prediction of their relative capture rates. The systematic increase of $\delta_{\rm cap}^{\rm stat}$ with nuclear mass explains why an isotope list sufficient for elastic scattering becomes incomplete near $\delta=300\,\keV$. Conversely, the fact that $^{73}{\rm Ge}$ has the largest stationary-target ceiling does not imply that it dominates the capture rate, since its dynamical weight can be strongly suppressed by its abundance and nuclear response. The full finite-temperature calculation in the next subsection includes these effects and finds that $^{29}{\rm Si}$, $^{61}{\rm Ni}$, $^{57}{\rm Fe}$, and $^{53}{\rm Cr}$ provide the largest contributions for the central benchmark.

\subsection{Finite-temperature Solar capture and the IceCube comparison}
\label{subsec:nsidthermal}

The stationary-target ceilings discussed in Section~\ref{subsec:nsidclosure} are useful for identifying which isotopes can potentially participate in capture, but they do not describe the physical Solar plasma. Solar nuclei have nonzero thermal velocities, and close to an endothermic threshold even a moderate target velocity can substantially modify the DM--nucleus relative speed. We therefore replace the stationary-target diagnostic by a finite-temperature capture calculation.

Let $\mathbf{w}$ denote the incoming DM velocity in the Solar frame and $\mathbf{v}_A$ the velocity of a target nucleus of species $A$. Their relative speed is
\begin{equation}
v_{\rm rel}=|\mathbf{w}-\mathbf{v}_A|.
\end{equation}
The local capture rate per incoming DM particle is then
\begin{equation}
\Omega_A^-(\mathbf{w},T)=n_A(r)\int d^3v_A\,f_A(\mathbf{v}_A;T)\,v_{\rm rel}\int_{v_{\chi_2}^{\rm out}<v_{\rm esc}(r)}d\sigma_A,
\label{eq:nsid_thermalcapture}
\end{equation}
where $n_A(r)$ is the number density of isotope $A$, $f_A(\mathbf{v}_A;T)$ is its Maxwell--Boltzmann velocity distribution at the local Solar temperature, and $v_{\chi_2}^{\rm out}$ is the speed of the outgoing DM state in the Solar frame. The scattering integral therefore contains only configurations for which the endothermic transition is kinematically allowed and the outgoing $\chi_2$ remains gravitationally bound. The latter condition is essential: an energetically allowed inelastic collision does not necessarily imply Solar capture.

Our numerical calculation integrates over the magnitude and direction of the target velocity, the outgoing center-of-mass scattering angle, the incoming halo speed, and the Solar radius. In the limit $T\rightarrow0$, the target-velocity distribution collapses to a stationary nucleus and the calculation reproduces the stationary-target capture kernel.

The importance of thermal motion is particularly transparent for $^{29}{\rm Si}$. Its stationary-target capture ceiling, $\delta_{\rm cap}^{\rm stat}=296.3\,\keV$, lies only $3.7\,\keV$ below the LZ benchmark $\delta=300\,\keV$. At the Solar center, after optimizing over the physical halo speed and scattering angle, the minimum outgoing DM speed in the stationary-target approximation exceeds the local escape velocity by only
\begin{equation}
\Delta v_{\rm bind}\equiv\left[v_{\chi_2}^{\rm out}\right]_{\rm min}-v_{\rm esc}(0)=8.68\,\kms.
\label{eq:nsid_binding_deficit}
\end{equation}
By comparison, the one-dimensional thermal velocity dispersion of $^{29}{\rm Si}$ near the Solar center is
\begin{equation}
\sqrt{\frac{T_c}{m_{29}}}\simeq66.6\,\kms.
\label{eq:nsid_si_thermal_speed}
\end{equation}
The velocity shift required to compensate the stationary-target binding deficit is therefore well within the thermal distribution. A nucleus with a suitable velocity component opposite to the incoming DM can increase $v_{\rm rel}$ and modify the final Solar-frame kinematics sufficiently to permit capture. Thermal motion therefore smooths the artificial sharp cutoff produced by the stationary-target approximation.

We first quantify this effect using only the three heaviest nuclei of the original elastic SD-neutron target set, $^{21}{\rm Ne}$, $^{25}{\rm Mg}$, and $^{29}{\rm Si}$. At $\delta=300\,\keV$, their finite-temperature capture rate is
\begin{equation}
C_\odot^{\rm baseline}\simeq1.02\times10^{17}\,\mathrm{s}^{-1},
\label{eq:nsid_sixthermal}
\end{equation}
with $^{29}{\rm Si}$ providing almost all of this contribution. Thus, although stationary $^{29}{\rm Si}$ cannot capture the $\delta=300\,\keV$ benchmark, the thermal motion of Solar nuclei reopens the process and removes the apparent cutoff at $\delta_{\rm cap}^{\rm stat}=296.3\,\keV$.

We then include the heavier odd-neutron isotopes identified in Section~\ref{subsec:nsidclosure}, whose larger reduced masses keep them kinematically accessible at $\delta=300\,\keV$. The central calculation gives
\begin{equation}
C_\odot(300\,\keV)=3.98\times10^{17}\,\mathrm{s}^{-1},
\label{eq:nsid_capture300_central}
\end{equation}
while varying the adopted nuclear-spin inputs gives
\begin{equation}
C_\odot(300\,\keV)=(1.77\text{--}9.72)\times10^{17}\,\mathrm{s}^{-1}.
\label{eq:nsid_capture300}
\end{equation}
The largest contributions in the central calculation come from $^{29}{\rm Si}$, $^{61}{\rm Ni}$, $^{57}{\rm Fe}$, and $^{53}{\rm Cr}$, which contribute approximately $9.1\times10^{16}$, $7.4\times10^{16}$, $6.8\times10^{16}$, and $6.0\times10^{16}\,\mathrm{s}^{-1}$, respectively. The $^{33}{\rm S}$ contribution is approximately $2.5\times10^{16}\,\mathrm{s}^{-1}$. The total rate is therefore shared among several nuclei rather than being controlled by a single rare isotope.

\begin{figure}[t]
\centering
\includegraphics[width=\columnwidth]{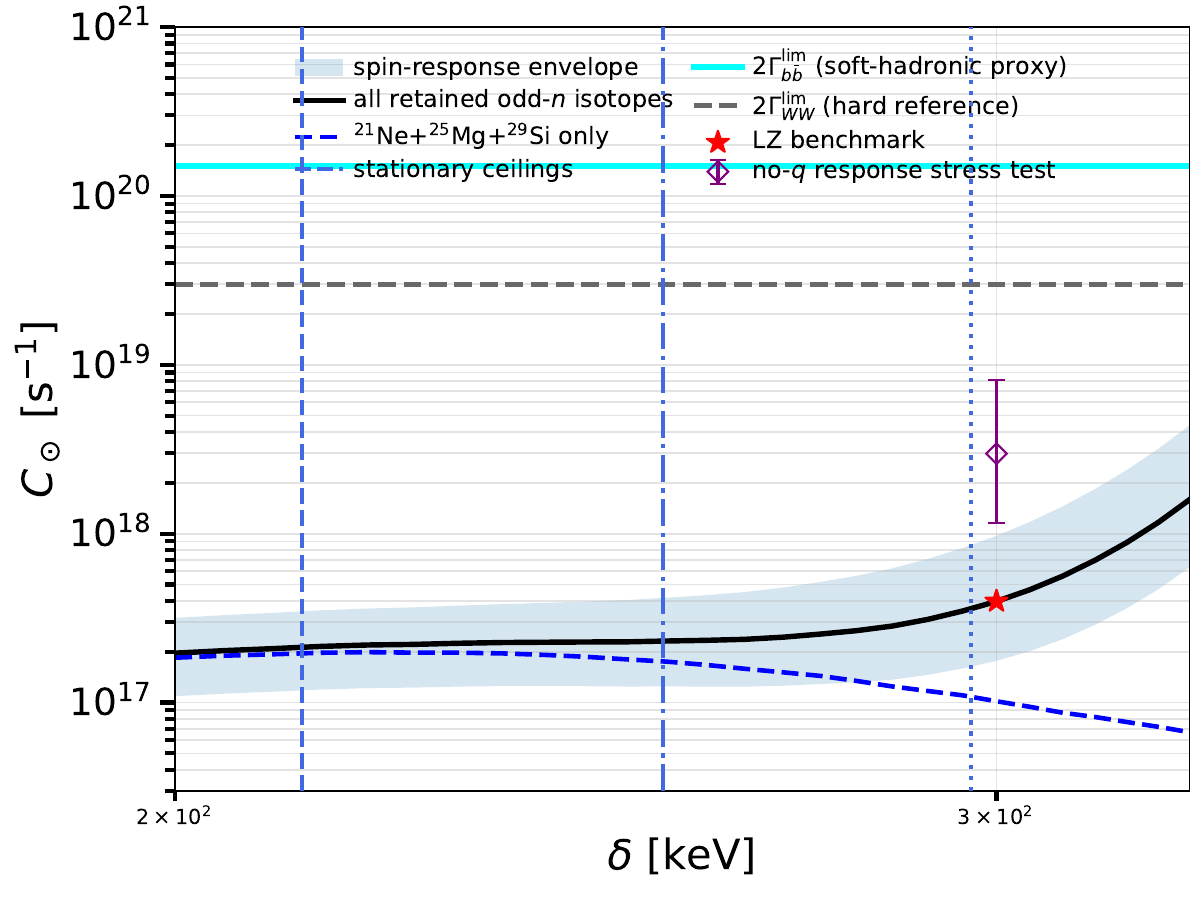}
\caption{Finite-temperature Solar capture rate $C_\odot$ along the LZ-normalized neutron-philic $\mathcal{O}_4$ locus at $m_\chi=1\,\TeV$. The black solid curve is the central prediction including all retained odd-neutron isotopes. The light-blue band shows the variation obtained from the adopted range of nuclear-spin inputs and is not a statistical uncertainty. The blue dashed curve includes only the baseline $^{21}{\rm Ne}+^{25}{\rm Mg}+^{29}{\rm Si}$ targets and therefore isolates the effect of thermal motion on the original elastic-era isotope set. The red star marks the central LZ benchmark at $\delta=300\,\keV$. The purple diamond shows the same benchmark after setting $F_{\rm SD,A}^2(q)=1$ for all Solar isotopes, thereby removing the finite-$q$ suppression in the Solar capture calculation; its vertical range shows the corresponding variation of the adopted nuclear-spin inputs. The blue vertical lines mark the stationary-target capture ceilings of $^{21}{\rm Ne}$, $^{25}{\rm Mg}$, and $^{29}{\rm Si}$. The cyan and gray horizontal lines show $2\Gamma_{b\bar b}^{\rm lim}$ and $2\Gamma_{WW}^{\rm lim}$, respectively, namely the capture rates that would saturate the corresponding IceCube annihilation-rate upper limits in the equilibrium limit $\Gamma_A=C_\odot/2$.}
\label{fig:nsidcapture}
\end{figure}

Figure~\ref{fig:nsidcapture} collects the different ingredients of the Solar calculation and is useful for separating their physical roles. The scan is performed along the LZ-normalized locus: for each value of $\delta$, the neutron cross section $\sigma_n^{\rm SD}$ is adjusted to reproduce the one-event LZ normalization of Figure~\ref{fig:nsidlz}. The vertical axis therefore shows how the corresponding Solar capture rate changes when the mass splitting is varied while the direct-detection signal is kept fixed.

The blue dashed curve includes only $^{21}{\rm Ne}$, $^{25}{\rm Mg}$, and $^{29}{\rm Si}$. The three blue vertical lines indicate their stationary-target capture ceilings. In a calculation with stationary nuclei, the contribution of each isotope would vanish once its corresponding ceiling is crossed. The fact that the blue dashed curve remains nonzero beyond these vertical lines, and in particular beyond the $^{29}{\rm Si}$ ceiling at $296.3\,\keV$, directly demonstrates the effect of the thermal velocity distribution of the Solar nuclei.

The black solid curve instead includes the complete set of retained odd-neutron isotopes. Its difference from the blue dashed curve quantifies the contribution of the additional heavier nuclei, such as $^{53}{\rm Cr}$, $^{57}{\rm Fe}$, and $^{61}{\rm Ni}$, whose stationary-target ceilings lie above the LZ benchmark. The light-blue band around the black curve is obtained by varying the adopted nuclear-spin inputs for these isotopes. It represents the estimated sensitivity of the capture calculation to the poorly known nuclear-spin responses and should not be interpreted as a statistical confidence interval.

The red star shows the central LZ benchmark at $\delta=300\,\keV$. Its value,
\begin{equation}
C_\odot=3.98\times10^{17}\,\mathrm{s}^{-1},
\end{equation}
corresponds to the black central curve and includes thermal target motion, the full retained odd-neutron isotope set, and the nominal finite-$q$ nuclear-response suppression.

The purple diamond tests separately the importance of the finite-momentum nuclear response. We repeat the calculation at the same benchmark value $\delta=300\,\keV$, keeping the same LZ-normalized $\sigma_n^{\rm SD}$ and all Solar and kinematic inputs fixed, but setting
\begin{equation}
F_{\rm SD,A}^2(q)=1
\end{equation}
for every Solar isotope. This removes all finite-$q$ suppression from the Solar nuclear responses and gives
\begin{equation}
C_\odot^{F_{\rm SD}=1}(300\,\keV)=2.98\times10^{18}\,\mathrm{s}^{-1}.
\label{eq:nsid_noq_central}
\end{equation}
The purple diamond therefore lies above the red star because finite-$q$ suppression is no longer reducing the Solar scattering rate. The vertical purple range is obtained by applying the same variation of the nuclear-spin inputs used for the light-blue band to this no-$q$ calculation. It is therefore not a statistical error bar, but a combined diagnostic of the sensitivity to the spin inputs after the finite-$q$ suppression has been removed.

This purple construction is deliberately one-sided. The terrestrial xenon response used to determine the LZ-normalized $\sigma_n^{\rm SD}$ is left unchanged, while the finite-$q$ suppression is removed only for the Solar nuclei. It should therefore not be interpreted as a second self-consistent nuclear model. Its purpose is to test whether an underestimate of the Solar finite-$q$ response could qualitatively alter the IceCube conclusion.

The two horizontal lines allow the capture calculation to be compared directly with IceCube without assuming that equilibrium is actually reached. The annihilation rate satisfies the general upper bound
\begin{equation}
\Gamma_A\leq\frac{C_\odot}{2}.
\label{eq:nsid_absolute_ann_bound}
\end{equation}
Consequently, a capture rate below $2\Gamma^{\rm lim}$ cannot violate an annihilation-rate upper limit $\Gamma^{\rm lim}$ even in the maximally efficient equilibrium case. The cyan horizontal line therefore corresponds to
\begin{equation}
C_\odot=2\Gamma_{b\bar b}^{\rm lim},
\end{equation}
while the gray dashed line corresponds to
\begin{equation}
C_\odot=2\Gamma_{WW}^{\rm lim}.
\end{equation}
Any capture curve lying below one of these lines is automatically below the corresponding IceCube constraint. The large vertical separation between the red star, the black curve, and these horizontal lines visually shows the substantial IceCube margin of the central benchmark.

The appropriate IceCube annihilation template is model dependent. For the thermal Higgsino, annihilation is dominated by hard electroweak final states and we therefore use the $WW$ limit. For the neutron-philic $\mathcal{O}_4$ benchmark, however, the effective operator responsible for nuclear scattering does not determine the annihilation final state. Different ultraviolet completions may lead to light quarks, heavy quarks, electroweak bosons, or more complicated cascades. We therefore use the IceCube $b\bar b$ result as a representative soft-hadronic proxy and the $WW$ result as a harder reference. A channel-exact constraint would require specifying the ultraviolet completion and propagating its annihilation spectrum through the Sun and the IceCube response.

At $m_\chi\simeq1\,\TeV$, we adopt
\begin{equation}
\Gamma_{b\bar b}^{\rm lim}\simeq7.5\times10^{19}\,\mathrm{s}^{-1},
\qquad
\Gamma_{WW}^{\rm lim}=1.5\times10^{19}\,\mathrm{s}^{-1}.
\label{eq:nsid_bb_limit_repeated}
\end{equation}
These correspond to the cyan and gray capture thresholds in Figure~\ref{fig:nsidcapture},
\begin{equation}
2\Gamma_{b\bar b}^{\rm lim}\simeq1.5\times10^{20}\,\mathrm{s}^{-1},
\qquad
2\Gamma_{WW}^{\rm lim}=3.0\times10^{19}\,\mathrm{s}^{-1}.
\end{equation}

For the central benchmark, Equation~\eqref{eq:nsid_absolute_ann_bound} gives
\begin{equation}
\Gamma_A\leq1.99\times10^{17}\,\mathrm{s}^{-1},
\label{eq:nsidicecube}
\end{equation}
which is approximately a factor $377$ below the adopted $b\bar b$ upper limit and a factor $75$ below the $WW$ reference. At the upper edge of the nuclear-spin variation, $C_\odot=9.72\times10^{17}\,\mathrm{s}^{-1}$, the corresponding equilibrium annihilation rate remains approximately a factor $154$ below the $b\bar b$ limit and a factor $31$ below the $WW$ reference.

The purple no-$q$ stress test also remains well below the IceCube templates. Its central value, $C_\odot^{F_{\rm SD}=1}=2.98\times10^{18}\,\mathrm{s}^{-1}$, implies
\begin{equation}
\Gamma_A\leq1.49\times10^{18}\,\mathrm{s}^{-1},
\label{eq:nsid_noff}
\end{equation}
which is a factor $50.4$ below the $b\bar b$ upper limit and a factor $10.1$ below the $WW$ reference. Applying the upper nuclear-spin variation simultaneously gives $C_\odot=8.08\times10^{18}\,\mathrm{s}^{-1}$ and therefore $\Gamma_A\leq4.04\times10^{18}\,\mathrm{s}^{-1}$, still a factor $18.6$ below the $b\bar b$ proxy.

The dominant uncertainty in the nominal result is therefore associated with the nuclear response of the rare heavy isotopes rather than with the multidimensional numerical integration. Repeating the calculation with three different production quadratures gives
\begin{equation}
C_\odot(300\,\keV)=3.972,\ 3.977,\ 3.972\times10^{17}\,\mathrm{s}^{-1},
\label{eq:nsid_quadrature_convergence}
\end{equation}
so grid refinement changes the total rate by only approximately $0.1\%$. The finite-temperature kernel also reproduces the stationary-target result in the $T\rightarrow0$ limit at machine precision. As an independent validation, a seeded $6\times10^5$-event Monte Carlo evaluation of the $^{29}{\rm Si}$ core thermal kernel agrees with the deterministic quadrature at the $0.12\%$ level, within the Monte Carlo statistical uncertainty.

We also test how strongly the conclusion depends on the neutron-philic condition $a_p\simeq0$. Relaxing this assumption to the isoscalar choice $a_p=a_n$ activates the heavy odd-proton isotopes $^{27}{\rm Al}$, $^{31}{\rm P}$, $^{35}{\rm Cl}$, $^{39}{\rm K}$, $^{51}{\rm V}$, $^{55}{\rm Mn}$, $^{59}{\rm Co}$, and $^{63,65}{\rm Cu}$. Using odd-group proton-spin estimates and AGSS09-like abundances, these nuclei add approximately $2.22\times10^{18}\,\mathrm{s}^{-1}$ to the capture rate, dominated by $^{55}{\rm Mn}$ and $^{59}{\rm Co}$. The resulting total is
\begin{equation}
C_\odot^{\rm isoscalar}\simeq2.62\times10^{18}\,\mathrm{s}^{-1},
\end{equation}
approximately a factor $6.6$ larger than the neutron-philic central value. Even in the equilibrium limit, however, the corresponding annihilation rate remains approximately a factor $57$ below the $b\bar b$ reference and a factor $11$ below the $WW$ reference. The neutron-philic choice therefore reduces Solar capture significantly, but the qualitative IceCube conclusion does not rely exclusively on this isospin assignment.

Finally, we test the nuclear input that produces the largest individual uncertainty, the $^{57}{\rm Fe}$ neutron-spin response. Taking the $p_{1/2}$ single-particle value $|\langle S_n\rangle|=1/6$ as a separate stress test raises the $^{57}{\rm Fe}$ contribution from $6.8\times10^{16}$ to $3.4\times10^{18}\,\mathrm{s}^{-1}$ and increases the upper capture-rate estimate to approximately $4.1\times10^{18}\,\mathrm{s}^{-1}$. The corresponding equilibrium bound is
\begin{equation}
\Gamma_A\lesssim2.0\times10^{18}\,\mathrm{s}^{-1},
\end{equation}
still approximately a factor $37$ below the $b\bar b$ reference and a factor $7.4$ below the $WW$ reference.

Compounding this extreme $^{57}{\rm Fe}$ spin choice with the independent no-$q$ stress test gives $C_\odot\simeq4.1\times10^{19}\,\mathrm{s}^{-1}$ and therefore $\Gamma_A\lesssim2.0\times10^{19}\,\mathrm{s}^{-1}$. This remains approximately a factor $3.7$ below the soft $b\bar b$ proxy but exceeds the adopted $WW$ reference. This last construction simultaneously combines two deliberately aggressive nuclear assumptions and compares them with a hard annihilation template that is not necessarily appropriate for the model. It should therefore not be interpreted as an exclusion. Rather, it identifies the $^{57}{\rm Fe}$ spin response as the nuclear input for which a dedicated shell-model calculation would most significantly improve the robustness of the Solar prediction.

\subsection{Relic density and ultraviolet considerations}
\label{subsec:nsid_uv_main}

For the neutron-philic benchmark we do not attempt to specify the cosmological mechanism that determines the DM relic abundance. The interaction responsible for the LZ signal need not also control freeze-out: the observed abundance can instead be generated through interactions entirely within the hidden sector, with only a much weaker portal connecting DM to the Standard Model. The Solar-capture calculation depends only on the low-energy transition interaction relevant for scattering on nuclei and is therefore insensitive to the details of this cosmological history.

It is nevertheless useful to discuss whether the transition axial interaction of Equation~\eqref{eq:nsid_lagrangian} can arise from a simple ultraviolet completion. A minimal axial-vector gauge completion is subject to two important constraints, associated respectively with the structure of the Majorana current and with perturbativity.

First, an off-diagonal axial current is not obtained in the usual pseudo-Dirac construction. For two Weyl fermions whose mass spectrum is dominated by a Dirac mass, the gauge current becomes off diagonal in the vector channel and diagonal in the axial channel, as in the pseudo-Dirac model discussed in Section~\ref{subsec:pseudo_model} and Appendix~\ref{app:pseudo}. Realizing instead a purely off-diagonal axial interaction, while suppressing the corresponding diagonal axial couplings, requires a different mass hierarchy. As discussed in Appendix~\ref{app:nsid}, this can be achieved with two Weyl fermions carrying opposite $U(1)_X$ charges, approximately equal Majorana masses generated by a dark Higgs, and a much smaller gauge-invariant Dirac mass. In this construction the splitting is
\begin{equation}
\delta=2|m_D|.
\end{equation}
Suppressing the diagonal axial interaction is important phenomenologically. A sizable diagonal coupling would generate elastic $\mathcal{O}_4$ scattering without the endothermic threshold, leading both to strong terrestrial SD-neutron constraints and to efficient Solar capture on lighter spin-carrying nuclei such as $^{3}{\rm He}$, $^{13}{\rm C}$, and $^{17}{\rm O}$.

Second, the same symmetry breaking that generates an axial-vector mediator mass also contributes to the DM mass. In the construction described above,
\begin{equation}
m_\chi\simeq M=\frac{y_\chi v_S}{\sqrt{2}},
\qquad
m_{Z'}=2g_\chi v_S,
\end{equation}
so that
\begin{equation}
y_\chi=\frac{2\sqrt{2}\,m_\chi g_\chi}{m_{Z'}}.
\end{equation}
Requiring the dark-sector Yukawa coupling to remain perturbative, $y_\chi\lesssim\sqrt{4\pi}$, gives
\begin{equation}
m_{Z'}\gtrsim\frac{2\sqrt{2}}{\sqrt{4\pi}}\,g_\chi m_\chi
\simeq0.8\,g_\chi m_\chi.
\label{eq:nsid_yukawa_bound}
\end{equation}
This is the familiar consistency condition for fermionic DM with axial-vector gauge interactions \cite{Kahlhoefer2016}. For $m_\chi=1\,\TeV$ and, illustratively, $g_\chi\simeq0.55$, perturbativity already requires
\begin{equation}
m_{Z'}\gtrsim0.44\,\TeV.
\end{equation}

The LZ normalization further fixes the combination of dark- and visible-sector couplings,
\begin{equation}
G_n=\frac{g_\chi a_n}{m_{Z'}^2}
=3.25\times10^{-6}\,\GeV^{-2},
\end{equation}
as given in Equation~\eqref{eq:nsid_Gn}. Combining this relation with the perturbativity bound on $g_\chi$ implies
\begin{equation}
a_n\gtrsim
\frac{2\sqrt{2}\,m_\chi}{\sqrt{4\pi}}\,G_n m_{Z'}
\simeq
1.1\left(\frac{m_{Z'}}{440\,\GeV}\right).
\label{eq:nsid_an_bound}
\end{equation}
For the neutron-philic relation of Equation~\eqref{eq:nsid_cancel}, this corresponds to order-one axial couplings to first-generation quarks. Such couplings for a mediator at the few-hundred-GeV scale are subject to strong collider constraints and require additional charged fields, including an appropriate Higgs sector, to obtain gauge-invariant Standard Model Yukawa interactions. A fully specified gauge completion is therefore substantially more involved than the low-energy interaction needed for the LZ and Solar-capture analysis.

For this reason, we treat the neutron-philic scenario as an operator-level benchmark. Its defining ingredients are two neutral Majorana states separated by $\delta\simeq300\,\keV$ and a dominant neutron-philic transition interaction producing the nonrelativistic $\mathcal{O}_4$ response with the normalization of Equation~\eqref{eq:nsid_benchmark}. The relic abundance is assumed to be generated by additional hidden-sector dynamics and is not used as a constraint on the low-energy benchmark. Possible ultraviolet realizations of the transition interaction are discussed further in Appendix~\ref{app:nsid}.

Leaving the ultraviolet completion unspecified also means that the annihilation final state is not uniquely determined by the operator responsible for the LZ recoil. Depending on the completion, annihilation may proceed into dark-sector states that subsequently decay to Standard Model particles, or directly into Standard Model fermions or bosons. The corresponding Solar neutrino spectrum is therefore model dependent. This motivates the treatment adopted in Section~\ref{subsec:nsidthermal}, where the IceCube $b\bar b$ upper limit is used as a representative soft-hadronic proxy and the $WW$ limit as a harder reference. Neither curve should be interpreted as the exact IceCube constraint on the operator-level benchmark. A channel-specific limit would require specifying the ultraviolet completion and propagating its annihilation spectrum through the Sun and the IceCube response, as discussed in Appendix~\ref{app:neutrino}.

Importantly, the Solar conclusion does not require such a specification. Independently of the annihilation channel or of whether capture--annihilation equilibrium is reached, the annihilation rate satisfies
\begin{equation}
\Gamma_A\leq\frac{C_\odot}{2}.
\end{equation}
The equality corresponds to the maximally efficient equilibrium limit. Velocity-suppressed annihilation, incomplete thermalization, or failure to reach equilibrium can only reduce the neutrino signal. The comparison in Section~\ref{subsec:nsidthermal} therefore uses $C_\odot/2$ as a conservative upper bound rather than assuming a particular relic-density mechanism or annihilation history.

A viable ultraviolet completion of the benchmark must consequently preserve the ingredients relevant for the Solar analysis: two neutral states separated by $\delta\simeq300\,\keV$, a dominant neutron-philic transition interaction at nuclear momentum transfer, and sufficiently suppressed diagonal SI, SD, or other coherent interactions that could reopen efficient capture on abundant Solar nuclei. The details of the hidden-sector dynamics responsible for the relic abundance are otherwise independent of the Solar-capture conclusion.

\section{Discussion and systematic limitations}
\label{sec:discussion}

The calculation is deliberately explicit but remains a semi-analytic phenomenological recast.  The most important limitations are therefore worth separating according to whether they can change a numerical endpoint or the qualitative conclusion.

\subsection{Solar model and capture systematics}

The high-$\delta$ capture tail depends on the Fe/Ni abundances, nuclear form factors, and the escape-speed profile.  We use the same truncated halo parameters as Ref.~\cite{PospelovRamani2026} together with an interpolated public BS05(AGS,OP) density, temperature, enclosed-mass, and H/He/C/N/O structure.  The capture quadrature is converged at better than the percent level at the benchmark splittings; for example, changing the Higgsino radial/velocity grid from $(30,40)$ to $(100,120)$ changes $C_\odot(377\,\keV)$ by less than $0.2\%$.  The residual Solar uncertainty is therefore dominated by the treatment of the heavy-element abundance profiles and nuclear response rather than numerical integration.  H/He/C/N/O are already interpolated from the SSM; a precision calculation should replace our common O-like radial profile for Ne--U by the complete isotope-by-isotope SSM abundance table.

The local DM density scales the capture rate linearly.  The halo velocity distribution also enters, but Solar gravitational acceleration reduces its importance compared with the LZ event, which is controlled directly by the terrestrial high-speed tail.  This difference is the basic reason Solar capture remains powerful at splittings that are close to or above the terrestrial endpoint.

\subsection{Orbital dynamics and loop-induced elastic scattering}

Our radial-orbit model uses a single effective turning point.  A complete treatment should evolve the joint distribution in orbital energy and angular momentum and sample the radius of every scatter.  This can shift the time required to reach a given $R_\chi$, especially for SD scattering on H.  The loop-induced SI rate is also theoretically uncertain because of amplitude-level cancellations.  Figure~\ref{fig:hloop} turns these two uncertainties into a physically transparent scan.  The key result is that the Higgsino exclusion survives even the singular limit in which all loop-induced elastic scattering is removed.

For the neutron-philic SD benchmark, the finite-temperature calculation gives a nonzero captured population.  Its subsequent orbital evolution can affect the degree of capture--annihilation equilibrium, but it cannot increase the annihilation rate above the conservative bound $C_\odot/2$.  At the central $300\,\keV$ point this bound is a factor about $377$ below the IceCube $b\bar b$ soft-hadronic proxy, and the upper edge of our representative spin envelope is a factor about $154$ below it.  Against the harder $WW$ reference, the corresponding factors are about $75$ and $31$.  Removing all Solar finite-$q$ suppression gives proxy/reference margins of about $50$ and $10$ for the central spins, while compounding that diagnostic with the upper spin envelope gives about $19$ and $3.7$.  These numbers are useful robustness diagnostics, but they are not a substitute for a spectrum-specific IceCube likelihood.  The precision priorities are improved heavy-isotope spin responses and, once a completion is specified, a detector-level recast of its annihilation spectrum.  Among the nuclear inputs $^{57}$Fe is the leading concern: replacing its odd-group spin by the single-particle stress value reduces the $b\bar b$-proxy margin to $37$ at the response-envelope level.  If that stress value is further compounded with the no-$q$ diagnostic, the rate remains a factor $3.7$ below the $b\bar b$ proxy but is $1/0.74\simeq1.35$ times the $WW$ hard-template limit.  This deliberately compounded construction is not a statistical exclusion, but it shows that the strongest universal statement is the nominal/representative-envelope viability; an exact hard-spectrum test requires both a dedicated $^{57}$Fe elastic finite-$q$ response and the actual annihilation spectrum.

For the pseudo-Dirac model the state label must ultimately be evolved together with the orbit.  Exothermic down-scattering changes both the internal state and orbital energy, thermally moving nuclei can re-excite $\chi_1\to\chi_2$ close to a stationary-target threshold, and the two state distributions need not have the same $V_{\rm eff}$.  We therefore do not interpret $0.418R_\odot$ as a final stopping radius.  Instead, Section~\ref{subsec:pseudo_reexcitation} evaluates thermally averaged up- and down-scattering rates on a sequence of fixed radial orbits.  The resulting fixed-orbit calculation reaches a maximum $\Gamma_A\simeq\revtextD{4.4}\times10^{14}\,\mathrm{s}^{-1}$, still a factor \revtextD{$\simeq1.7\times10^5$} below the adopted IceCube $b\bar b$ limit and approximately \revtextD{$3.4\times10^4$} below the stronger $WW$ template.  The semi-analytic energy-loss estimate of Section~\ref{subsec:pseudo_cooling} suggests further contraction toward a characteristic rate-crossing scale near $0.14R_\odot$ and gives an illustrative present-day rate of order $10^{11}\,\mathrm{s}^{-1}$ or less.  Because that estimate uses radial orbits, Fe-only recoil losses, and a one-zone population mapping, we do not treat either number as a prediction of the final distribution.  The viability statement instead rests on the fixed-orbit envelope, which is already many orders of magnitude below both published templates.

\subsection{Neutrino spectra and detector likelihood}

The IceCube limits in Equations~\eqref{eq:icecubeproxy} and \eqref{eq:icecubebbproxy} are channel dependent, so the physical annihilation topology must be kept separate from the capture calculation.  For the Higgsino, the relevant hard continuum is the $WW+ZZ$ electroweak mixture; our deliberately conservative construction that discards all $ZZ$ neutrinos already shows that this channel uncertainty cannot reopen the LZ region.  For the pseudo-Dirac model, the correct source is the branching-fraction-weighted $u,d,s,c,b,t$ mixture.  The light-flavor tree-level cascade is softened in the Sun, while heavy flavors are harder and electroweak radiation gives a nonzero high-energy component even for light-quark primaries \cite{IbarraTotzauerWild2014,Bauer2021}.  The fixed-orbit maximum nevertheless lies factors of about $1.7\times10^5$ and $3.4\times10^4$ below the IceCube upper limits for the $b\bar b$ and $WW$ channels, respectively, so the pseudo-Dirac viability conclusion is insensitive to this spectral refinement.

For the neutron-philic operator the final state is not fixed at all until a UV completion is chosen.  We therefore regard $b\bar b$ as a soft-hadronic proxy and $WW$ as a hard reference rather than claiming that they rigorously bracket every possible spectrum.  The central and representative spin-envelope rates are safely below both.  The deliberately compounded single-particle-$^{57}$Fe plus no-$q$ stress construction, however, illustrates why the distinction matters: it remains below the $b\bar b$ proxy but can exceed the $WW$ hard-template rate.  A channel-complete IceCube likelihood should therefore accompany an improved nuclear-response calculation before assigning a precision exclusion to that extreme corner.  This qualification does not affect the robust Higgsino exclusion or the many-orders-of-magnitude pseudo-Dirac margin.

\subsection{When the pseudo-Dirac conclusion changes}

The absence of a strong IceCube bound is specific to the minimal off-diagonal vector model.  A diagonal scalar or axial interaction can both thermalize $\chi_1$ and open additional annihilation modes.  A diagonal process $\chi_1\chi_1\to\mathrm{SM}$ directly removes the two-state kinetics with re-excitation.  Dark-sector reactions such as $\chi_1\chi_1\leftrightarrow\chi_2\chi_2$ can also repopulate the excited state if they overcome the endothermic threshold.  These extensions are well motivated in ultraviolet completions and should be constrained separately; Appendix~\ref{app:pseudo} explains which additional operators can appear.

\begin{table*}[t]
\caption{Representative benchmark results for the three endothermic models.  The neutron-philic SD cross section is the central normalization of our public high-$q$ xenon response estimate; the official LZ $\mathcal{O}_4$ significance is independent of that normalization.}
\label{tab:summary}
\begin{ruledtabular}
\footnotesize
\begin{tabular}{lccc}
Quantity & Thermal Higgsino & Pseudo-Dirac vector & Neutron-philic SD\\
\hline
$m_\chi$ & $1.08$--$1.1\,\TeV$ & $1.0\,\TeV$ & $1.0\,\TeV$\\
$\delta_{\rm LZ}$ & $377.1\,\keV$ & $297\,\keV$ & $300\,\keV$\\
LZ scattering normalization & $\sigma_n^{\widetilde H}=7.43\times10^{-39}\,\cmsq$ & $\sigma_N=6.5\times10^{-43}\,\cmsq$ & $\sigma_n^{\rm SD}=3.5\times10^{-39}\,\cmsq$\\
LZ local preference & $3.3\sigma$ Higgsino-like $\mathcal{O}_1^s$ & $3.4\sigma$ inelastic $\mathcal{O}_1^v$ & $3.4\sigma$ for $\mathcal{O}_4^s$\\
$C_\odot$ at benchmark & \revtextD{$1.03\times10^{23}\,\mathrm{s}^{-1}$} & \revtextD{$1.53\times10^{20}\,\mathrm{s}^{-1}$} & $3.98\times10^{17}\,\mathrm{s}^{-1}$\\
Solar suppression mechanism & none; efficient capture & two-state bottleneck & \revtextD{endothermic threshold; rare odd-$n$ targets}\\
Representative Solar result & $\delta_{\rm lim}=512$--$565\,\keV$ & $\Gamma_{A,\max}^{\rm FO}=\revtextD{4.4}\times10^{14}\,\mathrm{s}^{-1}\ll\Gamma_{b\bar b}^{\rm lim}$ & $C_\odot/2=1.99\times10^{17}\,\mathrm{s}^{-1}$ (central)\\
Thermal mechanism & electroweak freeze-out & $\chi_1\chi_2\to q\bar q$ & \revtext{secluded dark sector; App.~\ref{app:nsid}}\\
\end{tabular}
\end{ruledtabular}
\end{table*}

\section{Conclusions}
\label{sec:conclusion}

The LZ $248\,\keV$ candidate is particularly interesting for endothermic DM because a mass splitting of a few hundred keV naturally shifts the recoil spectrum toward high energies while suppressing ordinary low-energy events. Solar capture provides a powerful complementary test, since gravitational acceleration allows halo DM to reach substantially larger collision speeds inside the Sun. Our main result is that the connection between an LZ signal and a Solar-neutrino constraint is strongly model dependent. The three benchmarks considered here illustrate three qualitatively different possibilities: efficient capture followed by efficient annihilation, efficient capture followed by suppressed annihilation, and suppression already at the level of Solar capture.

For the thermal Higgsino we confirm the essential result of Ref.~\cite{PospelovRamani2026}. With the weak-charge interaction $\sigma_n^{\widetilde H}=G_F^2\mu_n^2/(2\pi)=7.43\times10^{-39}\,\cmsq$, our recoil-level reconstruction using public LZ information reproduces the event for a neutral-state splitting near $377\,\keV$, for which we find $C_\odot=1.03\times10^{23}\,\mathrm{s}^{-1}$. The Sun therefore captures the LZ benchmark very efficiently. The resulting IceCube lower bound on the splitting is approximately $\delta\gtrsim512.7\,\keV$ if loop-induced elastic cooling is completely neglected, $558.6\,\keV$ when only the central SD loop interaction drives the subsequent orbital evolution, and $564.8\,\keV$ when the central SI and SD loop interactions efficiently thermalize the captured population. Using the physical $WW+ZZ$ annihilation mixture does not reopen the LZ-preferred region. We therefore conclude that the standard thermal Higgsino constituting all of the DM is incompatible with the LZ interpretation under the Solar and halo assumptions adopted here.

The minimal pseudo-Dirac vector model gives a very different Solar history. At $m_\chi=1\,\TeV$ and $\delta=297\,\keV$ we obtain $C_\odot=1.53\times10^{20}\,\mathrm{s}^{-1}$, so capture itself is efficient. The suppression instead arises from the two-state dynamics: capture produces $\chi_2$, exothermic scattering rapidly converts it to $\chi_1$, while the leading annihilation process requires one particle of each state. We explicitly checked that the stationary-target Ni closure radius, $0.418R_\odot$, cannot be interpreted as a final stopping radius because thermal Solar nuclei can re-excite $\chi_1$ into $\chi_2$. Including this effect increases the annihilation rate substantially relative to a calculation with no re-excitation, but the largest rate found over our fixed-orbit scan is only $\Gamma_A\simeq4.4\times10^{14}\,\mathrm{s}^{-1}$. This remains below the IceCube $b\bar b$ and $WW$ reference limits by factors of approximately $1.7\times10^5$ and $3.4\times10^4$, respectively. A complementary semi-analytic cooling calculation indicates that repeated excitation--de-excitation cycles can contract extended orbits toward a characteristic scale $r_{\rm cool}\simeq0.14R_\odot$, where further up-scattering becomes extremely slow. Because this estimate does not evolve the full distribution in orbital energy, angular momentum, and internal state, we use it only as a robustness diagnostic. The pseudo-Dirac viability conclusion does not depend on assigning a precise final radius, since the explicit fixed-orbit calculation already remains many orders of magnitude below the IceCube limits.

The neutron-philic endothermic SD benchmark differs from both previous cases because the Solar capture rate itself is strongly suppressed. LZ finds a $3.4\sigma$ local preference for inelastic $\mathcal{O}_4^s$ at $m_\chi=1\,\TeV$ and $\delta=300\,\keV$, for which our recoil normalization gives $\sigma_n^{\rm SD}\simeq3.5\times10^{-39}\,\cmsq$. We find that the stationary $^{29}{\rm Si}$ threshold is not a physical Solar cutoff: thermal nuclear velocities reopen capture close to threshold, while heavier odd-neutron isotopes remain kinematically accessible at $\delta=300\,\keV$. Including both effects gives the central result $C_\odot=3.98\times10^{17}\,\mathrm{s}^{-1}$, with $(1.77$--$9.72)\times10^{17}\,\mathrm{s}^{-1}$ obtained from the adopted variation of the nuclear-spin inputs. Even taking the upper end of this range and assuming the maximally efficient relation $\Gamma_A=C_\odot/2$, the annihilation rate remains below the IceCube $b\bar b$ soft-hadronic proxy and the $WW$ hard reference by factors of approximately $154$ and $31$, respectively. Removing all finite-$q$ suppression in the Solar responses and simultaneously taking the upper nuclear-spin variation still leaves factors of approximately $19$ and $3.7$. We also tested the neutron-philic assumption itself: changing to an isoscalar interaction raises the capture rate by a factor $6.6$, but the nominal result remains below both IceCube reference curves.

The largest uncertainty in the neutron-philic calculation is associated with the nuclear response of the rare heavy isotopes, particularly $^{57}{\rm Fe}$, rather than with the numerical integration. Taking the single-particle value $|\langle S_n\rangle|=1/6$ for $^{57}{\rm Fe}$ as a separate stress test increases the capture rate substantially. Only when this aggressive choice is compounded with the independent assumption $F_{\rm SD,A}^2(q)=1$ for all Solar nuclei does the equilibrium rate approach the IceCube templates: it remains a factor approximately $3.7$ below the $b\bar b$ proxy while exceeding the harder $WW$ reference by approximately $35\%$. This construction is not a statistical uncertainty or an exclusion, since it simultaneously replaces two poorly controlled nuclear inputs and compares the result with an annihilation spectrum that is not specified by the operator-level model. It instead identifies the main ingredients required for a precision test: a dedicated finite-$q$ spin-structure calculation for $^{57}{\rm Fe}$ and the neutrino spectrum of a specified ultraviolet completion.

A complementary example has recently been provided by Higgs-coupled minimal DM \cite{SmirnovGriffithBeacom2026}. In that framework the LZ recoil, thermal relic abundance, and electroweak representation are correlated, and the resulting Solar constraint depends strongly on the multiplet: the lighter $3{\rm M}2{\rm D}$ and $5{\rm M}4{\rm D}$ Standard-Halo-Model benchmarks exceed the quoted IceCube $WW$ limit, the $7{\rm M}6{\rm D}$ case lies close to it, while the heavier $9{\rm M}8{\rm D}$, $11{\rm M}10{\rm D}$, and $13{\rm M}12{\rm D}$ configurations remain compatible in their analysis. The lighter representations can also evade the Solar bound if the LZ event is produced by a small high-velocity DM component, illustrating further the model and halo dependence of Solar-neutrino constraints on endothermic interpretations.

Our conclusions are therefore model dependent but robust within the assumptions tested here. The thermal Higgsino interpretation is excluded because even the limiting case with no loop-induced elastic cooling requires a splitting substantially larger than the LZ-preferred value. The minimal pseudo-Dirac vector benchmark remains viable because efficient capture does not translate into efficient annihilation once the two-state dynamics is included. The neutron-philic $\mathcal{O}_4$ benchmark also remains viable for the central calculation and for the representative nuclear-response variations because its Solar capture rate is strongly suppressed. The same high-recoil event can therefore correspond to efficient capture and annihilation, efficient capture with an internal-state annihilation bottleneck, or strongly suppressed capture controlled by the endothermic isotope hierarchy. A Solar-neutrino exclusion must consequently be derived from the complete interaction, Solar target content, post-capture evolution, and annihilation spectrum rather than inferred from the terrestrial scattering cross section alone. Future LZ exposure, a channel-complete IceCube likelihood, and improved finite-$q$ spin responses for the relevant rare Solar isotopes can sharpen these conclusions.

\begin{acknowledgments}
MDM acknowledges support from the research grant {\sl TAsP (Theoretical Astroparticle Physics)} funded by Istituto Nazionale di Fisica Nucleare (INFN).
H.S. acknowledges support from the Collaborative
Research Center SFB1258 and from the Deutsche
Forschungsgemeinschaft (DFG, German Research Foundation) under Germany’s Excellence Strategy - EXC2094 - 390783311.
\end{acknowledgments}

\appendix

\section{Thermal Higgsino: particle content, interactions, thermal history, and elastic loops}
\label{app:higgsino}

\subsection{Electroweakino sector and pseudo-Dirac limit}

The Higgsino is not an isolated Majorana fermion but part of the MSSM electroweakino sector.  In the neutral basis
\begin{equation}
 \psi^0=\left(-i\widetilde B,-i\widetilde W^0,\widetilde H_d^0,\widetilde H_u^0\right),
\end{equation}
where $\widetilde B$ is the bino, $\widetilde W^0$ the neutral wino, and $\widetilde H_d^0$ and $\widetilde H_u^0$ the neutral down- and up-type Higgsino Weyl fields.  The neutralino mass term is
\begin{equation}
 \mathcal{L}_{\rm mass}^{0}=-\frac{1}{2}\left(\psi^0\right)^T\mathcal{M}_N\psi^0+\mathrm{h.c.},
\end{equation}
with
\begin{equation}
\mathcal{M}_N=
\left(\begin{smallmatrix}
M_1 & 0 & -m_Zs_Wc_\beta & m_Zs_Ws_\beta\\
0 & M_2 & m_Zc_Wc_\beta & -m_Zc_Ws_\beta\\
-m_Zs_Wc_\beta & m_Zc_Wc_\beta & 0 & -\mu\\
m_Zs_Ws_\beta & -m_Zc_Ws_\beta & -\mu & 0
\end{smallmatrix}\right).
\label{eq:neutralinomatrix}
\end{equation}
where $s_\beta\equiv\sin\beta$ and $c_\beta\equiv\cos\beta$, with $\tan\beta=v_u/v_d$ the ratio of MSSM Higgs vacuum expectation values.  For $|M_1|,|M_2|\gg|\mu|$, the two light neutral states are approximately the symmetric and antisymmetric combinations of $\widetilde H_d^0$ and $\widetilde H_u^0$.  We denote these physical states by $\chi_1$ and $\chi_2$ everywhere in the paper, with $m_{\chi_2}>m_{\chi_1}$ and $\delta=m_{\chi_2}-m_{\chi_1}>0$.  In the strict gaugino-decoupling limit they combine into a Dirac fermion of mass $|\mu|$ and the diagonal vector current vanishes for each Majorana component; finite bino/wino exchange breaks the accidental Higgsino number symmetry and generates the small neutral Majorana splitting in Equation~\eqref{eq:higgsino_split}.  The same pseudo-Dirac Higgsino limit and its implications for inelastic DD have been studied in Refs.~\cite{NagataShirai2015,KrallReece2018,Graham2025}.  The charged Higgsino $\chi^\pm$ remains nearby, with the familiar electroweak loop splitting of order a few hundred MeV in the pure-Higgsino limit.

The gauge interactions follow from the Higgsino kinetic terms,
\begin{equation}
 \mathcal{L}_{\rm kin}=i\widetilde H_u^\dagger\bar\sigma^\mu D_\mu\widetilde H_u
 +i\widetilde H_d^\dagger\bar\sigma^\mu D_\mu\widetilde H_d.
\end{equation}
where $D_\mu$ is the SM electroweak covariant derivative and $\bar\sigma^\mu=(1,-\boldsymbol{\sigma})$ acts on the two-component Weyl spinors.  After diagonalization, the leading neutral $Z$ interaction is off diagonal as in Equation~\eqref{eq:higgsino_z}, while $W^\pm$ couples the neutral and charged Higgsinos.  The off-diagonal $Z$ current is the key connection to the LZ event: a halo $\chi_1$ can scatter as $\chi_1 A\to\chi_2 A$, whereas an elastic tree-level neutral-current process $\chi_1 A\to\chi_1 A$ is absent in the pure pseudo-Dirac limit.  These electroweak interactions also control freeze-out coannihilation and present-day annihilation, while elastic ground-state scattering is regenerated at loop level \cite{Hisano2005,Hisano2011,ChenHill2020}.

\subsection{When the Higgsino is a thermal relic}

At freeze-out the temperature is approximately $T_f\sim m_\chi/20\sim50\,\GeV$.  Both the neutral splitting of a few hundred keV and the charged-neutral splitting of order hundreds of MeV are therefore much smaller than $T_f$.  The two neutral states and the charged pair are simultaneously populated, and the relic abundance is determined by a coupled coannihilation system.  Electroweak annihilation and coannihilation into gauge bosons and fermions produce the observed abundance for a nearly pure Higgsino mass close to $1.08$--$1.1\,\TeV$ in standard radiation-dominated cosmology \cite{Beneke2015,KrallReece2018,Graham2025}.  This is what we mean by a \emph{thermal Higgsino} throughout the paper.  A lighter nonthermal Higgsino, a subdominant Higgsino component, or a nonstandard cosmological history need not obey the same Solar-capture interpretation and should be treated separately.

\subsection{$W^+W^-$ and $ZZ$ annihilation fractions}

For a nearly pure neutral Higgsino, present-day annihilation into electroweak gauge bosons is generated mainly by charged-Higgsino exchange for $W^+W^-$ and neutral-electroweakino exchange for $ZZ$.  Because the charged and neutral Higgsino states are nearly degenerate, the present-day annihilation problem also receives Sommerfeld corrections and is most consistently treated as a coupled electroweak system.

For the IceCube channel mapping we adopt the representative continuum rates quoted near the thermal Higgsino mass in Ref.~\cite{Rodd2024},
\begin{align}
 \langle\sigma v\rangle_{WW}&\simeq8\times10^{-27}\,\cmcubeds,\\
 \langle\sigma v\rangle_{ZZ}&\simeq5\times10^{-27}\,\cmcubeds.
 \label{eq:appWWZZratio}
\end{align}
Normalizing these two representative continuum rates gives the convenient reference weights
\begin{equation}
 B_{WW}^{\rm ref}\simeq0.62,
 \qquad
 B_{ZZ}^{\rm ref}\simeq0.38.
 \label{eq:appWWZZ}
\end{equation}
These numbers should not be read as exact Higgsino branching ratios.  They are normalized reference weights for the two dominant continuum channels and are used only to quantify the IceCube channel-mapping uncertainty.  The continuum rates can vary at the order-ten-percent level when the neutral and charged splittings and Sommerfeld treatment are changed \cite{Rodd2024,Beneke2015}.  For the present purpose the important feature is more robust than the exact percentages: both $WW$ and $ZZ$ are hard electroweak final states.  Our ``$WW$ only'' limit is deliberately more pessimistic than a physical mixed-channel calculation because it assigns zero IceCube response to every annihilation into $ZZ$.

\subsection{Loop-induced SI scattering and its cancellation uncertainty}

Once the tree-level off-diagonal transition is kinematically closed, an almost pure Higgsino has an irreducible elastic SI interaction generated by electroweak loops.  In a heavy-WIMP effective theory the nucleon amplitude has the schematic decomposition
\begin{equation}
 \mathcal{M}_{\rm SI}=\mathcal{M}^{(0)}_q+\mathcal{M}^{(0)}_g
 +\mathcal{M}^{(2)}_q+\mathcal{M}^{(2)}_g+\cdots,
 \label{eq:SIamplitude_components}
\end{equation}
where $\mathcal{M}^{(0)}_q$ and $\mathcal{M}^{(0)}_g$ denote the scalar quark and gluon contributions, while $\mathcal{M}^{(2)}_q$ and $\mathcal{M}^{(2)}_g$ denote the corresponding spin-two (twist-two) contributions.  The spin-zero terms arise from scalar quark/gluon operators and the spin-two terms from twist-two operators after electroweak and QCD matching.  The individual one- and two-boson contributions are considerably larger than the final answer.  For an electroweak doublet the scalar and spin-two amplitudes partially cancel, making the residual cross section unusually sensitive to perturbative matching, hadronic matrix elements, and $m_W/m_\chi$ power corrections, where $m_W$ is the $W$-boson mass \cite{Hisano2011,ChenHill2020}.  This explains both the small central value near
\begin{equation}
 \sigma_{\rm el,loop}^{\rm SI,N}\sim4\times10^{-50}\,\cmsq
\end{equation}
and the highly asymmetric uncertainty: changing hadronic matrix elements, perturbative matching, and the $m_W/m_\chi$ power correction can move the residual amplitude by a large relative amount, with the lower edge approaching a cancellation zero.  A recent next-to-leading-order (NLO) MSSM analysis explicitly finds parameter regions in which the Higgsino--nucleon SI amplitude can become extremely small or vanish because of these cancellations \cite{Bisal2026SI}.  Ref.~\cite{PospelovRamani2026} quotes the same $4\times10^{-50}\,\cmsq$ central scale and adopts $3.16\times10^{-49}\,\cmsq$ as an upper boundary based on the heavy-Higgsino analysis of Ref.~\cite{ChenHill2020}.  We retain that value for direct comparison, but it should not be interpreted as an experimental upper limit or a statistically defined confidence interval.  It is a theory-side reference value in a cancellation-dominated prediction whose lower side can approach zero.  Our wider scan is intended to show directly how the Solar cooling endpoint depends on this uncertain elastic amplitude.

A finite bino or wino admixture can also generate a tree-level Higgs-exchange amplitude.  Its magnitude and sign depend on $M_1$, $M_2$, $\mu$, $\tan\beta$, and electroweakino phases, and it can interfere with the irreducible loop amplitude.  In the LZ Higgsino interpretation the required few-hundred-keV neutral splitting generically points to very heavy gauginos, so this additional amplitude is usually suppressed, but it illustrates why the elastic SI rate is more ultraviolet sensitive than the tree-level off-diagonal $Z$ interaction.  We therefore treat the SI scan as a phenomenological envelope for Solar cooling rather than a Gaussian theoretical error bar.

This uncertainty is unusually important for Solar dynamics because cooling is cumulative.  A cross section too small to be observable in a conventional DD search can still change the captured radius over billions of orbits.  Our numerical scan therefore does not assume that $4\times10^{-50}\,\cmsq$ is exact: it spans $10^{-54}$--$7\times10^{-49}\,\cmsq$ and shows explicitly where the population transitions from stalled to thermalized.

\subsection{Loop-induced SD scattering}

The elastic SD interaction is generated by axial structures, including loop corrections to the diagonal neutralino-$Z$ vertex and electroweak box diagrams.  Earlier electroweak calculations give a reference proton cross section around $5\times10^{-47}\,\cmsq$ for a heavy Higgsino \cite{Hisano2011}, while more recent studies of Higgsino-like MSSM neutralinos find that one-loop vertex and box effects can modify the SD nucleon cross sections by order tens of percent and, in selected mixed scenarios, by approximately $50\%$ \cite{Bisal2024}.  The exact number depends on the residual gaugino admixture and renormalization prescription.  For this reason Figure~\ref{fig:hloop} scans $0$, $0.5$, $1$, and $1.5$ times the reference SD value rather than assigning a narrow Gaussian error.

The key point is that the SI and SD uncertainties affect the \emph{time required to shrink the orbit}; they do not change the large tree-level inelastic capture cross section responsible for the initial capture.  The Solar endpoint therefore approaches the conservative no-elastic result smoothly as both loop amplitudes are reduced.

\section{Pseudo-Dirac vector dark matter: detailed model and ultraviolet completion}
\label{app:pseudo}

\subsection{Mass eigenstates and off-diagonal current}

Pseudo-Dirac fermions and their inelastic phenomenology have been studied extensively since the original inelastic-DM construction, including collider, direct-detection, cosmological, and vector-mediator realizations \cite{TuckerSmith2001,TuckerSmith2005,DeSimone2010,Baryakhtar2020,DallaValleGarcia2025,Foguel2025}.  Writing the Dirac fermion as $\Psi=(\xi,\eta^\dagger)^T$, the mass matrix is
\begin{equation}
 \mathcal{M}=\begin{pmatrix}m_L&m_D\\m_D&m_R\end{pmatrix}.
\end{equation}
For $|m_{L,R}|\ll m_D$, the eigenstates are two Majorana fermions with masses
\begin{equation}
 m_{1,2}\simeq m_D\mp\frac{m_L+m_R}{2},
\end{equation}
so $\delta\simeq m_L+m_R$.  The approximate Dirac-number symmetry is restored as $m_L,m_R\to0$, which makes a small splitting technically natural.  The Dirac vector current becomes, up to phase conventions,
\begin{equation}
 \bar\Psi\gamma^\mu\Psi=i\bar\chi_2\gamma^\mu\chi_1+\mathcal{O}\left(\frac{m_L-m_R}{m_D}\right),
\end{equation}
whereas $\bar\chi_i\gamma^\mu\chi_i=0$ identically for Majorana fields.  This is the origin of both endothermic DD and the absence of a leading diagonal vector annihilation channel today.

\subsection{Coannihilation and quark branching fractions}

For universal vector quark couplings and $m_V\gg m_\chi$, the coannihilation rate into quark flavor $q$ is proportional to
\begin{equation}
 w_q=N_c\sqrt{1-r_q}\left(1+\frac{r_q}{2}\right),\qquad r_q=\frac{m_q^2}{m_\chi^2}.
 \label{eq:quarkweight}
\end{equation}
where $N_c=3$ is the number of QCD colors, $m_q$ is the quark mass, and $r_q=m_q^2/m_\chi^2$ is the corresponding phase-space ratio.  Thus
\begin{equation}
 B_q=\frac{w_q}{\sum_{q'}w_{q'}}.
\end{equation}
where $B_q$ is the branching fraction to $q\bar q$ and the sum in the denominator runs over all open quark flavors.  At $m_\chi=1\,\TeV$ all six flavors are open and the phase-space correction is tiny: the numerical values are those in Equation~\eqref{eq:pdmBR}.  This democratic mixture is the branching assumption used throughout the Solar-neutrino discussion.  Different quark charge assignments would change both the nucleon matching and the neutrino spectrum and should not be mapped onto our benchmark by a simple overall rescaling.

\subsection{Representative ultraviolet completion}

Equation~\eqref{eq:pdm_vector} is a simplified model.  One representative ultraviolet completion gauges a baryon-like $U(1)_X$ symmetry.  Let $\Psi_{L,R}$ carry equal charge $q_\chi$ so that the $V_\mu\bar\Psi\gamma^\mu\Psi$ coupling is vectorial, and let SM quarks carry a universal vector charge.  Introduce a dark scalar $S$ with charge $-2q_\chi$.  The renormalizable dark-sector terms can contain
\begin{align}
 \mathcal{L}_{\rm UV}\supset&\;\bar\Psi i\gamma^\mu D_\mu\Psi-m_D\bar\Psi\Psi
 +|D_\mu S|^2-V(S)\\
 &-\frac{1}{2}y_LS\,\overline{\Psi_L^c}\Psi_L
 -\frac{1}{2}y_RS\,\overline{\Psi_R^c}\Psi_R+\mathrm{h.c.}
 \label{eq:pdmUV}
\end{align}
Here $D_\mu$ is now the covariant derivative including $U(1)_X$, $g_X$ is its gauge coupling, $q_\chi$ is the $U(1)_X$ charge of $\Psi$, $V(S)$ is the dark-scalar potential, and $y_L$ and $y_R$ are the symmetry-breaking Majorana Yukawa couplings.  After $S$ develops the vacuum expectation value $\langle S\rangle=v_S/\sqrt{2}$, the gauge boson obtains $m_V\sim g_Xv_S$ and the Majorana masses are $m_{L,R}=y_{L,R}v_S/\sqrt{2}$.  The small pseudo-Dirac splitting can therefore originate from small symmetry-breaking Yukawa couplings while the vector interaction remains much larger.

A gauged baryon current is anomalous with only the SM fermions, so a complete theory requires additional heavy spectator fermions to cancel the mixed gauge anomalies.  Those states can be placed above the mediator scale and integrated out for the low-energy phenomenology considered here.  Alternatively, different anomaly-free $U(1)_X$ charge assignments can be used, at the cost of changing the relative quark and lepton couplings.  The simple direct--relic relation in Equation~\eqref{eq:direct_relic} is therefore a property of the universal-vector contact benchmark, not a universal prediction of every pseudo-Dirac completion.

The ultraviolet completion also clarifies how the Solar conclusion could change.  Scalar mixing between $S$ and the SM Higgs can generate diagonal SI interactions; axial gauge charges can generate diagonal SD interactions; additional dark states can mediate $\chi_1\chi_1$ annihilation or $\chi_1\chi_1\leftrightarrow\chi_2\chi_2$ conversion.  Any of these effects can remove the two-state bottleneck.  The minimal model studied in the main text is defined by taking such extra interactions to be negligible at Solar energies.

\section{\texorpdfstring{Neutron-philic inelastic SD model and \NoCaseChange{{\color{black}CONSTRAINTS ON ITS ULTRAVIOLET COMPLETION}}}{Neutron-philic inelastic SD model and constraints on its ultraviolet completion}}
\label{app:nsid}

\subsection{\texorpdfstring{Majorana states and the transition axial current: \textcolor{black}{which mass hierarchy produces it}}{Majorana states and the transition axial current: which mass hierarchy produces it}}

The low-energy interaction in Equation~\eqref{eq:nsid_lagrangian} should be regarded as \revtext{a simplified low-energy description whose ultraviolet origin is constrained below,} rather than as a fundamental Proca interaction.  Its low-energy $\mathcal O_4$ response follows the standard nonrelativistic DM--nucleus EFT and SD nuclear-response literature \cite{Fitzpatrick2013,Anand2014,Menendez2012,Klos2013,GarnyIbarraPatoVogl2013}, while the Solar target complementarity and inelastic kinematics are closely related to the capture studies of Refs.~\cite{LiangWu2014,CatenaSchwabe2015,BlennowClementzHerreroGarcia2016,NguyenBlancoLinden2026}.

{\color{black}
Consider two neutral Weyl fermions $\xi$ and $\eta$ with $U(1)_X$ charges $q_\xi$ and $q_\eta$, gauge coupling $g_X$, and the general renormalizable mass terms of Equation~\eqref{eq:pdm_mass}, $-m_D\xi\eta-\tfrac12m_L\xi\xi-\tfrac12m_R\eta\eta+{\rm h.c.}$  The gauge current is $J^\mu=g_X\left(q_\xi\,\xi^\dagger\bar\sigma^\mu\xi+q_\eta\,\eta^\dagger\bar\sigma^\mu\eta\right)$.  Writing the interaction-basis fields as $\psi_a=\sum_iU_{ai}\chi_i$ in terms of the Majorana mass eigenstates $\chi_i$,
\begin{equation}
 J^\mu=g_X\sum_{ij}\left(U^\dagger Q_XU\right)_{ij}\chi_i^\dagger\bar\sigma^\mu\chi_j,
 \qquad Q_X={\rm diag}(q_\xi,q_\eta).
 \label{eq:nsid_rotated_current}
\end{equation}
For Majorana fields the four-component bilinear $\bar\chi_i\gamma^\mu\chi_j$ is antisymmetric in $(i,j)$ and $\bar\chi_i\gamma^\mu\gamma^5\chi_j$ is symmetric.  Since $\chi_i^\dagger\bar\sigma^\mu\chi_j$ is a chiral projection of these two structures, the real symmetric part of the Hermitian matrix $U^\dagger Q_XU$ generates axial couplings and the imaginary antisymmetric part generates vector couplings.  Which of the two appears off the diagonal is decided by the mass matrix.

\emph{Pseudo-Dirac hierarchy, $|m_D|\gg|m_{L,R}|$.}  Gauge invariance of $m_D\xi\eta$ requires $q_\eta=-q_\xi$.  The real symmetric mass matrix is diagonalized by a real rotation $O$, but its two eigenvalues have opposite sign, and the negative one must be made positive by the phase redefinition $\chi_2\to i\chi_2$.  Hence $U=O\,{\rm diag}(1,i)$ and
\begin{equation}
 U^\dagger Q_XU=\begin{pmatrix}a&ib\\-ib&c\end{pmatrix},
 \qquad
 \begin{pmatrix}a&b\\b&c\end{pmatrix}\equiv O^TQ_XO.
 \label{eq:nsid_pd_charge}
\end{equation}
The off-diagonal entry is imaginary and therefore \emph{vector}; the diagonal entries are real and therefore \emph{axial}, with $a+c=q_\xi+q_\eta=0$ and $a=c=0$ at maximal mixing.  This hierarchy reproduces exactly the off-diagonal vector current of the pseudo-Dirac model of Appendix~\ref{app:pseudo}, Equation~\eqref{eq:pdm_contact}, and cannot produce Equation~\eqref{eq:nsid_lagrangian}.  Departing from $q_\eta=-q_\xi$ or from maximal mixing does not help: it reintroduces the diagonal axial couplings $a$ and $c$, which give elastic $\mathcal O_4$ scattering of $\chi_1$ without an endothermic threshold.

\emph{Inverted hierarchy, $m_L\simeq m_R\equiv M\gg|m_D|$.}  If both eigenvalues have the same sign no phase is needed, $U=O$ is real, $U^\dagger Q_XU=O^TQ_XO$ is real symmetric, and the off-diagonal coupling is \emph{axial}.  Near degeneracy with the same sign requires $m_L\simeq m_R$; for $m_L=m_R=M$ the eigenstates are $(\xi\pm\eta)/\sqrt2$ with masses $M\pm m_D$, so $\delta=2|m_D|$.  With $q_\eta=-q_\xi\equiv-q$,
\begin{equation}
 O^TQ_XO=q\begin{pmatrix}0&1\\1&0\end{pmatrix},
 \label{eq:nsid_inverted_charge}
\end{equation}
which is purely off diagonal.  In four-component notation this is
\begin{equation}
 \mathcal{L}\supset g_\chi Z'_\mu\bar\chi_2\gamma^\mu\gamma^5\chi_1+\mathrm{h.c.},
 \qquad g_\chi\simeq g_X|q|,
 \label{eq:nsid_uvcurrent}
\end{equation}
up to the normalization convention of Equation~\eqref{eq:nsid_lagrangian}, with \emph{vanishing} diagonal axial couplings.  The construction is gauge consistent: $m_D\xi\eta$ is neutral and may be a small bare mass, whereas $\xi\xi$ and $\eta\eta$ carry charges $\pm2q$ and must arise from a dark Higgs $S$ of charge $-2q$ through $-\tfrac12y_LS\,\xi\xi-\tfrac12y_RS^\dagger\eta\eta+{\rm h.c.}$, giving $m_{L,R}=y_{L,R}v_S/\sqrt2$ for $\langle S\rangle=v_S/\sqrt2$.  The equality $m_L=m_R$ is enforced by a charge-conjugation symmetry $\xi\leftrightarrow\eta$, $S\leftrightarrow S^\dagger$, and the small splitting is technically natural because $m_D$ is the only term odd under $\xi\to-\xi$.  This is the sense in which Equation~\eqref{eq:nsid_lagrangian} can be embedded in a gauge theory.  The earlier description, ``Dirac masses and dark-Higgs-induced Majorana masses'' with a generic mixing matrix, corresponds to the pseudo-Dirac hierarchy and is withdrawn.\par}

\subsection{Quark matching and neutron-philic coupling}

At scales below $m_{Z'}$, Equation~\eqref{eq:nsid_contact} matches onto nucleon axial currents.  Writing
\begin{equation}
 a_N=\sum_{q=u,d,s}g_q^A\Delta q^{(N)},
 \label{eq:nsid_an}
\end{equation}
where $\Delta q^{(N)}$ are the nucleon axial charges, the neutron-only limit is obtained by arranging $a_p\simeq0$ while maintaining $a_n\neq0$.  As an illustrative two-flavor choice, take $g_s^A=0$, $\Delta u^{(p)}\simeq0.84$, $\Delta d^{(p)}\simeq-0.43$, and use isospin to set $\Delta u^{(n)}\simeq-0.43$, $\Delta d^{(n)}\simeq0.84$.  Then
\begin{equation}
 \frac{g_d^A}{g_u^A}\simeq-\frac{\Delta u^{(p)}}{\Delta d^{(p)}}\simeq1.95
 \label{eq:nsid_cancel}
\end{equation}
sets $a_p\simeq0$ while leaving $a_n\simeq1.21g_u^A$.  This relation is a phenomenological neutron-philic benchmark, not a unique gauge-charge assignment; strange-quark and renormalization-group contributions can be included in a precision matching analysis.

Using Equation~\eqref{eq:nsid_sigman}, the benchmark $\sigma_n^{\rm SD}=3.47\times10^{-39}\,\cmsq$ corresponds to an effective neutron coefficient
\begin{align}
 G_n&\equiv\frac{g_\chi a_n}{m_{Z'}^2}
 \simeq3.25\times10^{-6}\,\GeV^{-2},\\
 \Lambda_n&\equiv G_n^{-1/2}\simeq5.54\times10^2\,\GeV.
 \label{eq:nsid_Gn}
\end{align}
Here $G_n$ is the low-energy axial-axial coefficient and $\Lambda_n$ is the corresponding effective interaction scale.  \revtext{For a gauge completion the quark couplings required by LZ scale as $a_n=G_nm_{Z'}^2/g_\chi$: they are at the percent level only if the mediator is light \emph{and} $g_\chi$ is of order unity.  The next subsection shows that this combination is not available once the DM mass is generated consistently.}

\subsection{\texorpdfstring{\textcolor{black}{Perturbativity of the axial gauge completion}}{Perturbativity of the axial gauge completion}}
\label{app:nsid_pert}

{\color{black}
In the inverted hierarchy the TeV-scale mass is a dark-Higgs Yukawa mass, $M=y_\chi v_S/\sqrt2$ with $y_L=y_R\equiv y_\chi$, and the mediator mass is $m_{Z'}=2|q|g_Xv_S=2g_\chi v_S$.  Eliminating $v_S$,
\begin{equation}
 y_\chi=\frac{2\sqrt2\,m_\chi g_\chi}{m_{Z'}}.
 \label{eq:nsid_yukawa}
\end{equation}
A bare Majorana mass cannot replace the Yukawa term because $\xi\xi$ is charged.  Requiring $y_\chi\lesssim\sqrt{4\pi}$ gives $m_{Z'}\gtrsim0.8\,g_\chi m_\chi$, i.e.\ $m_{Z'}\gtrsim0.44\,\TeV$ for $g_\chi=0.55$; equivalently, for a given mediator mass, $g_\chi\lesssim1.25\times10^{-3}\,(m_{Z'}/\GeV)$.  This is the axial-vector consistency condition of Ref.~\cite{Kahlhoefer2016}.  The earlier illustrative benchmark, $m_{Z'}=50\,\GeV$ and $g_\chi=0.555$, corresponds to $y_\chi\simeq31$ and is withdrawn.  The heavy-DM expression
\begin{equation}
 \langle\sigma v\rangle_{Z'Z'}\simeq
 \frac{g_\chi^4}{16\pi m_\chi^2}
 \frac{(1-r)^{3/2}}{(1-r/2)^2},\qquad
 r=\frac{m_{Z'}^2}{m_\chi^2},
 \label{eq:nsid_uvfreeze}
\end{equation}
is in any case incomplete when $y_\chi$ is large, because the longitudinal $Z'$ mode and the physical dark Higgs $s$ then contribute comparably through $\chi_1\chi_1\to ss$ and $sZ'$.

Combining Equation~\eqref{eq:nsid_yukawa} with the LZ normalization $G_n=3.25\times10^{-6}\,\GeV^{-2}$ gives a lower bound on the visible couplings.  With $g_\chi$ at its ceiling,
\begin{equation}
 a_n=\frac{G_nm_{Z'}^2}{g_\chi}\gtrsim\frac{2\sqrt2\,m_\chi}{\sqrt{4\pi}}\,G_nm_{Z'}
 \simeq2.6\times10^{-3}\left(\frac{m_{Z'}}{\GeV}\right),
 \label{eq:nsid_an_lower}
\end{equation}
so $a_n\gtrsim1.1$ at $m_{Z'}=440\,\GeV$, corresponding through Equation~\eqref{eq:nsid_cancel} to $g_u^A\simeq0.9$ and $g_d^A\simeq1.8$.  Lowering $m_{Z'}$ reduces the required $a_n$ only linearly while reducing the allowed $g_\chi$, and hence the $Z'Z'$ freeze-out rate, as $g_\chi^4$; at $m_{Z'}=200\,\GeV$ one still needs $a_n\gtrsim0.5$ with $g_\chi\lesssim0.25$.  Order-one axial couplings of a $Z'$ in this mass range to first-generation quarks are subject to strong dijet constraints, and axial couplings to SM quarks require the SM Higgs sector to carry $U(1)_X$ charge for the SM Yukawa couplings to be gauge invariant, inducing $Z$--$Z'$ mixing \cite{Kahlhoefer2016}.  We conclude that a light secluded axial mediator is not a consistent completion of the neutron-philic benchmark at the LZ normalization, and that a heavy one is at best marginal.  We therefore treat the benchmark at the operator level and assume that the relic density is set in the dark sector.

A different route avoids the axial gauge current altogether.  For Majorana fields the tensor bilinear $\bar\chi_i\sigma^{\mu\nu}\chi_j$ is antisymmetric in $(i,j)$, exactly like the vector bilinear, so a tensor--tensor contact interaction
\begin{equation}
 \mathcal{L}_{\rm eff}^{T}=\sum_qG_q^T\left(\bar\chi_2\sigma^{\mu\nu}\chi_1\right)\left(\bar q\sigma_{\mu\nu}q\right)
 \label{eq:nsid_tensor}
\end{equation}
is automatically purely off diagonal, with no diagonal partner to suppress.  Its leading nonrelativistic limit is again $\mathbf S_\chi\cdot\mathbf S_N$, now weighted by the nucleon tensor charges $\delta q^{(N)}$ rather than the axial charges $\Delta q^{(N)}$, so a neutron-philic combination can be arranged in the same way as in Equation~\eqref{eq:nsid_cancel}.  Such an operator arises, for example, from $t$-channel exchange of a heavy scalar coupling $\chi$ to quarks after Fierz rearrangement.  The thermal history of a tensor completion is then controlled by the coannihilation $\chi_1\chi_2\to q\bar q$ or by additional dark-sector states, and its LZ recoil spectrum receives momentum-dependent corrections beyond leading order.  We do not develop it here; it is noted to make clear that the operator-level benchmark is not tied to the axial gauge construction.\par}

\subsection{Gauge consistency and additional states}

A massive axial-vector mediator cannot be a complete theory by itself.  In a renormalizable $U(1)_X$ realization, the $Z'$ mass is generated by a dark Higgs field or an equivalent symmetry-breaking sector.  \revtext{In the inverted hierarchy above the same field generates the TeV-scale Majorana masses, which is the origin of the perturbativity constraint of Equation~\eqref{eq:nsid_yukawa}; the small splitting is instead a gauge-invariant bare Dirac mass.}  Gauge anomalies involving the SM quark charges must also cancel.  This can be achieved either by a more extended anomaly-free charge assignment or by additional spectator fermions.  These consistency requirements are standard for axial-vector simplified DM models and can produce additional collider signatures \cite{Kahlhoefer2016}.

The ultraviolet completion is therefore less unique than the Solar phenomenology.  What is required for the latter is only: (i) two neutral states separated by $\delta\simeq300\,\keV$; (ii) a dominant neutron-philic transition \revtext{spin-dependent} interaction at nuclear momentum transfer; and (iii) no additional large interaction that reopens capture on abundant Solar proton-spin or coherent targets.  \revtext{In the inverted axial construction, condition (iii) includes the vanishing of the diagonal axial charges of $\chi_1$ and $\chi_2$, which follows from $q_\eta=-q_\xi$ and $m_L=m_R$; in a tensor construction it is automatic.}  Any completion satisfying these three conditions inherits the strong endothermic Solar-capture suppression quantified in Section~\ref{subsec:nsidclosure}; the residual rate is controlled by Solar thermal motion and the rare heavy odd-neutron isotope responses.

\section{Capture integral, Solar profiles, and numerical convergence}
\label{app:capture}

The numerical integrator evaluates Equations~\eqref{eq:gould}--\eqref{eq:ctotal} on radial and asymptotic-speed grids.  For the Higgsino and pseudo-Dirac first-capture calculations we use the stationary-target form of the standard Gould kernel: for each $(r,u)$ pair we compute $w$, verify the endothermic threshold, construct $E_R^\pm$, impose the capture condition, and integrate the nuclear response over recoil energy.  For the neutron-philic benchmark, whose $300\,\keV$ point lies directly at the stationary $^{29}$Si endpoint, we instead use Equation~\eqref{eq:nsid_thermalcapture} and average over Maxwell--Boltzmann Solar target velocities while imposing the final Solar-frame binding condition.  The halo-speed integral uses the exactly truncated distribution in Equation~\eqref{eq:fsun}; no unphysical tail above $v_{\rm esc}^{\rm Gal}+v_\odot$ is retained.

The Solar interpolation is described in Section~\ref{subsec:solarinputs}.  The public BS05(AGS,OP) table is sampled in the distributed file \texttt{data/bs05\_agsop\_sampled.csv}.  We use monotone cubic interpolation for $M(r)$, $T(r)$, $\rho(r)$, and the directly tabulated light-element mass fractions, and derive the escape speed from Equation~\eqref{eq:vesc_ssm}.  Thus the density and gravitational-potential profiles entering capture are not independent fitting functions.  For Ne--U the remaining approximation is compositional: their normalizations follow Table~\ref{tab:solarcomp} and their radial dependence follows the oxygen diffusion profile.

For the neutron-philic endothermic calculation the isotope inputs are collected explicitly in Table~\ref{tab:nsidisotopes}.  The central spin values and ranges for $^{21}$Ne, $^{25}$Mg, and $^{29}$Si follow the elastic Solar SD-neutron analysis of Ref.~\cite{NguyenBlancoLinden2026}; selected heavier values and model spreads are taken from the compilation of Ref.~\cite{Bednyakov2005}.  For $^{33}$S, $^{43}$Ca, $^{53}$Cr, $^{57}$Fe, and $^{61}$Ni, where we do not have a dedicated high-$q$ shell-model response, the central value is the standard odd-group-model estimate \cite{Bednyakov2005,GoriKnapenLin2025}
\begin{equation}
 \langle S_n\rangle_{\rm OGM}\simeq-\frac{\mu_A}{3.826},
 \label{eq:ogmspin}
\end{equation}
with the measured magnetic moment $\mu_A$ in nuclear magnetons \cite{Stone2005}.  For these OGM-only entries we vary $|\langle S_n\rangle|$ by $\pm50\%$ around the central estimate; for $^{57}$Fe, whose anomalously small magnetic moment makes the OGM particularly fragile, we use the wider range $0.5$--$2$ times the central magnitude.  The resulting envelope is deliberately illustrative rather than statistical.

{\color{black}
The $^{57}$Fe entry deserves a separate comment because its odd-group estimate is unusually fragile.  The ground state is $1/2^-$; in a simple $p_{1/2}$-neutron picture one has $\langle S_n\rangle=-1/6$ and a Schmidt moment $+0.638\,\mu_N$, whereas the measured moment is only $+0.0906\,\mu_N$ \cite{Stone2005}.  Inverting the magnetic moment with the odd-group formula assigns this entire suppression to the neutron spin, even though configuration mixing and proton/neutron core contributions can contribute substantially.  Realistic shell-model calculations of the $^{57}$Fe $14.4\,\keV$ M1 transition explicitly include both proton and neutron excitations and demonstrate that a one-particle interpretation is inadequate \cite{Avignone2018}.  That calculation is not the elastic ground-state finite-$q$ spin structure function needed for the present capture problem, so it cannot simply be substituted for our response.  We therefore retain the factor-two envelope around the odd-group value as a transparent representative uncertainty and carry $|\langle S_n\rangle|=1/6$ separately as a single-particle stress value, explicitly not as an upper bound.  A dedicated $pf$-shell calculation of the elastic finite-$q$ response would supersede both prescriptions.
}

The lighter members of the usual elastic target set, $^{3}$He, $^{13}$C, and $^{17}$O, have stationary ceilings far below $300\,\keV$ and their thermally assisted tails are negligible at the benchmark compared with the nuclei shown in the table; they are therefore omitted from the production endothermic sum rather than being assigned an artificial exact cutoff.

\begin{table*}[t]
\caption{Nuclear inputs and central isotope-by-isotope capture contributions for the neutron-philic benchmark at $m_\chi=1\,\TeV$ and $\delta=300\,\keV$.  Here $f_{\rm iso}$ is the terrestrial isotopic fraction used in the Solar element abundance, $X_{\rm iso}(0)=X_{\rm elem}(0)f_{\rm iso}$ is the central effective Solar mass fraction, and the spin column gives $|\langle S_n\rangle|$ followed by the range used for our illustrative envelope.  ``NBL'' denotes Ref.~\cite{NguyenBlancoLinden2026}, ``BS'' the compilation of Ref.~\cite{Bednyakov2005}, and ``OGM'' the magnetic-moment estimate in Equation~\eqref{eq:ogmspin}.  The finite-$q$ falloff is the common transparent prescription described in the text; dedicated isotope-specific structure functions would supersede it.  \revtextD{$^{\dagger}$The $^{57}$Fe odd-group estimate is the least reliable entry: its measured moment is seven times smaller than the $p_{1/2}$ single-particle value, so the tabulated range is not an upper bound.  The single-particle value $|\langle S_n\rangle|=1/6$, which would raise this contribution to $3.4\times10^{18}\,\mathrm{s}^{-1}$, is carried separately as a stress value rather than an upper bound in Section~\ref{subsec:nsidthermal}.}}
\label{tab:nsidisotopes}
\begin{ruledtabular}
\scriptsize
\begin{tabular}{lcccccc}
Isotope & $f_{\rm iso}$ & $X_{\rm iso}(0)$ & $|\langle S_n\rangle|$ (range) & spin input & $\delta_{\rm cap}^{\rm stat}$ [keV] & $C_A(300\,\keV)$ [$10^{16}\,{\rm s}^{-1}$]\\
\hline
$^{21}$Ne & $2.70\times10^{-3}$ & $3.73\times10^{-6}$ & $0.230$ $(0.170$--$0.290)$ & NBL & $212.9$ & $1.34\times10^{-4}$\\
$^{25}$Mg & $1.00\times10^{-1}$ & $7.76\times10^{-5}$ & $0.300$ $(0.220$--$0.380)$ & NBL & $254.4$ & $1.10$\\
$^{29}$Si & $4.70\times10^{-2}$ & $3.46\times10^{-5}$ & $0.165$ $(0.130$--$0.200)$ & NBL & $296.3$ & $9.09$\\
$^{33}$S  & $7.50\times10^{-3}$ & $2.57\times10^{-6}$ & $0.168$ $(0.084$--$0.252)$ & OGM & $338.4$ & $2.48$\\
$^{43}$Ca & $1.35\times10^{-3}$ & $9.64\times10^{-8}$ & $0.344$ $(0.172$--$0.516)$ & OGM & $445.3$ & $1.49$\\
$^{47}$Ti & $7.44\times10^{-2}$ & $2.59\times10^{-7}$ & $0.210$ $(0.210$--$0.357)$ & BS & $488.6$ & $1.93$\\
$^{49}$Ti & $5.41\times10^{-2}$ & $1.88\times10^{-7}$ & $0.290$ $(0.290$--$0.500)$ & BS & $510.4$ & $2.58$\\
$^{53}$Cr & $9.50\times10^{-2}$ & $1.76\times10^{-6}$ & $0.124$ $(0.062$--$0.186)$ & OGM & $554.2$ & $5.98$\\
$^{57}$Fe & $2.119\times10^{-2}$ & $3.05\times10^{-5}$ & $0.0237$ $(0.0119$--$0.0474)$\revtextD{$^{\,\dagger}$} & OGM & $598.4$ & $6.84$\\
$^{61}$Ni & $1.140\times10^{-2}$ & $8.92\times10^{-7}$ & $0.196$ $(0.098$--$0.294)$ & OGM & $642.9$ & $7.41$\\
$^{67}$Zn & $4.04\times10^{-2}$ & $7.75\times10^{-8}$ & $0.230$ $(0.230$--$0.357)$ & BS & $710.3$ & $0.698$\\
$^{73}$Ge & $7.73\times10^{-2}$ & $2.05\times10^{-8}$ & $0.230$ $(0.230$--$0.496)$ & BS & $778.6$ & $0.148$\\
\end{tabular}
\end{ruledtabular}
\end{table*}

As a separate truncation check we also bound the first ultra-heavy odd-neutron species omitted from the production sum.  We add $^{207}$Pb and $^{235}$U using the deliberately maximal assignment $|\langle S_n\rangle|=J$ and simultaneously remove all finite-$q$ suppression, $F_{\rm SD,A}^2(q)=1$.  This intentionally overestimates their contribution.  At $300\,\keV$ the two isotopes together give only $1.52\times10^{17}\,\mathrm{s}^{-1}$, of which $^{207}$Pb supplies more than $99\%$.  This is $1.9\%$ of the already aggressive no-$q$-suppression, upper-spin diagnostic for the production target set.  Thus truncating the central isotope list at $^{73}$Ge is subleading to the nuclear-response uncertainty that we expose explicitly.

We checked numerical convergence at the two LZ benchmarks.  For the Higgsino at $\delta=377\,\keV$, radial/speed grids $(30,40)$, $(60,80)$, and $(100,120)$ give
\begin{align}
 C_\odot&=1.03301\times10^{23},\quad1.03401\times10^{23},\nonumber\\
 &\hspace{0.7cm}1.03410\times10^{23}\,\mathrm{s}^{-1}.
\end{align}
For the pseudo-Dirac benchmark at $297\,\keV$ the same sequence gives
\begin{align}
 C_\odot&=1.53039\times10^{20},\quad1.53034\times10^{20},\nonumber\\
 &\hspace{0.7cm}1.53059\times10^{20}\,\mathrm{s}^{-1}.
\end{align}
For the neutron-philic finite-temperature benchmark we vary simultaneously the radial, halo-speed, nuclear-speed, target-angle, and center-of-mass-angle quadratures.  The three grids $(24,36,6,8,14)$, $(36,54,8,10,18)$, and $(48,72,10,12,24)$ give
\begin{align}
 C_\odot(300\,\keV)&=3.9719\times10^{17},\quad3.9765\times10^{17},\\
 &\hspace{0.7cm}3.9715\times10^{17}\,\mathrm{s}^{-1}.
\end{align}
The spread is about $0.13\%$.  As an independent kinematic check, for $^{29}$Si at fixed incident speed we lower the target temperature to $10^{-10}\,{\rm K}$; the finite-temperature kernel then reproduces the stationary-target value with ratio $0.9999999999999994$.  These tests verify that the nonzero capture above the stationary $^{29}$Si ceiling is a physical thermal-tail effect rather than a quadrature artifact.
As a method-level cross-check that does not reuse the angular quadrature, we also evaluate the $^{29}$Si kernel at the Solar center with a seeded direct Monte Carlo over $6\times10^5$ target velocities and final-state scattering directions.  At the near-threshold test point used in the package, the Monte Carlo gives $\langle g\sigma_{\rm cap}\rangle_{\rm MC}/\langle g\sigma_{\rm cap}\rangle_{\rm quad}=0.9988\pm0.0055$, in agreement with the deterministic thermal kernel.  A uniform $\pm20\%$ rescaling of the heavy-element abundances in our common-profile approximation changes the neutron-philic capture rate by the same $\pm20\%$, well inside the much broader spin-response envelope.  The numerical quadrature error is therefore sub-percent and far smaller than the heavy-element, nuclear-response, and halo-density systematics.  Production figures use a lighter grid after this convergence test; quoted benchmark values use the converged calculation.

Because the capture rate drops through many orders of magnitude near the kinematic endpoint, exclusion crossings are bracketed on the physical positive curve and interpolated logarithmically; we do not extrapolate a polynomial through the cutoff.

\section{Orbit-averaged post-capture cooling}
\label{app:orbit}

After capture, the DM population is described by a distribution in orbital energy and angular momentum, not by a single radius.  Our semi-analytic treatment reduces this distribution to an effective highly eccentric orbit so that the dependence on the elastic scattering rate can be shown explicitly.  Such a particle spends most of each period near apocenter, where its radial speed is small, but accumulates most of its scattering optical depth during the rapid passage through the dense inner Sun.

For a radial orbit with turning point $r_{\rm max}$, the period is
\begin{equation}
 T(r_{\rm max})=4\int_0^{r_{\rm max}}\frac{dr}{v(r;r_{\rm max})}.
 \label{eq:period}
\end{equation}
where $T(r_{\rm max})$ is the orbital period and $v(r;r_{\rm max})$ is the radial-orbit speed defined in Equation~\eqref{eq:radialv}.  The factor of four accounts for the inward and outward passages through both sides of the radial orbit.  The local loop-induced elastic collision rate with species $A$ is $n_A(r)v\,d\sigma_{A,\rm el}^{\rm loop}$, while the fractional kinetic-energy loss per recoil is $E_R/E_\chi$.  The local specific-energy loss contains an additional factor $1/2$ because $E_\chi=m_\chi v^2/2$.  Averaging that local loss over the four radial legs of a complete period therefore gives
\begin{align}
 \left\langle\frac{dE}{dt}\right\rangle
 &=-\frac{2}{T(r_{\rm max})}\int_0^{r_{\rm max}}dr\,
 \sum_A n_A(r)v^2(r;r_{\rm max})\nonumber\\
 &\hspace{1.8cm}\times\sigma_{A,\rm slow}^{\rm loop}[v(r;r_{\rm max})].
 \label{eq:edotavg}
\end{align}
where $\sigma_{A,\rm slow}^{\rm loop}$ is defined in Equation~\eqref{eq:slowcross} and the angle brackets denote an average over one complete orbit.  The factor is $2/T$, not $4/T$: the four orbit legs are multiplied by the local $1/2$ kinetic-energy factor.  This is the normalization used in the numerical implementation and agrees with Ref.~\cite{PospelovRamani2026}.  Consequently, all Higgsino cooling results quoted here correspond to Equation~\eqref{eq:edotavg} with the $2/T$ prefactor.  Since the orbital specific energy per unit DM mass is $E=\Phi(r_{\rm max})$, where $\Phi$ is the Solar gravitational potential, the turning point evolves according to
\begin{equation}
 \frac{dr_{\rm max}}{dt}=\left(\frac{d\Phi}{dr}\right)^{-1}_{r_{\rm max}}
 \left\langle\frac{dE}{dt}\right\rangle.
 \label{eq:rdot}
\end{equation}
For the elastic-cooling uncertainty scan we initialize the orbit near $r_{\rm max}=0.20R_\odot$, the tree-level inelastic-stall scale relevant to the Higgsino IceCube endpoint.  This number is not a universal capture radius.  It follows from the kinematic condition in Equation~\eqref{eq:stallradius}: for uranium, the last conservative inelastic target used by Ref.~\cite{PospelovRamani2026}, our BS05 potential gives $r_U^{\rm kin}\simeq0.20R_\odot$ for $\delta\simeq0.50\,\mathrm{MeV}$.  The no-elastic curve in Figure~\ref{fig:hann} uses the full $\delta$-dependent $r_U^{\rm kin}(\delta)$; the fixed $0.20R_\odot$ starting point is used only for the subsequent elastic-cooling kernels, for which the relevant endpoint range changes the initial radius by only a few percent.  For the pseudo-Dirac model we do not identify any single $r_A^{\rm kin}$ with a final distribution; Section~\ref{subsec:pseudo_reexcitation} instead includes thermally assisted re-excitation in the explicit fixed-orbit re-excitation calculation.

The SI slowing term sums coherently over all nuclei with the Helm form factor.  The SD term is dominated by H in our simplified treatment.  Numerically, the turning-point integrals contain the familiar square-root behavior $dt=dr/v_r$ as $r\to r_{\rm max}$.  We remove this endpoint behavior with the substitution $r=r_{\rm max}\sin^2\theta$ and evaluate the orbit with Gauss--Legendre quadrature.  The SI slowing kernel is additionally cached in velocity bins of $10\,\kms$; comparison with finer binning changes the final cooling radius far below the loop and SSM uncertainties.

The radial approximation has two competing limitations.  It maximizes central passages for a given apocenter, but it also replaces a distribution of orbital energies by one effective value.  A Monte Carlo treatment should draw the scattering angle and recoil at every collision, update both orbital energy and angular momentum, and reconstruct $n_\chi(r,t)$ directly.  The present method is best interpreted as a semi-analytic map between elastic cross section and characteristic radius, which is exactly the uncertainty explored in Figure~\ref{fig:hloop}.

\section{Two-state pseudo-Dirac evolution and fixed-orbit re-excitation calculation}
\label{app:twostate}

Equations~\eqref{eq:N1}--\eqref{eq:N2} are the appropriate population equations for the minimal off-diagonal vector model, but in the exact Solar problem their coefficients are functionals of the orbital distributions.  A particle is specified by orbital energy $E$, angular momentum $L$, and internal state.  Every up- or down-scatter changes all three.  A complete calculation should therefore evolve a transition matrix in $(E,L,i)$ space, including the Solar nuclear velocity distribution, in the spirit of the inelastic-DM Monte Carlo treatment of Refs.~\cite{Blennow2018,BlennowErratum2019}.  We do not replace that calculation by a single claimed final radius.

For comparison with the previous zero-re-excitation limit, if $\Gamma_{12}=0$ and $\Gamma_{21}$ is much faster than secular population evolution, the excited state is quasi-steady,
\begin{equation}
 N_2\simeq\frac{C_\odot}{\Gamma_{21}+C_{12}N_1}.
 \label{eq:N2steady_zero}
\end{equation}
Defining
\begin{equation}
 y=\frac{C_{12}N_1}{\Gamma_{21}},\qquad
 \kappa=\frac{C_{12}C_\odot t}{\Gamma_{21}},
\end{equation}
one obtains
\begin{equation}
 -y-2\ln(1-y)=\kappa,
 \qquad
 \Gamma_A=C_\odot\frac{y}{1+y}.
 \label{eq:zero_reexc_solution}
\end{equation}
This remains a useful limiting check, but it is not the physical approximation used for an extended pseudo-Dirac orbit once thermally assisted re-excitation is allowed.

The nonzero-$\Gamma_{12}$ fixed-orbit test used in the main text evaluates the orbit-averaged rates in Equation~\eqref{eq:pdm_orbit_transition} for a sequence of fixed radial apocenters.  The Solar nuclei are Maxwellian with one-dimensional dispersion
\begin{equation}
 \sigma_{v,A}(r)=\sqrt{\frac{T(r)}{m_A}},
\end{equation}
and the thermal average is performed over the exact relative-speed distribution for a fixed DM speed.  Near the Ni stationary-target closure radius this procedure yields
\begin{equation}
 \Gamma_{12}=1.10\times10^{-10}\,\mathrm{s}^{-1},
 \qquad
 \Gamma_{21}=1.92\times10^{-7}\,\mathrm{s}^{-1}.
\end{equation}
The up-scattering rate is numerically threshold-sensitive, so we explicitly checked the quadrature.  Increasing the orbit/relative-speed Gauss--Legendre orders from $(64,56)$ to $(80,64)$ and $(120,88)$ gives $\Gamma_{12}=1.12$, $1.10$, and $1.10\times10^{-10}\,\mathrm{s}^{-1}$, respectively, while $\Gamma_{21}$ is stable at $1.921\times10^{-7}\,\mathrm{s}^{-1}$.  The quoted benchmark is therefore not a quadrature artifact.

Because the resulting annihilation rate is always much smaller than $C_\odot/2$, annihilation removes a negligible fraction of the accumulated particles and $N_1+N_2\simeq C_\odot t$ is self-consistent.  The rapid-transition quasi-steady solution then becomes
\begin{equation}
 N_2\simeq\frac{C_\odot+\Gamma_{12}N_1}{\Gamma_{21}+C_{12}N_1},
 \qquad N_1\simeq C_\odot t,
\end{equation}
which directly gives Equation~\eqref{eq:pdm_gamma_reexc}.  The largest benchmark fixed-orbit rate after the coherent-species update, $4.39\times10^{14}\,\mathrm{s}^{-1}$, satisfies $2\Gamma_A/C_\odot\simeq5.7\times10^{-6}$, validating the neglect of annihilation in the accumulated total number at the level relevant here.  As a separate check of the quasi-steady reduction, we directly integrate Equations~\eqref{eq:N1}--\eqref{eq:N2} with fixed coefficients at the Ni reference orbit and at $r_{\rm max}=0.62R_\odot$.  The resulting present-day annihilation rates differ from Equation~\eqref{eq:pdm_gamma_reexc} by only $0.11\%$ and $1.4\%$, respectively.  The dominant uncertainty is therefore the evolution of the orbital distribution, not the algebraic treatment of the two-state populations at fixed orbit.

Two qualifications are important.  First, exothermic down-scattering releases the splitting energy and can change the orbit substantially or eject a marginally bound particle.  We neglect state-changing ejection in the fixed-orbit calculation; retaining all particles tends to maximize the available annihilating population.  \revtextC{Section~\ref{subsec:pseudo_cooling} quantifies this: the maximum kinetic gain per down-scatter is about $250\,\keV$ against several MeV of binding, so ejection does not occur for the orbits of interest.}  Second, the fixed radial orbit does not update $E$ and $L$ after every transition.  Section~\ref{subsec:pseudo_cooling} adds a semi-analytic energy-drift diagnostic: a stationary-target cycle loses the sum of the two recoil energies and an effective radial-orbit equation suggests contraction toward the scale at which $\Gamma_{12}t_\odot\sim1$.  Because this estimate uses a representative Fe recoil loss, radial orbits, and a one-zone overlap prescription, it is not a reconstruction of $n_1(E,L)$ and $n_2(E,L)$.  The fixed-orbit curve should therefore be interpreted as the robust envelope, while Equation~\eqref{eq:pdm_gamma_today} is only an illustrative present-day estimate.

Thermal detailed balance applies only after genuine thermalization.  In that regime $n_2/n_1\sim\exp(-\delta/T_c)$, which is approximately $6\times10^{-100}$ at the benchmark.  For a nonthermal eccentric orbit, by contrast, the relevant quantity is the orbit-averaged $\Gamma_{12}$ in Equation~\eqref{eq:pdm_orbit_transition}.  This distinction is the reason the main text no longer uses the Boltzmann factor to justify setting $\Gamma_{12}=0$ for the extended-orbit case.

\section{Neutrino spectra and a full IceCube recast}
\label{app:neutrino}

A channel-exact IceCube limit requires a source spectrum for every final state.  For the Higgsino one should generate the $WW$ and $ZZ$ yields with a common electroweak/Sommerfeld calculation.  For the pseudo-Dirac model one should generate all six $q\bar q$ channels with the branching fractions in Equation~\eqref{eq:pdmBR}.  For the neutron-philic benchmark the final state is not fixed by the LZ operator; it must be supplied by a chosen ultraviolet completion.  Solar propagation must include interactions of the primary products before decay.  Long-lived light hadrons are strongly degraded and produce an important stopped-meson neutrino component at MeV energies \cite{RottSiegalGaskinsBeacom2013,BernalMartinAlboPalomaresRuiz2013}.  This does not make the high-energy yield identically zero: higher-order electroweak radiation from TeV-scale light-fermion primaries produces weak gauge bosons and hence a hard neutrino component potentially visible to IceCube \cite{IbarraTotzauerWild2014}.  Modern neutrino-yield calculations include these electroweak corrections, and the ten-year IceCube analysis notes their significant impact on high-mass hadronic limits \cite{IceCubeSolar2025,Bauer2021}.  The main text therefore uses the IceCube $b\bar b$ upper limit as a soft-hadronic proxy and the $WW$ upper limit as a hard comparison, not as a rigorous bracket for every completion.

The resulting neutrino density matrices are propagated through the Sun including charged- and neutral-current interactions, tau regeneration, matter effects, and flavor oscillations, followed by vacuum propagation to Earth \cite{Blennow2008,Charon2020}.  Electroweak radiation is relevant for TeV-scale primaries \cite{Bauer2021}.

If $\mathcal{A}_{\rm eff}^\alpha(E_\nu)$ denotes the IceCube effective area for neutrino flavor $\alpha\in\{e,\mu,\tau\}$ and neutrino energy $E_\nu$, the predicted event count can be written schematically as
\begin{equation}
 N_{\rm sig}=\frac{\Gamma_A}{4\pi D_\odot^2}
 \sum_{f,\alpha}B_f\int dE_\nu\,
 \mathcal{A}_{\rm eff}^\alpha(E_\nu)\frac{dN_{\nu_\alpha,f}}{dE_\nu}.
 \label{eq:IceCubeCount}
\end{equation}
Here $N_{\rm sig}$ is the expected number of detected signal events.  A likelihood upper limit $N_{\rm sig}^{\rm lim}$ on that event count, quoted at the chosen confidence level, then gives
\begin{equation}
 \Gamma_A<\Gamma_A^{\rm lim}
 =\frac{4\pi D_\odot^2N_{\rm sig}^{\rm lim}}
 {\displaystyle\sum_{f,\alpha}B_f\int dE_\nu\,
 \mathcal{A}_{\rm eff}^\alpha(E_\nu)dN_{\nu_\alpha,f}/dE_\nu}.
 \label{eq:fulllimit}
\end{equation}
The simple rescalings used in the main text correspond to changing the denominator of Equation~\eqref{eq:fulllimit}.  For the Higgsino this is an order-one correction; for the pseudo-Dirac benchmark it is negligible compared with the many-orders-of-magnitude suppression in $\Gamma_A$.  For the neutron-philic benchmark the final state is a model-dependent input rather than a capture uncertainty.  Light-quark-dominated spectra are softened by Solar hadronic interactions but retain a hard electroweak-radiative component, while electroweak or heavy-flavor spectra are harder.  A completion-specific detector recast is therefore required for a precision bound.  This separation is useful because the absolute bound $\Gamma_A\leq C_\odot/2$ allows any alternative spectral template to be tested without rerunning the Solar capture calculation.

\bibliography{main}

\end{document}